\documentclass[11pt,a4paper]{article}

\usepackage{setspace}

\usepackage{cite}

\usepackage[left=1.5cm,right=1.5cm,top=3.5cm,bottom=3.5cm]{geometry}

\usepackage{float}
\usepackage[latin1]{inputenc}
\usepackage{amsmath}
\usepackage{amsfonts}
\usepackage{amssymb}
\usepackage{mathtools}
\usepackage{mathrsfs}
\usepackage{amsmath}
\usepackage{comment}
\usepackage{graphicx}
\usepackage{epsf,epsfig}
\usepackage{esint}
\usepackage{soul}

\usepackage{xcolor}
\usepackage{colortbl}
\definecolor{dark-red}{rgb}{0.80,0.12,0.12} 
\definecolor{dark-green}{rgb}{0,0.60,0.30} 
\definecolor{dark-blue}{rgb}{0,0.1,0.7} 

\usepackage{hyperref}
\hypersetup{
    colorlinks=true,
    linkcolor=dark-red,
    filecolor=blue,      
    urlcolor=cyan,
    citecolor=dark-green
  }

\numberwithin{equation}{section}
\makeatletter 
\@addtoreset{equation}{section} 
\makeatother

\usepackage[english]{babel}

\usepackage{mdframed}

\DeclareMathAlphabet{\mathpzc}{OT1}{pzc}{m}{it}

\usepackage{tikz}
\usetikzlibrary{trees}
\usetikzlibrary{snakes}
\usetikzlibrary{arrows}
\usepackage{color}
\usetikzlibrary{arrows.meta}
\usetikzlibrary{decorations.pathmorphing,arrows.meta,shadows,calc}

\def\intsumfour{\mathop{\sum \kern-1.2em \int}\limits_{1234}}
\def\intsumfive{\mathop{\sum \kern-1.2em \int}\limits_{01234}}

\def\dlta {\tilde \delta}
\def \be  	{\begin{equation}}
\def \ee  	{\end{equation}}
\def \ba  	{\begin{eqnarray*}}
\def \ea  	{\end{eqnarray*}}

\def\beq 	{\begin{equation}}
\def\eeq	{\end{equation}}
\def\bea 	{\begin{eqnarray*}}
\def\eea	{\end{eqnarray*}}
\def\ulmb {\underline{\lambda}}

\def\k{\kappa}

\def\lm{l}

\begin{document}

\hfill
	\vspace{18pt}
	\begin{center}
		{\Large 
			\textbf{
                          Multi-particle correlators with higher KK modes I: \\[.2cm]  a bootstrap approach
                        }}
		
	\end{center}

		\vspace{8pt}
	\begin{center}
		{\textsl{Francesco Aprile}}
		\vspace{.0cm}
				
		\textit{\small Departamento de Fisica Teorica \& IPARCOS,  Facultad de Ciencias Fisicas, \\Universidad Complutense de Madrid, 28040 Madrid, Spain} \\ 
		
		\vspace{.5cm}
		
		{\textsl{Stefano Giusto}}
		\vspace{.0cm}

		\textit{\small Dipartimento di Fisica,  Universit\`a di Genova, Via Dodecaneso 33, 16146, Genoa, Italy} \\  \vspace{0pt}
		
		\textit{\small I.N.F.N. Sezione di Genova,
			Via Dodecaneso 33, 16146, Genoa, Italy}\\
			
			\vspace{.5cm}

	{\textsl{Rodolfo Russo\, and\, Jo\~ao Vilas Boas}}
		\vspace{.0cm}
				
		\textit{\small School of Mathematical Sciences, Queen Mary University of London, \\Mile End Road, London, E1 4NS, United Kingdom} \\ 
					
	\end{center}

	\vspace{12pt}

	\begin{center}
		\textbf{Abstract}
	\end{center}
	
	\vspace{4pt} 
	\begin{center}
	\begin{minipage}{15.2cm}
	{\small
	\baselineskip=15pt
	\parskip=5pt
	\noindent 
		We bootstrap tree-level supergravity four-point correlators on AdS$_5\times$S$^5$ 
		with one external half-BPS double-particle operator and three half-BPS 
		single-particle operators. Our only input is the consistency of the operator 
		product expansion of $SU(N)$ ${\cal N} = 4$ super Yang-Mills theory at large $N$ 
		and large 't Hooft coupling. Even though the leading order OPE  does 
		not close on double-particle operators, but involves triple-particle operators,  
		the CFT data of the double-particle operators, both long and protected, 
		is sufficient to uniquely fix the correlators. We then verify that our results for the \emph{four}-point correlators with one    double-particle and three single-particle operators are reproduced by the appropriate double-particle limit 
		of the \emph{five}-point tree-level correlators of 
		single-particle operators, with arbitrary Kaluza-Klein levels, recently conjectured in
		\href{https://arxiv.org/abs/2507.14124}{2507.14124}. Our study thus provides further evidence for the latter result.
		
		}
	\end{minipage}
	\end{center}	
		
		\vspace{1cm}

		\thispagestyle{empty}

		\vfill
		\vskip 5.mm
		\hrule width 5.cm
		\vskip 2.mm
		{
			\noindent {\scriptsize e-mails: faprile@ucm.es, stefano.giusto@ge.infn.it, r.russo@qmul.ac.uk, j.m.v.boas@qmul.ac.uk}
		}

		\setcounter{footnote}{0}
		\setcounter{page}{0}

		
		\baselineskip=17pt
		\parskip=5pt
		
		\newpage

{		
\hypersetup{linkcolor=black}
\setcounter{tocdepth}{3}
\tableofcontents
}

\vspace{.7cm}

\section{Introduction}

The study of holographic correlators in ${\cal N}=4$ Super Yang-Mills (SYM), at large $N$ and large 't Hooft coupling, has
been the subject of intense work since the early days of the AdS/CFT correspondence 
\cite{Maldacena:1997re,Witten:1998qj}. In this regime, the dual gravitational description is well approximated by supergravity and various techniques have been developed to evaluate correlators of half-BPS primary operators ${\cal O}_{p}$ dual to the single-particle Kaluza-Klein (KK) modes of the sphere. 
The first result becoming available for all KK  levels was the one for three-point correlators \cite{Lee:1998bxa}. 
This was derived by computing an AdS$_5$ effective action up to cubic couplings,  
and served as a non-trivial test of the AdS/CFT correspondence because of 
a non-renormalization theorem from weak to strong coupling. 
However, it soon became  clear that applying this approach at higher points is in general 
too cumbersome to do in practice, and in fact,  only a handful of four-point  correlators 
\cite{Arutyunov:2000py,Arutyunov:2002fh,Arutyunov:2003ae,Berdichevsky:2007xd,Uruchurtu:2008kp,Uruchurtu:2011wh}
were computed by using the AdS$_5$ effective action \cite{Arutyunov:1999fb}.

A breakthrough in the study of holographic correlators at four points came from the 
work of Rastelli-Zhou \cite{Rastelli:2016nze}.  The essence of their method \cite{Rastelli:2017udc} 
was to bootstrap $\langle{\cal O}_{p_1} {\cal O}_{p_2}{\cal O}_{p_3} {\cal O}_{p_4}\rangle$ at 
tree-level by writing an ansatz of physical Witten diagrams (exchange and contact) and 
in order to fix the numerical coefficients, impose superconformal symmetry, in particular, the  
partial non-renormalization theorem \cite{Eden:2000bk,Heslop:2002hp}. This theorem 
predicts that any four-point correlator of half-BPS scalar primaries admits a special 
rewriting in terms of a reduced correlator. It is a powerful constraint because the AdS$_5$ 
effective action has no explicit information about it.  As it turned out, the bootstrap problem  
has a unique solution for all Kaluza-Klein levels (up to an overall constant fixed in \cite{Aprile:2018efk}).
The solution is particularly simple in Mellin space. 
Further studies of the Rastelli-Zhou formula led to the discovery of an accidental 
10d conformal symmetry \cite{Caron-Huot:2018kta} and to the AdS$\times$S 
Mellin space formalism  \cite{Aprile:2020luw}.

However, even in the supergravity regime, four-point correlators do not capture all the CFT data of 
local operators.  First of all, the BPS spectrum not only contains the single-particle operators dual to 
the elementary fields on the gravity side, but also multi-particle operators that can be obtained by taking 
the OPE of the single-particle fields mentioned above. Importantly, since single-particle 
operators \cite{Aprile:2020uxk} exist as long as their dimension $\Delta\leq N$, heavy half-BPS operators in the CFT, 
whose dimension scales with the central charge $\sim N^2$, are necessarily built out of 
(superposition of) mutually BPS multi-particles. 

A way to extract information about multi-particle operators is to study the OPE in multi-point 
single-particle correlators. Bootstrap techniques have been applied to multi-point correlators as well,
and significant progress was made at five points \cite{Goncalves:2019znr,Goncalves:2023oyx,VilasBoas:2025vvw}, 
in which case the successful strategy was to impose, on top of superconformal symmetry, 
properties of the Mellin amplitudes under factorization \cite{Goncalves:2014rfa}. Following these 
developments, and taking advantage of the AdS$\times$S Mellin space formalism, a closed formula for  
the tree-level \emph{five}-point correlators of single-particle operators for all KK levels has recently 
been proposed by the authors of \cite{Fernandes:2025eqe}. Then, a bootstrap approach has been successfully 
applied to derive the first six-point correlator $\langle{\cal O}_2 {\cal O}_2 {\cal O}_2 {\cal O}_2 {\cal O}_2 {\cal O}_2\rangle$ 
in the supergravity regime~\cite{Alday:2023kfm}. Not surprisingly, such higher-point correlators 
are substantially more involved than the well-studied four-point correlators, as the kinematics alone are already more complicated. For example, at five points, there are five independent conformal invariant cross ratios, 
and at six-points there are nine cross ratios.

Here we take an alternative approach which preserves the simple kinematics of \emph{four-point correlators}, 
but still probes the dynamical CFT data encoded in multi-point correlators: we study correlators involving 
also \emph{multi-particle local operators}. Our general program is to compute four-point correlators with both 
single- and multi-particle operators and extract from them new information on the holographic 
regime of ${\cal N}=4$ SYM. As expected, these correlators are directly related to  multi-point 
single-particle correlators in the limit in which two or more single-particle operators are brought together, 
yielding a bound state built out of single-particle operators. In this paper, we focus on the simplest 
class of such a new class of correlators, i.e. the \emph{holographic four-point correlators with three 
half-BPS single-particle operators and one double-particle half-BPS bound state}.
\begin{equation}\label{intro4pt}
\langle\, [{\cal O}_{r_1}{\cal O}_{r_2}]({z}_1) {\cal O}_{p_2}({z}_2){\cal O}_{p_3}({z}_3) {\cal O}_{p_4} ({z}_4)\rangle\;.
\end{equation}
Of course, these results are directly related to the five-point correlators we mentioned above,
\begin{equation}\label{intro5pt}
\langle {\cal O}_{r_1}\!({z}_0) {\cal O}_{r_2}\!({z}_1) {\cal O}_{p_2}({z}_2){\cal O}_{p_3}({z}_3) {\cal O}_{p_4} ({z}_4)\rangle.
\end{equation}
Since the authors of \cite{Fernandes:2025eqe} have already given a closed-form 
expression for the latter, in this paper, we will directly use our results at four points to 
provide a further test of their formula in the appropriate double-particle OPE limit.

At this point, it is worth emphasizing that four-point correlators with half-BPS bound states provide a natural set 
of observables for several different reasons. As mentioned above, even in the BPS sector, operators 
with $\Delta > N$ are necessarily multi-particle bound states and thus the study of correlators involving such operators \cite{Giusto:2019pxc,Ceplak:2021wzz,Bissi:2021hjk,Ma:2022ihn,Abajian:2023jye,Aprile:2024lwy,Bissi:2024tqf,Aprile:2025hlt} 
is a crucial piece of information for all holographic CFTs. In particular heavy operators with $\Delta \sim N^2$ are 
dual to backreacted geometries (for ${\cal N}=4$ SYM see the prototypical case discussed in~\cite{Lin:2004nb}) 
which makes this class of correlators particularly interesting in the holographic analysis of asymptotically AdS black holes.
Secondly, all four-point correlators of half-BPS operators satisfy the same partial non-renormalization 
theorem \cite{Eden:2000bk,Heslop:2002hp}, and so the constraints from superconformal symmetry can be made
manifest from the start, differently from the current status for the multi-point correlators.\footnote{Also, the superconformal blocks at four points are well 
understood \cite{Doobary:2015gia}, and this facilitates the study of new CFT data.}

In this paper, we will compute the leading tree-level contribution to 
$\langle [{\cal O}_{r_1}{\cal O}_{r_2}]{\cal O}_s {\cal O}_p {\cal O}_q \rangle$ by
making use of the so-called ``double-particle bootstrap". 
Before introducing what this is, and why it is alternative to 
the approach that leads to \cite{Fernandes:2025eqe}, we would like to mention another method 
that can be used for our purposes. This is the method of ``coherent-state geometries"
put forward in  \cite{Ceplak:2021wzz,Giusto:2019pxc,Aprile:2024lwy,Turton:2025svk,Aprile:2025hlt,Turton:2025cnn, Turton:2024afd}. 
The first explicit examples of dynamical correlators involving multi-particle operators in ${\cal N}=4$ SYM were derived 
through this geometric approach~\cite{Aprile:2024lwy,Aprile:2025hlt}. However, the method was applied 
only at the lowest KK level and, in this case, the first non-trivial correlator involves two double-particle 
operators, since for $r_i=p_i=2$ the correlator in~\eqref{intro4pt} is protected and determined by 
its free-theory contribution. We will discuss in a separate work~\cite{KKII} how to extend the geometric approach 
to include the KK descendants of the single-particle states and show that it yields exactly the same results obtained in this paper for~\eqref{intro4pt} with $r_i=2$ and  $p_i>2$.
Here, we decided to keep the discussion self-contained on the CFT side, focusing
on the four-point correlators~\eqref{intro4pt}, their five-point ancestors, and the 
double-particle bootstrap approach described below.

{\bf The double-particle bootstrap.} 
With this terminology, we shall  refer to the collection of constraints on four-point correlators 
that can be determined from the OPE and the CFT data (anomalous dimensions and three-point couplings) of double-particle operators \cite{Aprile:2018efk}. The basic observation is that even 
though the space of double-particle operators is degenerate, their CFT data can be unmixed from known 
four-point single particle correlators, see e.g.~\cite{Aprile:2017xsp}, 
and then used to constrain multi-point correlators. Since the leading-order OPE of two single-particle operators closes 
on double-particle operators, the first application of the double-particle bootstrap 
goes back to the study of one-loop amplitudes of four single-particle operators 
\cite{Aprile:2017bgs,Alday:2017xua,Aprile:2017qoy,Alday:2019nin,Aprile:2019rep}. 
The idea of the double-particle bootstrap is, of course, more general and can be translated to 
the study of correlators with one external double-particle operator and three single-particle operators, 
as we now describe.

The four-point correlators in \eqref{intro4pt} have the following large $N$ expansion, 
\begin{equation}\label{introequa5pt}
\langle [{\cal O}_{r_1}{\cal O}_{r_2}]{\cal O}_s {\cal O}_p {\cal O}_q \rangle
= \ \underline{ O(\tfrac{1}{N^{1}}) }_{\ \emph{disconnected} }  \ +\     \underline{ O(\tfrac{1}{N^3}) \ +\  \ldots  }_{\ \emph{connected}}\,.
\end{equation}
The leading-order OPE (disconnected diagrams)
common to $ [{\cal O}_{r_1}{\cal O}_{r_2}]\times{\cal O}_s$ and ${\cal O}_p\times {\cal O}_q$ 
does not close on double-particle operators because of the triple-particle operators. However, 
we will prove that there exists a window of CFT data that is fully determined by the double-particle 
operators, and it will turn out that this window is enough to solve (and over-constrain) the 
leading connected contribution, $O(\tfrac{1}{N^3})$, at strong 't Hooft coupling. The result is 
always expressible as a sum of contact AdS diagrams, i.e.~D-functions.

A useful tool in the double-particle bootstrap is a formula for the superblock decomposition of any free-theory diagram in ${\cal N}=4$ SYM, given in \cite{Aprile:2025nta}. The authors of  \cite{Aprile:2025nta} also  
showed the consistency of the double-particle OPE for some families of four-point correlators, 
e.g.~$\langle {\cal O}^2_2{\cal O}^2_2 {\cal O}_p {\cal O}_q \rangle$ at $O(\frac{1}{N^2})$
(see also \cite{Bissi:2024tqf}). Our approach here generalises the one in  \cite{Aprile:2025nta} 
to $\langle [{\cal O}_{r_1}{\cal O}_{r_2}]{\cal O}_s {\cal O}_p {\cal O}_q \rangle$ at order $O(\frac{1}{N^3})$.

{\bf This paper is organized as follows:} In section \ref{section_background} we review some 
background material, in particular, we set up notation for the four-point correlators, and
we review some known results about the double-particle spectrum \cite{Aprile:2019rep},
that will be crucial to understand the double-particle bootstrap. 

In section \ref{section_analitic_boot} 
we explain how to derive the OPE predictions that the double-particle bootstrap can make 
regarding $\langle [{\cal O}_{r_1}{\cal O}_{r_2}]{\cal O}_s {\cal O}_p {\cal O}_q \rangle$, 
and then implement an algorithm that uses these predictions to bootstrap 
the correlators of interest. We will discuss in detail  the simple test cases, 
$\langle {\cal O}^2_2{\cal O}_2 {\cal O}_p {\cal O}_p \rangle$. 
Then, we give results for more general  families of correlators,  such as 
\begin{equation}
\langle {\cal O}^2_r{\cal O}_2 {\cal O}_p {\cal O}_{p} \rangle\qquad;\qquad 
\langle [{\cal O}_2{\cal O}_4] {\cal O}_2 {\cal O}_p {\cal O}_{p} \rangle\qquad;\qquad 
\langle {\cal O}^2_2{\cal O}_s {\cal O}_p {\cal O}_{p+s-2} \rangle\,.
\end{equation}
It will become clear how these are representative of the general case. 

In section \ref{sec_MellinAmps} we discuss our bootstrap results from the point of view of 
Mellin amplitudes \cite{Mack:2009mi}.  While Mellin amplitudes ${\cal M}_{p_1p_2p_3p_4}$ 
for the four-point correlators of single-particle operators are well understood, in particular, 
their flat space limit \cite{Penedones:2010ue}, our motivation here is to explore \emph{a} 
definition of Mellin amplitudes when the external operators are double-particle, and test 
our proposal. For concreteness, we will focus on the two cases
\begin{equation}\label{intro_summary_tree}
{\cal M}_{[2^2],p_2,p_3,p_4}\quad \forall\,p_2,p_3,p_4\qquad\qquad ;
\qquad\qquad  {\cal M}_{[r^2],2,p,p}\quad \forall\,r,p.
\end{equation}
We will compute the flat space limit of these tree-level amplitudes and show
that the limit of large Mellin variables has a uniform scaling for all different correlators  in \eqref{intro_summary_tree},
and moreover, it is sub-leading by one power compared to the flat space limit of the
single-particle correlators at tree-level, ${\cal M}_{p_1p_2p_3p_4}$.

In section \ref{doublelimitsection}, we introduce the double-particle limit on the single-particle 
five-point correlators, and use our bootstrap results to test the five-point formula of \cite{Fernandes:2025eqe}. 
The double-particle limit is engineered by taking the limit of coincident points of two single-particle operators, 
first in the R-symmetry space, so both operators are made out of the same polarization vector, and then in space-time.
On the five-point correlator, the latter translates into a limit on D-functions. Mathematically, 
this limit is obvious in position space, while in Mellin space it is performed by applying twice the 
first Barnes' lemma~\cite{Ma:2022ihn}. Physically, two things must happen in the double-particle 
limit: the partial non-renormalization theorem must be satisfied and the reduced correlator obtained 
in the limit must coincide with the one computed by the double-particle bootstrap. Both these statements 
are very non-trivial because, firstly, there is no classification of superconformal invariants 
at five-points, and in fact the formula proposed in \cite{Fernandes:2025eqe} uses a skeleton 
expansion, and secondly, the double-particle bootstrap is alternative to the bootstrap 
approach used in \cite{Fernandes:2025eqe}.

In section \ref{concl_discuss}, we give a summary of our results 
and we discuss generalisations and future directions. 
Finally, we included three appendices that contain 
some further technical details on the bootstrap approach introduced in this paper, 
and the OPE analysis of related higher-point and multi-particle correlators.

\section{Background}\label{section_background}

\subsection{Four-point correlators of half-BPS multiplets}\label{section_4pt_gen}

Half-BPS scalar primary operators in ${\cal N}=4$ SYM belong to the symmetric traceless representation 
of the R-symmetry group, $su(4)$, whose Dynkin labels are  $[0,p,0]$. 
These operators are protected and have exact dimension $\Delta=p$.
A simple construction of the space of all half-BPS operators can be given in free theory.  
Starting from the elementary scalar field in the ${\cal N}=4$ multiplet in the adjoint of $SU(N)$,  
$\phi^{I=1,\ldots, 6}(\vec{X})$, introduce a \emph{null} vector $Y^I$ and define the index-free scalar field
\begin{equation}\label{elementaryscalar}
\phi({z})= Y^I \phi_I(\vec{X})\,\qquad;\qquad {z}\equiv(\vec{X},\vec{Y})\;.
\end{equation}
The single-trace operators 
\begin{equation}
T_p({z}):={\rm Tr}\big(\,\phi({z})^{p}\big)\;,
\end{equation}
have dimension $\Delta=p$ and correspond to the highest weight
state of the (symmetric traceless) rep with Dynkin labels $[0,p,0]$. 
More generally, one can build half-BPS multi-trace operators,
\begin{equation}
T_{p_1,p_2,\ldots,p_n}(z)= T_{p_1}(z)T_{p_2}(z)\ldots T_{p_n}(z)\,\qquad,\qquad \Delta=\sum_{i} p_i\,,
\end{equation}
and these also belong to the rep $[0,p,0]$ where $\Delta=p$.
The space of all half-BPS operators is thus degenerate since for a given dimension $\Delta$ there 
are as many operators as the partitions of $\Delta$. The single-particle basis is that particular subspace of 
operators that is dual to the Kaluza-Klein scalar fields of the S$^5$ compactification 
of type-IIB supergravity on AdS$_5$ \cite{Kim:1985ez}. The precise definition of the single-particle 
basis was given in \cite{Aprile:2020uxk}: these are the operators ${\cal O}_p$ that are orthogonal 
to all multi-trace operators in the two-point function metric. Their general form is
\begin{equation}
{\cal O}_p = T_p + \sum_{\underline{q}\,\vdash p} c_{\underline{q}}(p,N) \, T_{\underline{q}}\;,
\end{equation}
where $\underline{q}$ is a partition of $p$, and  an \emph{explicit} formula for the $c_{\underline{q}}$ 
was given in \cite[\emph{eq}.\!~(18)]{Aprile:2020uxk}. The multi-particle admixture is sub-leading 
in the large $N$ limit, thus at leading order ${\cal O}_p$ reduces to $T_p$. However, 
beyond the leading order, the multi-particle admixture is a crucial piece of information for 
precision holography \cite{Turton:2025svk}. It is worth mentioning that, by definition, the single-particle 
operators exist only for $p\leq N$, after which the operators are necessarily multi-particle operators. 
This property has long been expected from the AdS/CFT duality and the results 
of \cite{McGreevy:2000cw}, i.e.~that ``particles should not be bigger than the sphere that contains it".

The elementary propagator $\langle\phi(z_1)\phi(z_2)\rangle$, with $z_i=(\vec{X}_1,\vec{Y}_1)$, is given by 
\begin{equation}
\langle\phi^a_b(z_1)\phi^n_m(z_2)\rangle= \left(\delta^a_m \delta^b_m-\frac{1}{N}\delta^a_b\delta^n_m\right)g_{12} \qquad;\qquad
g_{12}= \frac{\vec{Y}^{2}_{12}}{\vec{X}^{2}_{12}}\qquad;\qquad \vec{Y}_{12}^{2}\equiv \vec{Y}_1\cdot \vec{Y}_2\;.
\end{equation}
Now, let ${\cal O}_{\Delta}$ be \emph{any} half-BPS operator of dimension $\Delta$, the two-point function is
\begin{equation}
\langle {\cal O}_{\Delta}(z_1) {\cal O}_{\Delta}(z_2)\rangle  = g_{12}^\Delta\  f_{O,\Delta}(N) \qquad;\qquad |{\cal O}_{\Delta}|^2\equiv  f_{O,\Delta}(N)\;,
\end{equation}
where $|{\cal O}_{\Delta}|^2$ will be our shorthand for the two-point function normalization $f_{O,\Delta}(N)$.
If there is a space of degenerate operators, then there is also a two-point function metric. 

Our conventions for the four-point correlators will be as follows. 
First, we will extract a certain prefactor of free propagators
and two-point function normalizations, that we call ${\cal P}_{\vec{p}}$. 
This yields
\begin{equation}\label{basic_c_1}
\langle {\cal O}_{p_1}(z_1) {\cal O}_{p_2}(z_2) {\cal O}_{p_3}(z_3) {\cal O}_{p_4}(z_4)\rangle = {\cal P}_{\vec{p}}[\,g_{ij}\,]\  {\cal C}_{\vec{p}}(U,V,\hat{\sigma},\hat{\tau})\;,
\end{equation}
where ${\cal C}_{\vec{p}}(U,V,\hat{\sigma},\hat{\tau})$ is the \emph{normalized} correlator that is a function of  the spacetime cross ratios $U,V$, and the R-symmetry cross ratios $\hat{\sigma},\hat{\tau}$:
\begin{equation}\label{def_cross_ratiosFA}
U=\frac{\vec{X}^{2}_{12}\vec{X}^{2}_{34}}{  \vec{X}^{2}_{13} \vec{X}^{2}_{24} } \qquad;\qquad  
V=\frac{\vec{X}^{2}_{14} \vec{X}^{2}_{23} }{  \vec{X}^{2}_{13}\vec{X}^{2}_{24} }  \qquad;\qquad  
\hat{\sigma}=\frac{ \vec{Y}^{2}_{13} \vec{Y}^{2}_{24} }{ \vec{Y}^{2}_{12} \vec{X}^{2}_{34} }\qquad;\qquad 
\hat{\tau}=\frac{ \vec{Y}^{2}_{14} \vec{Y}^{2}_{23} }{ \vec{Y}^{2}_{12} \vec{Y}^{2}_{34} }\,.
\end{equation}
The {normalized} correlator ${\cal C}_{\vec{p}}(U,V,\hat{\sigma},\hat{\tau})$ depends 
 non trivially on $N$ and the 't Hooft coupling.  Additionally, the ``partial non-renormalization" theorem 
 of \cite{Eden:2000bk,Heslop:2002hp} implies that
\begin{equation}\label{notation4pt_intro}
{\cal C}_{\vec{p}}(U,V,\hat{\sigma},\hat{\tau}) 
= {\cal G}^{\emph free}_{\vec{p}}(U,V,\hat{\sigma},\hat{\tau})\, +\, {\cal I}(U,V,\hat{\sigma},\hat{\tau}) {\cal H}_{\vec{p}}(U,V,\hat{\sigma},\hat{\tau})\;,
\end{equation}
where $ {\cal I}(U,V,\hat{\sigma},\hat{\tau})$ is fixed by superconformal symmetry, and reads
\begin{equation}\label{intriligator}
 {\cal I}(U,V,\hat{\sigma},\hat{\tau})= V + \hat{\sigma}^2 U V + \hat{\tau}^2 U + \hat{\sigma} V (-1-U+V)+\hat{\tau} (1-U-V)+\hat{\sigma}\hat{\tau} U ( -1+U-V)\,.
\end{equation}
The function ${\cal H}_{\vec{p}}(U,V,\hat{\sigma},\hat{\tau})$ is the so called ``reduced correlator" 
or ``dynamical function", since it is the only contribution depending on the 't Hooft coupling. 
Note that ${\cal H}_{\vec{p}}(U,V,\hat{\sigma},\hat{\tau})$ is a polynomial in the $\hat{\sigma},\hat{\tau}$ 
cross ratios, of degree
\begin{equation}\label{degree_extre}
\kappa=-2-\Sigma +\!\!\!\sum_{ij=12,13,14} \!\!\!\!\!\!\min\left(\tfrac{p_i+p_j}{2},\Sigma-\tfrac{p_i+p_j}{2}\right)\,,
\end{equation}
where $\Sigma=\frac{p_1+p_2+p_3+p_4}{2}$ and the shift of $-2$ is due to ${\cal I}(U,V,\hat{\sigma},\hat{\tau})$. 

It will be useful to introduce the symmetric cross ratios, $x_{i=1,2}$ and $y_{j=1,2}$ defined as follows \cite{Doobary:2015gia},  
\begin{equation}\label{notation4pt_intro_symcr1}
U=x_1x_2,\quad V=(1-x_1)(1-x_2)\qquad;\qquad \hat{\sigma}=\frac{1}{y_1 y_2},\quad\hat{\tau}=\frac{(1-y_1)(1-y_2)}{y_1 y_2}\;.
\end{equation}
Note that  
\begin{equation}\label{notation4pt_intro_symcr2}
{\cal I}(x_1,x_2,y_1,y_2)=\frac{1}{(y_1y_2)^2} \prod_{i=1,2} (x_i-y_j) \,.
\end{equation}
In this form it is evident that ${\cal I}$ vanishes under the chiral algebra twist \cite{Beem:2013sza} 
and that the form of the correlator in \eqref{notation4pt_intro} satisfies the superconformal Ward identity, 
i.e.~for any ${\cal H}$ \cite{Nirschl:2004pa} we have $[(\partial_{x_i}+\partial_{y_j}){\cal C}]_{x_i=y_j}=0$.

\subsection{The space of double-particle operators}\label{REVdP}

The OPE of two single-particle operators in the supergravity regime contains double-particle 
operators and multi-particle operators.  In this section, we describe the space of all  (superconformal primary) 
double-particle operators, since they will soon become relevant to explain the double-particle bootstrap. 
We closely follow \cite{Aprile:2018efk}.

The (superconformal primaries) double-particle operators in the free theory have the  schematic form 
\be\label{dpoperator}
\mathcal{D}_{pq,\tau,l,[aba]} = \mathcal{O}_p \partial^l \Box^{\frac{1}{2}(\tau - p - q)} \mathcal{O}_q\, \qquad ( 2\leq p \leq q )\;.
\ee
In the following, we shall use $\vec{\tau}=(\tau_{},l,[a,b,a])$ to denote the collection of the quantum numbers. 

Focusing first on the non-BPS primaries, for a given $\vec{\tau}$, there is a degenerate space. 
We count such operators with the same quantum numbers by the number of pairs $(pq)$ 
that fill in the following rectangle \cite{Aprile:2018efk}~\footnote{Compared 
to~\cite{Aprile:2018efk}, $i_{\rm here}=i_{\rm there}-1$}
\begin{align}\label{ir}
 {R}_{\vec{\tau}}:=\left\{(p,q):\begin{array}{l}
	p=i+a+2+r\\q=i+a+2+b-r\end{array}\text{ for } \begin{array}{l}
	i=\,0,\ldots,(t-2)\\ r=0,\ldots,(\mu -1)\end{array}\,
	\right\}
\end{align}
with $t$ and $\mu$ defined as
\be\label{multiplicity}
t\equiv \frac{(\tau-b)}{2}-a,\qquad\qquad 
\mu \equiv \left\{\begin{array}{ll}
\bigl\lfloor{\frac{b+2}2}\bigr\rfloor \quad &a+l \text{ even,}\\[.2cm]
\bigl\lfloor{\frac{b+1}2}\bigr\rfloor \quad &a+l \text{ odd,}
\end{array}\right.\,
\ee
where $\lfloor x \rfloor$ is the greatest integer smaller than or equal to $x$. 
The distinction made in \eqref{multiplicity} has to do with the absence of 
operators of the form \eqref{dpoperator} with $p=q$ when $a+l$ is odd.
The total number of pairs in $R_{\vec{\tau}}$ is always $\mu(t-1)$, and this number equals the total 
number of non-BPS degenerated operators with quantum numbers $\vec{\tau}$.

For all operators,  therefore also for the double-particle operators, 
there is a unitarity bound to satisfy, and for the rep $[a,b,a]$ this bound implies $\tau\ge 2a+b+2$. 
Let us see what the minimum value of $\tau$ is for operators described by $R_{\vec{\tau}}$ and how it compares with the unitarity bound. 
To start with, note that the value of $p+q$ computed from \eqref{ir} does not depend on $r$, thus $r$ labels a partition of $p+q$. 
Let us also note then that for $i=0$ the value of $p+q$ gives $p+q=2a+b+4$, which coincides 
with the minimum twist for a long double-particle in the rep $[a,b,a]$. (This value is then incremented by 
adding $\Box^{\frac{1}{2}(\tau-p-q)}$, which means incrementing the value of $i$ in $R_{\vec{\tau}}$.)

Now, at the unitarity bound $\tau=2a+b+2$, the double-particle 
operators are protected and so these are not included in the class $R_{\vec{\tau}}$ described above. 
Also in this case, there is a space of degenerate primaries of the form
\begin{equation}\label{doper_semis}
\mathcal{D}_{pq,[aba]} = \mathcal{O}_p \partial^l \mathcal{O}_q\,, \qquad 2a+b+2=p+q\qquad ( 2\leq p \leq q )
\end{equation}
and the degenerate operators are labelled by a partition of $2a+b+2$.

Let us give some examples, some of which will be useful in later sections. In $[0,0,0]$ there are no protected 
double-particle operators. The long operators have $\mu=1$, the first operator has twist $4$ 
and the class $R_{\vec{\tau}}$ for higher values of the twist is described by the following filtration
\begin{equation}
\bigoplus_{\tau=4,6,\ldots} R_{\tau,l,[000]} = \Bigg\{ 
\{ \underbrace{{\cal O}_2 \partial^l {\cal O}_2}_{\tau=4} \}, 
\{ \underbrace{{\cal O}_2\partial^l \Box {\cal O}_2 , {\cal O}_3\partial^l {\cal O}_3}_{\tau=6} \},\ldots \Bigg\}\;.
\end{equation}
Similarly, in $[0,1,0]$, there are no protected double-particles, the long operators also have $\mu=1$ 
and the first double-particle is long and has twist $5$,
\begin{equation}\label{eq:010D}
\bigoplus_{\tau=5,7,\ldots} R_{\tau,l,[010]} = \Bigg\{ 
\{ \underbrace{{\cal O}_2 \partial^l {\cal O}_3}_{\tau=5} \}, 
\{ \underbrace{{\cal O}_2\partial^l \Box {\cal O}_3 , {\cal O}_3\partial^l {\cal O}_4}_{\tau=7} \},\ldots \Bigg\}\;.
\end{equation}
In $[1,0,1]$ we find protected double-particle operators at twist $4$, which are ${\cal O}_2\partial^l{\cal O}_2$ in the 
antisymmetric rep, and the long operators have again $\mu=1$, and they are described by
\begin{equation}
\bigoplus_{\tau=6,8,\ldots} R_{\tau,l,[101]} = \Bigg\{ 
\{ \underbrace{{\cal O}_3\partial^l {\cal O}_3}_{\tau=6} \}, \{ 
\underbrace{{\cal O}_3\partial^l \Box{\cal O}_3 , {\cal O}_4\partial^l {\cal O}_4}_{\tau=8} \}\ldots \Bigg\}\;.
\end{equation}
The case of $[0,2,0]$ has more features.  The protected operators have twist $4$, thus 
they are ${\cal O}_2\partial^l{\cal O}_2$ in the symmetric traceless rep. The long operators 
with even spin have $\mu=2$, and they are described by
\begin{equation}\label{bigO020}
\bigoplus_{\tau=6,8,\ldots} R_{\tau,l,[020]} = \Bigg\{ 
\{ \underbrace{{\cal O}_2 \partial^l {\cal O}_4 , {\cal O}_3\partial^l {\cal O}_3}_{\tau=6} \}, 
\{ \underbrace{{\cal O}_2\partial^l \Box {\cal O}_4 , 
{\cal O}_3\partial^l  {\cal O}_5 , {\cal O}_3\partial^l  \Box {\cal O}_3 ,
{\cal O}_4\partial^l {\cal O}_4}_{\tau=8} \},\ldots \Bigg\}\;.
\end{equation}
The long operators with odd spins have $\mu=1$: The difference compared to \eqref{bigO020} is 
that operators in \eqref{bigO020} of the form \eqref{dpoperator} 
with $p=q$ are absent. 
The reps $[0,b,0]$ will frequently appear in the next section.

\section{The double-particle bootstrap for $\langle [{\cal O}_{r_1} {\cal O}_{r_2}] {\cal O}^{\phantom{2}}_s {\cal O}_p {\cal O}_q\rangle$}
\label{section_analitic_boot}

In this section we explain how to determine ${\cal H}_{[r_1r_2]spq}$ at tree level, $O(\frac{1}{N^3})$,
from the knowledge of the free theory and the CFT data of double-particle operators 
\cite{Aprile:2018efk,Aprile:2019rep} -- both short and long representations.  As mentioned already, 
the computation of ${\cal H}_{[r_1r_2]spq}$ yields new four-point correlators in 
${\cal N}= 4$ SYM in the supergravity regime, i.e.~at large $N$ and large 't Hooft coupling.

\subsection{Overview of the method}\label{overV}

Consider the space of  (superconformal primary) double-particle operators ${\cal D}_{\vec{\tau} }$ 
with quantum numbers $\vec{\tau}=(\tau,l,[aba])$, see the review in \eqref{dpoperator}-\eqref{doper_semis}. 
Then, take the \emph{normalized three-point coupling} between a double particle operator ${\cal D}_{\vec{\tau}}$ 
and two half-BPS single particle operators,
\begin{equation}
 {\cal C}_{ {\cal O}_{r} {\cal O}_{s} {\cal D}_{\vec{\tau}} }= {\cal C}^{(0)}_{ {\cal O}_{r} {\cal O}_{s} {\cal D}_{\vec{\tau}} } + \frac{1}{N} {\cal C}^{(\frac{1}{2})}_{ {\cal O}_{r} {\cal O}_{s} {\cal D}_{\vec{\tau}} } + \frac{1}{N^2} {\cal C}^{(1)}_{ {\cal O}_{r} {\cal O}_{s} {\cal D}_{\vec{\tau}} } + \ldots\;,
\end{equation}
where on the RHS we explicitly wrote the large $N$ expansion. {The order in $\frac{1}{N}$ of the 
first non-trivial terms of this three-point coupling depends on $\tau$}
\begin{equation}\label{3ptssd}
\begin{tikzpicture}
\def\xinterm{2.5};
\def\xstop{5.5};
\draw (-2,0) node {${\cal C}_{ {\cal O}_{r} {\cal O}_{s} {\cal D}_{\vec{\tau}} }=$};
\draw (0,+.05) -- (\xstop,+.05) -- (\xstop,-.05) --(0,-.05);
\draw[dashed] (-1,+.05) --(-0.4,+.05); \draw[] (-0.4,+.05) --(0,+.05);
\draw[dashed] (-1,-.05) --(-0.4,-.05); \draw[] (-0.4,-.05) --(0,-.05);
\draw (0,+.05) -- (\xstop,+.05) -- (\xstop,-.05) --(0,-.05);
\draw[gray, line width=3pt] (\xstop-.2, +.1) -- (\xstop-.2+.4, 0) -- (\xstop-.2, -.1);

\draw[red,line width=1pt] (\xinterm, +.2) -- (\xinterm, -.2);
\draw  (\xinterm, -.4) node[scale=.7,blue] {$r+s$};
\draw[black,line width=1pt] (0, +.2) -- (0, -.2);

\draw  (-.5, -.4) node[scale=.5] {$\tau=2a+b+2$};

\draw (\xinterm+1.3, -.2) node[scale=.5] {$\underbrace{\rule{4.5cm}{0pt}}$};
\draw (\xinterm+1.3, -.5) node[scale=.7] {$O(1)$};

\draw (\xinterm-1.2, -.2) node[scale=.5] {$\underbrace{\rule{4cm}{0pt}}$};
\draw (\xinterm-1.2, -.5) node[scale=.7] {$O(\frac{1}{N^2})$};

\end{tikzpicture}
\end{equation}
In particular, when $\tau$ is above the 
threshold $\tau\ge r+s$, the three-point function is leading order, which means $O(1)$, 
otherwise it is $\frac{1}{N^2}$ suppressed.  
To understand this scaling consider for example $\langle {\cal O}_{r} {\cal O}_{s} [{\cal O}_{r}{\cal O}_{s}] \rangle$. 
In this case the leading order contribution comes from the (disconnected) two-point correlators 
$\langle {\cal O}_{r} {\cal O}_{r}\rangle$ and $\langle {\cal O}_{s} {\cal O}_{s}\rangle$, 
thus the normalized three-point function is $O(1)$. An important subtlety to 
keep in mind here is that the space of double-particle operators 
is degenerate in $\tau$, and therefore the ${\cal D}_{\tau,l}$ are linear combinations 
of the explicit basis of operators given in \eqref{dpoperator}. 

We shall now consider the normalized three-point coupling 
\begin{equation}
{\cal C}_{ [{\cal O}_{r_1}{\cal O}_{r_2}] {\cal O}_{s} {\cal D}_{\vec{\tau}} }
\end{equation}
where an external single-particle operator
has been replaced by a half-BPS double-particle operator.
For illustration, we shall consider the example of $r_1=r_2=r$.  The scaling with $N$ now is the following,
\begin{equation}\label{3ptsdd_figure}
\begin{tikzpicture}
\def\xinterm{2.5};
\def\xstop{5.5};
\draw (-2,0) node {${\cal C}_{ {\cal O}^2_{r} {\cal O}_{s} {\cal D}_{\vec{\tau}} }=$};
\draw (0,+.05) -- (\xstop,+.05) -- (\xstop,-.05) --(0,-.05);
\draw[dashed] (-1,+.05) --(-0.4,+.05); \draw[] (-0.4,+.05) --(0,+.05);
\draw[dashed] (-1,-.05) --(-0.4,-.05); \draw[] (-0.4,-.05) --(0,-.05);
\draw[gray, line width=3pt] (\xstop-.2, +.1) -- (\xstop-.2+.4, 0) -- (\xstop-.2, -.1);

\draw[red,line width=1pt] (\xinterm, +.2) -- (\xinterm, -.2);
\draw  (\xinterm, -.4) node[scale=.7,blue] {$r+|s-r|$};
\draw[black,line width=1pt] (0, +.2) -- (0, -.2);

\draw  (-.5, -.4) node[scale=.5] {$\tau=2a+b+2$};

\draw (\xinterm+1.3, -.2) node[scale=.5] {$\underbrace{\rule{4.5cm}{0pt}}$};
\draw (\xinterm+1.3, -.5) node[scale=.7] {$O(\frac{1}{N})$};

\draw (\xinterm-1.2, -.2) node[scale=.5] {$\underbrace{\rule{4cm}{0pt}}$};
\draw (\xinterm-1.2, -.5) node[scale=.7] {$O(\frac{1}{N^3})$};

\end{tikzpicture}
\end{equation}
It is useful to understand the case $\tau\ge r+|s-r|$, by elaborating on our previous example. 
Let us take $\langle {\cal O}^2_{r} {\cal O}_{s} [{\cal O}_{r}{\cal O}_{X}] \rangle$ and let us observe that 
its leading order contribution comes from the  three-point function $\langle {\cal O}_{r} {\cal O}_{s} 
{\cal O}_{X}\rangle$ and the two point function $\langle {\cal O}_r {\cal O}_r\rangle$, 
where one of the ${\cal O}_r$ is taken from $[{\cal O}_r{\cal O}_X]$.  Now, $\langle {\cal O}_{r} {\cal O}_{s} 
{\cal O}_{X}\rangle$ is $\frac{1}{N}$ suppressed, thus \eqref{3ptsdd_figure} follows.  
Note also that $\langle {\cal O}_{r} {\cal O}_{s} {\cal O}_{X}\rangle$ exists as long as $X\ge |s-r|$.
 
In a four-point correlator where one external operator is a half-BPS 
double-particle operator and three are single-particle operators, the exchanged 
operators to be taken into account at leading order, $O(\tfrac{1}{N})$, 
are \emph{both {double}-particle and {triple}-particle operators}. This is obvious if we consider 
the OPE between a double-particle operator and a single-particle 
operator, say ${\cal O}^2_{r}$ and ${\cal O}_s$:  in this example, at twist  $\tau= 2r+s$ we certainly find 
$[{\cal O}_r{\cal O}_r{\cal O}_s]$.  

Let ${\cal T}_{\vec{\tau}}$ be a triple particle operator with generic quantum 
numbers $\vec{\tau}$. Repeating a reasoning similar to the one explained above, 
we find that the three-point coupling between two single particle operators, say ${\cal O}_{r} {\cal O}_{s}$ and 
any triple-particle operators ${\cal T}_{\vec{\tau}}$ behaves as 
 ${\cal C}_{ {\cal O}_{r} {\cal O}_{s} {\cal T}_{\vec{\tau}} }= O(\frac{1}{N})$  if the twist $\tau$ is greater that $r+s$
otherwise ${\cal C}_{ {\cal O}_{r} {\cal O}_{s} {\cal T}_{\vec{\tau}} }=O(\frac{1}{N^3})$. Schematically, 
\begin{equation}\label{3ptsdt_figure}
\begin{tikzpicture}
\def\xinterm{2.5};
\def\xstop{5.5};
\draw (-2,0) node {${\cal C}_{ {\cal O}_{r} {\cal O}_{s} {\cal T}_{\vec{\tau}} }=$};
\draw (0,+.05) -- (\xstop,+.05) -- (\xstop,-.05) --(0,-.05);
\draw[dashed] (-1,+.05) --(-0.4,+.05); \draw[] (-0.4,+.05) --(0,+.05);
\draw[dashed] (-1,-.05) --(-0.4,-.05); \draw[] (-0.4,-.05) --(0,-.05);
\draw[gray, line width=3pt] (\xstop-.2, +.1) -- (\xstop-.2+.4, 0) -- (\xstop-.2, -.1);

\draw[red,line width=1pt] (\xinterm, +.2) -- (\xinterm, -.2);
\draw  (\xinterm, -.4) node[scale=.7,blue] {$r+s$};
\draw[black,line width=1pt] (0, +.2) -- (0, -.2);

\draw  (-.5, -.4) node[scale=.5] {$\tau=2a+b+2$};

\draw (\xinterm+1.3, -.2) node[scale=.5] {$\underbrace{\rule{4.5cm}{0pt}}$};
\draw (\xinterm+1.3, -.5) node[scale=.7] {$O(\frac{1}{N})$};

\draw (\xinterm-1.2, -.2) node[scale=.5] {$\underbrace{\rule{4cm}{0pt}}$};
\draw (\xinterm-1.2, -.5) node[scale=.7] {$O(\frac{1}{N^3})$};

\end{tikzpicture}
\end{equation}
Replacing a single-particle operator with a double-particle operator, say ${\cal O}_r^2$,
we find that the scaling of ${\cal C}_{ {\cal O}^2_{r} {\cal O}_{s} {\cal T}_{\vec{\tau}} }$ 
with $N$ behaves as follows,
\begin{equation}
\begin{tikzpicture}
\def\xinterm{2.5};
\def\xstop{5.5};
\draw (-2,0) node {${\cal C}_{ {\cal O}^2_{r} {\cal O}_{s} {\cal T}_{\vec{\tau}} }=$};
\draw (0,+.05) -- (\xstop,+.05) -- (\xstop,-.05) --(0,-.05);
\draw[dashed] (-1,+.05) --(-0.4,+.05); \draw[] (-0.4,+.05) --(0,+.05);
\draw[dashed] (-1,-.05) --(-0.4,-.05); \draw[] (-0.4,-.05) --(0,-.05);
\draw (0,+.05) -- (\xstop,+.05) -- (\xstop,-.05) --(0,-.05);
\draw[gray, line width=3pt] (\xstop-.2, +.1) -- (\xstop-.2+.4, 0) -- (\xstop-.2, -.1);

\draw[red,line width=1pt] (\xinterm, +.2) -- (\xinterm, -.2);
\draw  (\xinterm, -.4) node[scale=.7,blue] {$2r+s$};
\draw[black,line width=1pt] (0, +.2) -- (0, -.2);

\draw  (-.5, -.4) node[scale=.5] {$\tau=2a+b+2$};

\draw (\xinterm+1.3, -.2) node[scale=.5] {$\underbrace{\rule{4.5cm}{0pt}}$};
\draw (\xinterm+1.3, -.5) node[scale=.7] {$O(1)$};

\draw (\xinterm-1.2, -.2) node[scale=.5] {$\underbrace{\rule{4cm}{0pt}}$};
\draw (\xinterm-1.2, -.5) node[scale=.7] {$O(\frac{1}{N^2})$};

\end{tikzpicture}
\end{equation}
A similar statement applies to a generic double-particle operator $[{\cal O}_{r_1}{\cal O}_{r_2}]$ and, in this case, the value of the threshold is $r_1+r_2+s$ instead of $2r+s$.

{\bf A window of predictions.} Coming back to $\langle [{\cal O}_{r_1}{\cal O}_{r_2}]{\cal O}_s {\cal O}_p {\cal O}_q \rangle$, its large $N$ expansion reads
\begin{equation}\label{general_scaling_dp}
\langle [{\cal O}_{r_1}{\cal O}_{r_2}]{\cal O}_s {\cal O}_p {\cal O}_q \rangle
= O(\tfrac{1}{N})\times \emph{free}   +    O(\tfrac{1}{N^3})\times\Big( \emph{free}+ \emph{dynamical}\Big)+\ldots \;.
\end{equation}
From our previous considerations on the three-point couplings, we understand that both 
double- and triple-particle operators can be exchanged in the leading-order term $O(\tfrac{1}{N})$. 
However, there is always a window on the twist axis where the 
double-particle operators are the only ones that contribute at order $O(\frac{1}{N})$, 
and therefore, for such a window, we can make predictions at the next order 
$O(\frac{1}{N^3})$ by using the double-particle bootstrap.
Schematically, the predictions are:
\begin{mdframed}
~\rule{0pt}{.2cm}\\
\begin{equation}\label{intro_predictions}
\tfrac{1}{N}{\cal C}^{(\frac{1}{2})}_{ [r_1r_2] s; {\cal D}_{\tau,l}} \times 
 \left\{ \begin{array}{ll} \displaystyle  \tfrac{1}{N^2} \eta_{\cal D}\, {\cal C}^{(0)}_{pq; {\cal D} }\,\log(U)&\rule{.5cm}{0pt} \emph{if}\ \ \ \ p+q\leq \tau < r_1+r_2+s \\[.4cm] 
 				  \displaystyle \tfrac{1}{N^2} {\cal C}^{(1)}_{pq; {\cal D} } &\rule{.5cm}{0pt} \emph{if}\ \ \ \ \tau_{\emph unitary}\leq \tau< p+q \end{array}\right.
\end{equation}
~\\
\end{mdframed}
Rather than describing how to compute both lines of \eqref{intro_predictions} in one go, 
in the next sections, we opted for a more pragmatic approach. 
In section \ref{sec_warmup_p} we will start 
by describing the bootstrap algorithm for simple test cases, and gradually 
arrive at more general cases in section \ref{more_correlators_sec}.
The simple test cases are
\begin{equation}\label{testcase_intro}
\langle  {\cal O}_2^2{\cal O}_2 {\cal O}_p{\cal O}_{p} \rangle\;.
\end{equation}
This family of correlators is N$^2$E, i.e.~they have $\kappa=0$ in \eqref{degree_extre}, 
and therefore their dynamical function ${\cal H}_{[2^2]2pp}$ does not depend on the R-symmetry 
cross ratios. We will see that the predictions in \eqref{list_correlators_intro} are consistent with an
ansatz of {D}-functions, and moreover, that they fix the ansatz uniquely.

In section \ref{more_correlators_sec} we will generalize the previous results  
in two directions: 
\begin{itemize}
\item[$\bullet$] By increasing the degree of extremality keeping ${\cal O}_2^2$. We will give the results for
\begin{equation}
 \langle {\cal O}_2^2\, {\cal O}_{p_2} {\cal O}_{p_3} {\cal O}_{p_4}\rangle\,.
\end{equation}
\item[$\bullet$] By considering more instances of composite double-particle operators. We will give the results for
\begin{equation}\label{list_correlators_intro}
\langle {\cal O}_3^2  {\cal O}_2 {\cal O}_p {\cal O}_p \rangle\,,\, 
 \langle[{\cal O}_2{\cal O}_4] {\cal O}_2 {\cal O}_p {\cal O}_p\rangle,
 \qquad
   \langle {\cal O}_r^2 {\cal O}_2 {\cal O}_p {\cal O}_p\rangle\,.
\end{equation}
\end{itemize}
The statement that the $O(\frac{1}{N^3})$  contribution is determined by 
{D}-functions will be true for all correlators of the form ${\cal H}_{[r_1r_2]spq}$.

\subsection{Algorithm at work in simple test cases}
\label{sec_warmup_p}

In this section, we explain the details of the double-particle bootstrap by considering the simplest test case,  
$\langle {\cal O}_2^2{\cal O}_2{\cal O}_3{\cal O}_3\rangle$, and then the whole family,
$\langle {\cal O}_2^2  {\cal O}_2{\cal O}_p{\cal O}_p \rangle$, for arbitrary values of $p$. 
Note that $p\ge 3$ since otherwise the correlator (when it exists) is protected.\footnote{Note in particular that $\langle  {\cal O}_2^2{\cal O}_2 
{\cal O}_p{\cal O}_{p+k} \rangle$ vanish for $k>0$ and that  $\langle  {\cal O}_2^2{\cal O}_2 
{\cal O}_2{\cal O}_{2} \rangle$ is protected.}
We shall start by collecting the OPE predictions. Out of this data, we will learn that the 
predictions are consistent with an ansatz of D-functions. Then, we will make an ansatz 
in position space and fix it \emph{uniquely}. For arbitrary values of $p$, 
we will repeat the same procedure. In addition, we will take advantage of the fact that D-functions 
in Mellin space are represented by rational functions to simplify the step in which we make the ansatz. 
In this way, we will be able to give a formula for ${\cal H}_{[2^2]2pp}$ with explicit $p$ dependence. 

At this stage, we still use Mellin space as a simplifying tool in the bootstrap algorithm. 
We will discuss tree-level \emph{Mellin amplitudes} with one double-particle operator 
later on, in section \ref{sec_MellinAmps}.

\begin{center}
{\underline{\bf Bootstrapping $\langle$[2$^2$]$,$2$,$3$,$3$\rangle$ at tree level}}
\end{center}

There are two independent orientations for this correlator to be considered, namely 
\begin{equation}
  \langle {\cal O}_2^2(z_1){\cal O}_2(z_2){\cal O}_3(z_3){\cal O}_3(z_4)\rangle\qquad;\qquad   
 \langle {\cal O}_2^2(z_1){\cal O}_3(z_2){\cal O}_2(z_3){\cal O}_3(z_4)\rangle.
\end{equation}
The common OPEs, taken w.r.t.~$X_1\rightarrow X_2$ and $X_3\rightarrow X_4$, are different in the 
two orientations, and so the predictions that we input in the bootstrap.

We will work with normalized correlators  in order to make the large $N$ 
scaling manifest. To this end, let us recall the (exact) two-point function normalizations:
\begin{equation}\label{eq:O2nor}
|{\cal O}_3|^2=\frac{ 3a^2 }{N}\left(1-\frac{3}{a}\right)\qquad;\qquad
|{\cal O}_2|^2= 2a\qquad;\qquad |{\cal O}^2_2|^2=2|{\cal O}_2|^4\left(1+\frac{2}{a}\right)\;,
\end{equation}
where we used $a=N^2-1$. The prefactors we shall use are
\begin{equation}\label{prefactors4233}
{\cal P}_{[2^2]233} = |{\cal O}_3|^2 |{\cal O}_2||{\cal O}_2^2| \  g_{12}^2 g_{14}g_{13}g_{34}^2\qquad;
\qquad {\cal P}_{[2^2]323} =  |{\cal O}_3|^2 |{\cal O}_2||{\cal O}_2^2| \   g_{12}^3 g_{14} g_{34}^2
\end{equation}
and in particular the normalization $ |{\cal O}_3|^2 |{\cal O}_2||{\cal O}_2^2|= 12 N^{6}(1+\ldots)$.

The normalized correlator ${\cal G}_{[2^2]233}$ in the free theory is
\beq\label{free3322d}
 {\cal G}_{[2^2]233}= \frac{6}{\sqrt{a+2} }\Bigg[ 1+\frac{2}{a} +\frac{4}{a}\Bigg[U\hat{\sigma}+\frac{U\hat{\tau}}{V}\Bigg] + \frac{2}{a} \frac{U^2\hat{\sigma}\hat{\tau}}{V} \Bigg]\,.
 \eeq
 
The other orientation  ${\cal G}_{[2^2]323}$ in the free theory is
\begin{equation}\label{free2332d}
 {\cal G}_{[2^2]323}= \frac{6}{\sqrt{a+2} }\Bigg[ U^2\hat{\sigma}^2+
 \frac{2}{a}\Bigg[ \frac{U\hat{\tau}}{V}+2 U\hat{\sigma}\Bigg] + \frac{2}{a}\Bigg[ U^2\hat{\sigma}^2 +2\frac{U^2\hat{\sigma}\hat{\tau}}{V}\Bigg] \Bigg]\;.
\end{equation}

{ \bf OPE predictions for ${\cal H}_{[2^2]323}$.} 
In this orientation we consider the OPE common to both ${\cal O}_2^2 {\cal O}_3$ and ${\cal O}_2 {\cal O}_3$. 
The exchanged operators relevant to make predictions for ${\cal H}_{[2^2]323}$ 
belong to the $su(4)$ rep $[0,1,0]$. Both OPEs contain triple-particle operators 
when $\tau\ge 7$.\footnote{These triple-particle operators have $O(1)$ three-point couplings 
with ${\cal O}_2^2 {\cal O}_3$, and $O(\frac{1}{N})$ three-point couplings with ${\cal O}_2 {\cal O}_3$.}
These triple-particle operators contribute already at order $O(\frac{1}{N})$ in the disconnect correlator,
and therefore we are forced to look at the spectrum of operators strictly below this threshold, 
because otherwise there is nothing we can say about ${\cal H}_{[2^2]323}$ without also knowing 
the triple-particle CFT data. Since the unitary bound, $\tau=2+2a+b$, for the $[0,1,0]$ rep is 
$\tau=3$, we are interested in operators with twist $\tau=3,5$.

\underline{At $\tau=5$}
the double-particle operators are non-degenerate, since only the first family 
of operators in the list~\eqref{eq:010D} exists for which we now use the shorthand notation
\begin{equation}
  \label{eq:Dtau5}
  {\cal D}_{5,l}  = {\cal O}_2\partial^l {\cal O}_3\,,
\end{equation}
and there are no triple-particle operators for this value of the twist. The anomalous 
dimension $ \eta_{{\cal D}_{5,l}} $ of these operators can be read from 
Eq.~8 of~\cite{Aprile:2017qoy}\footnote{Consider $t=2$ and $i=1$ in the notation of that paper, formulae (9) and (10).}
\begin{equation}
  \label{eq:andimt5}
  \eta_{{\cal D}_{5,l}} = -\frac{80}{(l+1) (l+4)} \;, \quad l=0,2,4,\ldots\;, \qquad
  \eta_{{\cal D}_{5,l}} = -\frac{80}{(l+4) (l+7)}\;, \quad l=1,3,5,\ldots\;.
\end{equation}
Following the general reasoning in \eqref{3ptssd}-\eqref{3ptsdt_figure},  
we find that 
\begin{equation}
{\cal C}_{[2^2]3; {\cal D}_{5,l} }\propto O(\tfrac{1}{N})\qquad;\qquad {\cal C}_{23; {\cal D}_{5}}\propto O(1)\;.
\end{equation}
Then from the first line in \eqref{intro_predictions} we have
\begin{equation}\label{pred_twist5}
{\cal H}_{[2^2]323}\Bigg|_{\frac{1}{N^3}}=\log(U) \sum_{l}   {\cal C}^{(\frac{1}{2})}_{[2^2]3 ;{\cal D}_{5,l} }
 \eta_{{\cal D}_{5,l}} \, {\cal C}^{(0)}_{23; {\cal D}_{5,l}} H^{4323}_{5,l,[010]}   +\ldots\;,
\end{equation}
where $ H^{p_1 p_2 p_3 p_4}_{\tau,l,[aba]} $ is the long superconformal block for the exchange 
of superprimary of twist $\tau$, spin $l$ in the $SU(4)$ irrep $[aba]$ (here we follow the notation 
of~\cite{Aprile:2025hlt}, see Appendix~D of that paper for more details).

For the actual computation of \eqref{pred_twist5}, we need 
${\cal C}^{(\frac{1}{2})}_{[2^2]3; {\cal D}_{5,l}}  {\cal C}^{(0)}_{23 ;{\cal D}_{5,l} }$. This is 
a $O(\frac{1}{N})$ free theory contribution, and since there is only one ${\cal D}_{5,l}$ we 
can read it off directly from performing the block decomposition of the term  
$U^2\hat{\sigma}^2$ in \eqref{free2332d}. The result of this decomposition is given
by the \emph{ready-to-use} formula  presented in \cite{Aprile:2025nta}, which we denote 
hereafter by the same coefficients $M^{\vec{p}}_{k,\gamma,\ulmb}$. 
The parameters of $M^{\vec{p}}_{k,\gamma,\ulmb}$ are the
external charges of the correlator of interest, $\vec{p}=\{p_1,p_2,p_3,p_4\}$,
then the two integers, $k$ and $\gamma$, which are read from the free 
theory diagram under study, and finally, the quantum numbers of the exchanged 
superconformal primary that are encoded in the same value of $\gamma$ and 
the Young diagram $\ulmb$. We refer to \cite[(16)]{Aprile:2025nta} for the explicit expression of $M^{\vec{p}}_{k,\gamma,\ulmb}$. 
This formula will appear frequently in the next sections, and therefore, it is useful to see 
how to determine the parameters in the example 
at hand.\footnote{Compared to~\cite[(16)]{Aprile:2025nta} 
we only added a sign: $M_{\rm here}=(-1)^{|\ulmb|} M_{\rm there}$.}

To compute ${\cal C}^{(\frac{1}{2})}_{[2^2]3; {\cal D}_{5,l}}  {\cal C}^{(0)}_{23 ;{\cal D}_{5,l} }$ 
we start from the free theory result ${\cal P}_{[2^2]323} {\cal G}_{[2^2]323}$, see~\eqref{prefactors4233} 
and \eqref{free2332d}, and, since we are interested in the contribution of the twist-five long operators, 
we focus on the term $g_{12}^2 g_{14}g_{13}g_{34}^2({6}\,U^2\hat{\sigma}^2)$. By comparing this part of 
the free result with the general expression in Eq.~(6) of~\cite{Aprile:2025nta}, we read $k=0$ (which here 
is related to the power of $\hat{\tau}$) and $\gamma=5$ (which here is related to the power of $\hat{\sigma}$). 
The Young diagram $\ulmb$ for the exchange of a long operator in the irrep 
$[\mu_1,\gamma-2\mu_1-2\mu_2-4,\mu_1]$ can be derived by the last row of 
table~(D.8) in~\cite{Aprile:2025hlt} and, since in our case $\tau=\gamma$ and the 
operators sit in the $[0,1,0]$ irrep, we have $\ulmb=[l+2,2]$ as function of the spin $l$. In summary, we find
\begin{equation}\label{pred2d3_uno}
{\cal C}^{(\frac{1}{2})}_{[2^2]3; {\cal D}_{5,l}}  {\cal C}^{(0)}_{23 ;{\cal D}_{5,l} }\Bigg|_{\frac{1}{N}}  
=  6 M^{4323}_{k=0,\gamma=5,[l+2,2]} = \frac{6}{10} {(l+1)(l+7)}\frac{(l+3)!(l+4)!}{(2l+7)!}\;.
\end{equation}
At this point, since we know the anomalous dimensions 
of the ${\cal D}_{5,l}$ from \cite{Aprile:2017qoy}, we can predict
\begin{equation}\label{pred_warmup_logU}
  {\cal C}^{(\frac{1}{2})}_{[2^2]3 ;{\cal D}_{5,l} }
 \eta_{{\cal D}_{5,l}} \, {\cal C}^{(0)}_{23; {\cal D}_{5,l}}\Bigg|_{\frac{\log(U)}{N^3}} =
  -\frac{12\times 4(l+7)(l+3)!^2}{(2l+7)!}\frac{1+(-1)^l}{2} -\frac{12\times 4(l+1)(l+3)!^2}{(2l+7)!}\frac{1-(-1)^l}{2} \;.
\end{equation}
It will become useful to rewrite the prediction by performing a 
one-variable spin resummation. Namely, by writing
\begin{equation}\label{pred_2d323log}
\sum_{l}   {\cal C}^{(\frac{1}{2})}_{[2^2]3 ;{\cal D}_{5,l} }
 \eta_{{\cal D}_{5,l}} \, {\cal C}^{(0)}_{23; {\cal D}_{5,l}} H^{4323}_{5,l,[010]} \Bigg|_{\frac{\log(U)}{N^3}}  
 = U^2 \Bigg[ -\sum_{\lm\ge 0} \frac{144(\lm+1)(\lm+2)}{(\lm+4)(\lm+5)(\lm+6)} x_1^\lm + \ldots \Bigg] \;,
\end{equation}
where the $\sum_{\lm} \# x_1^\lm$ is the aforementioned sum over spins in one variable, 
at twist $\tau=5$. (Terms omitted in $\ldots$ are higher order in $x_2$ and would mix 
with blocks of higher twist, this is why we are not considering them.)
Performing the sum, for example by using Wolfram Mathematica, one finds 
\begin{equation}\label{resum_twist5}
-\sum_{\lm\ge 0} \frac{144(\lm+1)(\lm+2)}{(\lm+4)(\lm+5)(\lm+6)} x_1^\lm =  \frac{48(30-21 x_1 +x_1^2)}{x_1^5} +\frac{144(10-12x_1+3x_1^2)\log(1-x_1)}{x_1^6}\;.
\end{equation}
A couple of observations are in order, 
\begin{itemize}
\item[$\bullet$] The result \eqref{resum_twist5} only contains $\log(U)\times \{1, \log(V)\}$ 
and no higher polylogs, thus the prediction is compatible with the 
correlator being determined by an ansatz of ${D}$-function, 
as for tree-level supergravity of single particle operators \cite{Dolan:2006ec,Rastelli:2017udc}. 
\item[$\bullet$] We used the anomalous dimensions of the operators 
${\cal D}_{5,l}$ computed in \cite{Aprile:2017qoy}, however we 
can rewrite the prediction in a way that uses four-point correlators,
and does not make reference to the explicit unmixed CFT data. 
To see this, consider the rewriting  
\begin{equation}\label{rewrite_pred50}
 {\cal C}^{(\frac{1}{2})}_{[2^2]3; {\cal D}_{5,l}}\eta_{{\cal D}_{5,l}} \, {\cal C}^{(0)}_{23 ;{\cal D}_{5,l} } = 
 \Bigg[ {\cal C}^{(\frac{1}{2})}_{[2^2]3; {\cal D}_{5,l}}{\cal C}^{(0)}_{23 ;{\cal D}_{5,l} }\Bigg]_{(3)}
\Bigg[ \frac{1}{{\cal C}^{(0)}_{23; {\cal D}_{5,l}} {\cal C}^{(0)}_{23; {\cal D}_{5,l}} }\Bigg]_{(2)}  
\Bigg[ {\cal C}^{(0)}_{23; {\cal D}_{5,l}}\eta_{{\cal D}_{5,l}} {\cal C}^{(0)}_{23; {\cal D}_{5,l}} \Bigg]_{(1)} \,.
\end{equation}
\end{itemize}
Each term on the RHS of \eqref{rewrite_pred50} corresponds directly to the block 
expansion of a four-point correlator. More precisely, 
$[\ldots]_{(3)}$ and $[\ldots]_{(2)}$ are free theory contributions, while
$[ \ldots ]_{(1)}$ is obtained (as done in \cite{Aprile:2019rep}) from the 
tree-level supergravity of the formula of Rastelli 
and Zhou \cite{Rastelli:2016nze}.
By using the notation $( {\cal C}_{\vec{p}}|_{[0,1,0],\tau=5,l} )$ to understand that we project the correlator ${\cal C}_{\vec{p}}$ onto 
the corresponding block coefficient, we immediately find from \eqref{rewrite_pred50} that  
\begin{equation}\label{rewrite_pred5}
 {\cal C}^{(\frac{1}{2})}_{[2^2]3; {\cal D}_{5,l}}\eta_{{\cal D}_{5,l}} \, {\cal C}^{(0)}_{23 ;{\cal D}_{5,l} } = 
 \frac{ \left( {\cal G}_{[2^2]323}^{(\frac{1}{2})} \Big|_{[010],\tau=5,l} \right) \times \left(  {\cal H}^{(1)}_{2323} \Big|_{\log(U),[010],\tau=5,l} \right) }{ 
\left(  {\cal G}^{(0)} _{2323} \Big|_{[010],\tau=5,l} \right)}\,. \rule{1cm}{0pt} 
\end{equation}
Note that the block expansion of ${\cal G}$, i.e.~the free theory part in our notation, is dramatically 
simplified by the formula in terms of $M^{\vec{p}}_{k,\gamma,\ulmb}$~\cite{Aprile:2025nta}, 
and the only extra information needed is the color factor.  

A rewriting like the one above will be preferred in the general case. The point is that,
in the general case, there is a space of degenerate double-particle 
operators exchanged (see again section \ref{REVdP}) and what will happen then is that  
a formula like \eqref{rewrite_pred5} will generalize simply by adding a sum over correlators.
In particular, a sum over all four-point correlators where the operators in question are exchanged.
This sum is nicely labelled by the rectangle $R_{\vec{\tau}}$ introduced in section \ref{REVdP}. Since 
we know the space of double-particle operators, it is not difficult to show that the aforementioned 
generalization is indeed correct, see e.g.~\cite[sect.~3]{Aprile:2019rep}. 
We will see more examples in the next sections.

\underline{At $\tau=3$} the only long operators are single-trace stringy states\footnote{{These states are semi-short in 
the free theory, so in order to derive the Young diagram $\ulmb$ in this case, one needs to use the semi-short 
dictionary in table~(D.8) in~\cite{Aprile:2025hlt}}.}. In the free theory  \eqref{free2332d}, they contribute 
in the block expansion of the terms  $2 U\hat{\sigma}+\frac{U\hat \tau}{V}$. In supergravity, their contribution has 
to be recombined and cancelled against a contribution coming from ${\cal H}_{[2^2]323}$, thus giving a 
prediction for ${\cal H}_{[2^2]323}$ at twist three. This is a well-known phenomenon discussed already 
in the past, see e.g. \cite{Dolan:2006ec}. We find
\begin{align}\label{notwist3}
{\cal H}_{[2^2]323}\Bigg|_{\frac{\log^0(U)}{N^3}} & = {-  \sum_{l} 12 \times \Bigg( 2 M^{4323}_{k=0,\gamma=3,[2+l]} + M^{4323}_{k=1,\gamma=3,[2+l]} \Bigg) H^{4323}_{3,l,[010]}  + \emph{higher\ twists}   }  \\
&=- \Bigg[  U \sum_{\lm} \frac{12(\lm+1)(\lm+6)}{(\lm+3)(\lm+4)}x_1^\lm + O(U^2)  \Bigg]\;.
\end{align}
In the second line, we have rewritten the one-variable sum, whose resummation gives the prediction
\begin{align}\label{pred_2d323nonlog}
{\cal H}_{[2^2]323}\Bigg|_{\frac{\log^0(U)}{N^3}} & =  -x_1x_2\Bigg[ \frac{12 (6-9x_1 +2 x_1^2)}{(-1+x_1) x_1^3} {+ \frac{12\times 6 (-1+x_1)\log(1-x_1)}{x_1^4}} \Bigg] + O((x_1 x_2)^2)\,.
\end{align}
Let us remark that this prediction is again compatible with a correlator built out of D-functions. 
This is so because exchanged operators with twist below the threshold of long double-particle operators  
are accounted for by single poles in the Mellin amplitude. Performing the integration in Mellin space, 
the result in position space would be hypergeometric functions whose transcendental content is $\{1,\log(V)\}$.\footnote{In the 
same non-log$(U)$ sector, ${\rm Li}_2$ is found when looking at the double poles in Mellin space, 
but since we are at twist $5$, thus below the double-poles threshold, we can only see  $\{1,\log(V)\}$, and this is indeed what we find.}

 { \bf OPE predictions for ${\cal H}_{[2^2]233}$.} 
 In this orientation, the relevant OPE is the one common to both   
 ${\cal O}_3{\cal O}_3$ and ${\cal O}_2{\cal O}_2^2$ and 
 the exchanged operators relevant to make predictions 
 belong to the $su(4)$ rep $[0,2,0]$.  
 Both OPEs contain triple-particle operators starting from $\tau\ge 6$.  
 Since the unitary bound is $\tau=2+2a+b=4$, 
 the only prediction that we can make about this correlator, without 
 knowledge of triple-particles, is the one for the twist four sector.

The relevant operators are the semishort double-particle operators  
at the $[0,2,0]$ unitarity bound, schematically, ${\cal O}_2\partial^l {\cal O}_2$.
What we have to do is to compute the prediction for their 
contribution to the correlator and check whether this agrees with the 
block decomposition of the free theory part, see \eqref{free3322d}.
If this is not the case, we have to fix the discrepancy by using multiplet recombination. 
Compared to the reasoning that leads to the absence of twist three 
stringy states in \eqref{notwist3}, the use of multiplet recombination here is 
more interesting because the protected double-particle operators at the 
unitarity bound are in the spectrum of supergravity,
but their contribution to the correlator might not coincide with the 
block coefficient extracted from the free theory part of the correlator itself. 
This is another instance of the phenomenon analyzed 
in \cite{Aprile:2025nta,Aprile:2025hlt,Aprile:2019rep} and first discussed in \cite{Doobary:2015gia}. 
In the end, from multiplet recombination we generate a non-trivial prediction for  ${\cal H}_{[2^2]233}$.

We are looking for the three-point correlators of the protected operators  
${\cal O}_2\partial^l {\cal O}_2$ with the external operators, ${\cal O}_2{\cal O}_2^2$ and ${\cal O}_3{\cal O}_3$. 
Even though these three-point couplings are protected, they are not immediately computable. 
However, we find them by repeating the reasoning proposed in \cite{Doobary:2015gia}. For this purpose consider the following ``Gram" matrix, 
\begin{equation}\label{020_Gram_M}
\rule{2cm}{0pt}\left( \begin{array}{cc} 
\langle {\cal O}_2{\cal O}_2 {\cal O}_2 {\cal O}_2\rangle & \langle {\cal O}_2{\cal O}_2 {\cal O}_3 {\cal O}_3\rangle  \\[.2cm] 
\langle {\cal O}^2_2{\cal O}_2 {\cal O}_2 {\cal O}_2\rangle & \langle  {\cal O}^2_2{\cal O}_2{\cal O}_3{\cal O}_3 \rangle \end{array}\right)\Bigg|_{[020],\tau=4}\;.
\end{equation}
The operator ${\cal O}_2\partial^l {\cal O}_2$ is exchanged in each correlator 
of the Gram matrix, but since there is only one ${\cal O}_2\partial^l {\cal O}_2$, 
columns/rows of \eqref{020_Gram_M} must be linearly dependent, 
and in particular, the determinant must vanish. 
We refer to \cite{Aprile:2025nta,Aprile:2025hlt} for more details. 
From the vanishing of the determinant, we can obtain the information about the 
desired three-point functions: Call ${\cal S}^{}_{}$ the semishort prediction that we are interested in, 
\begin{equation}
{\cal S}_{[2^2]233,[020],\tau=4,l}\Bigg|_{\frac{1}{N^3}}\equiv C^{(\frac{1}{2})}_{[2^2]2,{\cal O}_2\partial^l{\cal O}_2} C^{(1)}_{33,{\cal O}_2\partial^l{\cal O}_2} \,.
\end{equation}
Then, the vanishing of the Gram determinant \eqref{020_Gram_M} gives,
\begin{equation}\label{prediction_Stwist4}
\!\!\!\!{\cal S}_{[2^2]233,[020],\tau=4,l}\Bigg|_{\frac{1}{N^3}}=\frac{ \left( {\cal G}^{(\frac{1}{2})}_{[2^2]222} \Big|_{[0,2,0],\tau=4,l} \right) \times\left( {\cal G}^{(1)}_{2233} \Big|_{[0,2,0],\tau=4,l} \right)  }{  \left( {\cal G}^{(0)}_{2222} \Big|_{[0,2,0],\tau=4,l}\right) }\;,
\end{equation}
where again the notation $( {\cal C}_{\vec{p}}|_{[0,2,0],\tau=4,l} )$ means that 
we project ${\cal C}_{\vec{p}}$ onto the corresponding block coefficient.  
Note that we are looking for a $O(\frac{1}{N^3})$ prediction. In fact, the denominator  
in \eqref{prediction_Stwist4} has leading order $O(1)$, while the numerator contributes 
with $ O(\frac{1}{N})\times O(\frac{1}{N^2})$.  Extracting the block coefficients from 
  the various free theory correlators is efficiently done with the help of $M^{\vec{p}}_{k,\gamma,\ulmb}$ as before. 
  For the reader's convenience, we collect below the relevant correlators,
  \begin{align}\label{eq:sfr1}
    \mathcal{P}_{[2^2]222} & =8 a^2 \sqrt{a+2}\,g^2_{12} g_{13} g_{14} g_{34} \;, \quad
    \mathcal{G}_{[2^2]222} =\frac{4 \sqrt{a+2}}{a}\left[1+U \hat{\sigma}+\frac{U \hat{\tau}}{V}\right]\;,  
 \end{align}
 \begin{align}   
    \mathcal{P}_{22pp} & = 2a |{\mathcal O}_p|^{2p}  g_{12}^2  g_{34}^{2p}\;, \quad \label{eq:sfr2}
    \mathcal{G}_{22pp} =  1 + \delta_{p,2} \left( U\hat{\sigma}^2 +\frac{U^2\hat{\tau}^2}{V^2}\right)+ \frac{2p}{a} \left( U\hat{\sigma} + \frac{U\hat{\tau}}{V} + (p-1)\frac{U^2\hat{\sigma}\hat{\tau}}{V}  \right)\,.
  \end{align}
Let us stress that for N$^2$E single-particle correlators, the free theory contains 
only three (out of five possible) connected propagator structures and this is due 
to the orthogonality of single-particle operators \cite{Aprile:2020uxk}. 
With this information, we find,
\begin{equation}\label{semishortO2O2}
{\cal S}_{[2^2]233,[020],\tau=4,l}\Bigg|_{\frac{1}{N^3}}= {
\frac{ \left( 4 M^{4222}_{0,4,[l+2]} + 4 M^{4222}_{1,4,[l+2]} \right) \times\left(12 M^{2233}_{1,4,[l+2]} \right)  }{  
\left( M^{2222}_{0,4,[l+2]} + M^{2222}_{2,4,[l+2]}\right) } } = \frac{12\times 8 (l+2)!(l+3)!}{(2l+6)!}\frac{1+(-1)^l}{2}\,.
\end{equation}
At this point, we go back to the free theory part of the correlator, given in \eqref{free3322d}, and 
we look for the block coefficient with the same quantum numbers as the protected semishorts.
We shall call it $A_{[2^2]233,[020],\tau=4,l}$. We find 
\begin{equation}
{
A_{[2^2]233,[020],\tau=4,l}\Bigg|_{\frac{1}{N^3}}= 6\times 4\Big(M^{4233}_{0,4,[l+2]}+ M^{4233}_{1,4,[l+2]}\Big)=\frac{12\times 4 (l+3)!(l+4)!}{(2l+6)!}\frac{1+(-1)^l}{2}\,,}
\end{equation}
where the first parenthesis follows from the terms linear in $U$ of~\eqref{free3322d}. 
Since this does not coincide with the result in \eqref{semishortO2O2} 
we deduce that there is a non trivial prediction for ${\cal H}_{[2^2]233}$.
The prediction is therefore
\begin{align}
{\cal H}_{[2^2]233}\Bigg|_{ \frac{\log^0(U)}{N^3} }& = \sum_{l} \Big( - A_{[2^2]233}+{\cal S}_{[2^2]233} \Big)\Bigg|_{\frac{1}{N^3}}\!\!\!\!\!H_{[020],4,l}  +\ldots 
 \, = \, -U\Bigg[ \sum_{l} \frac{24(\lm+1) }{(\lm+3)} x_1^\lm  +O(U^2) \Bigg]
\end{align}
whose resummation gives
\begin{equation}\label{pred_2d233nonlog}
{\cal H}_{[2^2]233}\Bigg|_{ \frac{\log^0(U)}{N^3} } =  -x_1 x_2 \Bigg[  \frac{ 24 (2-x_1) }{ (1-x_1)x^2_1 } + \frac{ 48 \log(1-x_1)}{x_1^3} \Bigg] + O((x_1 x_2)^2)\;.
\end{equation}
Also in this case, we find that the prediction is compatible with a correlator described by D-correlators.

{\bf Bootstrap  algorithm.} 
We are now ready to implement the position space bootstrap.  
Since the CFT predictions provide evidence that D-functions are sufficient to describe the correlator,  we make the following ansatz,
\begin{equation}\label{eq:ans2q233}
{\cal H}^{\emph ansatz}_{[2^2]323} = \frac{ N_2\, {\cal P}^{(1)}(x_1,x_2) }{(x_1-x_2)^{d-1} } + \frac{ N_{1,U}\, \log(U)  + N_{1,V}\, \log(V) }{ V(x_1-x_2)^{d-2} }  + \frac{ N_{0} }{ V(x_1-x_2)^{d-2} } 
\end{equation}
where $N_2$, $N_{1,U}$, $N_{1,V}$ and $N_0$ are polynomials in $x_1,x_2$ with 
arbitrary coefficients. 

At this point we need a guess for the exponent $d$, which will fix the degree of the polynomials $N_2$, $N_{1}$, $N_0$. 
Our initial guess is to take $d$ as for a correlator with single particle 
operators, therefore $d=(2+2)+3+2+3=12$. We can, of course, validate our guess a posteriori. 
Before proceeding, let us note that the factor of $V$ in the denominator can be inferred from 
the one-variable resummations derived in \eqref{pred_2d323nonlog}.

The totality of available constraints is:
\begin{itemize}
{\item[$\bullet$]  ${\cal H}^{\emph ansatz}_{[2^2]323} |_{\frac{\log(U)}{N^3}}$ starts at twist $5$,  ${\cal H}^{\emph ansatz}_{[2^2]323} |_{\frac{\log^0(U)}{N^3} }$ starts at twist $3$,  and
${\cal H}^{\emph ansatz}_{[2^2]233}|_{\frac{\log^0(U)}{N^3}}$ starts at twist $4$. The corresponding predictions are\footnote{
Crossing here acts as follows:
${\cal H}_{[2^2]233}(x_1,x_2)={\cal H}_{[2^2]323}(\frac{1}{x_1},\frac{1}{x_2})$} 
\begin{equation}
{\cal H}^{\emph ansatz}_{[2^2]323} \Big|_{\frac{\log(U)}{N^3}} = \eqref{pred_2d323log} \qquad;\qquad {\cal H}^{\emph ansatz}_{[2^2]323} \Big|_{\frac{\log^0(U)}{N^3}}=\eqref{pred_2d323nonlog} \qquad;\qquad  {\cal H}^{\emph ansatz}_{[2^2]233} \Big|_{\frac{\log^0(U)}{N^3}}=\eqref{pred_2d233nonlog} \notag \;,
\end{equation}
}
\item[$\bullet$] We impose that there are no spurious poles at $x_1=x_2$. 
\item[$\bullet$] We impose crossing symmetry, 
\begin{equation}
\frac{U^2}{V^2} {\cal H}^{\emph ansatz}_{[2^2]323}(1-x_1,1-x_2)=  {\cal H}^{\emph ansatz}_{[2^2]323}(x_1,x_2) \;.
\end{equation}
\end{itemize}

As a result, we find a unique function. Moreover, we find that 
the output value of $d$ is actually $d=10$. In other words, our initial guess, $d=12$, is reduced by $2$.   
The position space result is
\begin{equation}\label{eq:cs2q233}
{\cal H}_{[2^2]233}\Big|_{\frac{1}{N^3}}= U^2\,( -36 - 12U \partial_U)\circ \overline{D}_{2422}
\end{equation}
and takes a very simple form in Mellin space:  
\begin{equation}\label{eq:ms2q233}
{\cal H}_{[2^2]233}\Big|_{\frac{1}{N^3}}= 
24\oiint_{-i\infty}^{+\infty}\!\!\frac{dsdt}{(2\pi i)^2} U^{s+2} V^t\Gamma[-s]^2\Gamma[-t]^2 \Gamma[-u]^2  \frac{\frac{1}{2}(3+s)}{(1+s)(1+t)(1+u)}\;,
\end{equation}
where $u=-s-t-4$.

{\bf Comments.} Repeating the bootstrap with  $d=10$ from the beginning, we discover that the totality 
of constraints is actually over-constraining with respect to the parameters of the ansatz.  
In particular, the $\log^1(U)$ prediction comes out for free once the $\log^0(U)$ predictions 
and the rest of the constraints are imposed.

\begin{center}
{\underline{\bf Bootstrapping $\langle$[2$^2$]$,$2$,$p$,$p$\rangle$ at tree level}}
\end{center}

The study of the correlators $\langle {\cal O}_2^2  {\cal O}_2{\cal O}_p{\cal O}_p\rangle$ 
for $p\ge 4$ closely follows the bootstrap algorithm detailed in the 
previous section. There are, however, a couple of important novelties in the reasoning that are worth explaining:
~\\[.5cm]
$\bullet$ We have to account for the
degeneracy of double-particle operators.  
~\\[.2cm]
$\bullet$
We have to show that triple-particle operators do not spoil 
the window of predictions we are interested in.\\[-.3cm]

We will discuss both points in the following, for all the various $su(4)$ reps. Before doing so, let us start 
by giving the free theory. Consider the prefactors (we are assuming $p\ge 3$)
\begin{equation}\label{prefactorswith_22_NNE}
{\cal P}_{[2^2]2pp} =  |{\cal O}_p|^2 |{\cal O}_2||{\cal O}_2^2|\  g_{12}^2 g_{14}g_{13}g_{34}^{p-1}
\qquad;\qquad
{\cal P}_{[2^2]p2p} = |{\cal O}_p|^2 |{\cal O}_2||{\cal O}_2^2| \ g_{12}^3 g_{14}g^{p-3}_{24}g_{34}^{2},
\end{equation}
with the two-point normalizations given by $ |{\cal O}_p|^2 |{\cal O}_2||{\cal O}_2^2| = 4p N^{p+3}(1+\ldots)$.
We have
\begin{equation}
 {\cal G}_{[2^2]2pp}=\frac{2p}{\sqrt{a+2}}\Bigg[
 1+\frac{2}{a} +\frac{2}{a}\Bigg[(p-1)U\hat{\sigma}+(p-1)\frac{U\hat{\tau}}{V}+  \frac{(p-2)(p-1)}{2}\frac{U^2\hat{\sigma}\hat{\tau}}{V}  
 \Bigg]\Bigg]\,,
\end{equation}
and for the other orientation,
\begin{equation}\label{free_22p2p}
 {\cal G}_{[2^2]p2p}= 
 \frac{2p}{\sqrt{a+2}}\Bigg[ 
 U^2\hat{\sigma}^2+
 \frac{2}{a}\Bigg[ \frac{(p-2)(p-1)}{2}\frac{U\hat{\tau}}{V}+(p-1) U\hat{\sigma}
 + U^2\hat{\sigma}^2 +(p-1)\frac{U^2\hat{\sigma}\hat{\tau}}{V}\Bigg]
  \Bigg]\,.
\end{equation}

{\bf OPE predictions for }${\cal H}_{[2^2]p2p}$. The relevant exchanged operators belong to 
the $su(4)$ rep $[0,p-2,0]$ whose unitarity bound is $\tau(=2a+b+2)=p$. 
The threshold for triple particle operators is $\tau\ge p+4$, therefore, we can 
make predictions for $\tau=p$ and $\tau=p+2$. From this point of view, the situation is parallel 
to the one for ${\cal H}_{[2^2]323}$. Schematically, we can make the following predictions: 
\begin{equation}\label{table_pred1}
\begin{array}{|c|c|c|}
\hline
\ \log^1(U)\ & \rule{0pt}{.6cm} \ \tau=p+2 \ &\ \   {\cal C}^{(\frac{1}{2})}_{[2^2]p;[2p]}\eta_{[2p]} \, {\cal C}^{(0)}_{2p ;[2p] }\ \ \\[2ex]
\hline
\ \log^0(U)\ & \rule{0pt}{.6cm}\tau=p  & 
 \ \ {\cal S}^{(\frac{3}{2})}_{[2^2]p2p}- A^{(\frac{3}{2})}_{[2^2]p2p} \ \  \\[2ex]
\hline
\end{array}
\end{equation}
where we shall see that ${\cal S}^{(\frac{3}{2})}_{[2^2]p2p}=0$.

It is a part of our claim to prove that triple-particle operators 
do not contribute to the predictions \eqref{table_pred1} at $O(\frac{1}{N^3})$. 
The case of ${\cal C}_{[2^2]323}$ was special because 
at low twist there are no triple-particle operators. However, 
this is not the general case. 
Now, the triple-trace contributions to table \eqref{table_pred1} have schematically the form 
\begin{equation}\label{discarding_tripltrace}
{\cal C}_{[2^2]p; {\cal T}_{p+2}}\frac{ \eta_{} }{N^2}  {\cal C}_{2p {\cal T}_{p+2}}  \log^1(U) \qquad;\qquad 
{\cal C}_{[2^2]p; {\cal T}_{p}}{\cal C}_{2p; {\cal T}_{p}} \log^0(U)\;,
\end{equation}
where ${\cal T}_{\vec{\tau}}$ is a generic triple particle operator of twist $\tau$. 
We want to focus on the scaling of the three-point couplings, and for this reason, we 
included already a $\frac{1}{N^2}$ for the anomalous dimension. 
Consider first  the $\log(U)$ prediction in \eqref{discarding_tripltrace}. Then,  
\begin{equation}
{\cal C}_{2p; {\cal T}_{p+2}}\propto O(\tfrac{1}{N})\qquad;\qquad {\cal C}^{}_{[2^2]p {\cal T}_{p+2}}\propto O(\tfrac{1}{N^2})\;.
\end{equation} 
This is the expected behavior because on the LHS the twist $\tau=p+2$ is above threshold for ${\cal C}_{2p;\#}$, while 
on the RHS the twist $\tau=p+2$ is below the threshold for $ {\cal C}^{}_{[2^2]p;\#}$. 
(This is what we explained in section \ref{overV}.) 
Putting the two three-point couplings together yields, 
\begin{equation}
{\cal C}_{[2^2]p; {\cal T}_{p+2}}\frac{ \,\,\eta^{} }{N^2} \, {\cal C}_{2p {\cal T}_{p+2}}  \log^1(U)\ \propto \ O(\tfrac{1}{N^5})\;.
\end{equation}
Similarly, the $\log^0(U)$ prediction in \eqref{discarding_tripltrace} is $O(\frac{1}{N^5})$ because
\begin{equation}
{\cal C}^{}_{[2^2]p {\cal T}_{p}}\propto O(\tfrac{1}{N^2})\qquad;\qquad {\cal C}_{2p; {\cal T}_{p}}\propto O(\tfrac{1}{N^3})\;.
\end{equation}
We conclude that at order $O(\frac{1}{N^3})$ there are only double-particle operators exchanged in the window of twists
$\tau=p$ and $\tau=p+2$. 

Let us now proceed by computing the predictions, quoted already in \eqref{table_pred1}.  
The computation will also show how the degeneracy of double-particle operators is 
taken into account, which in \eqref{table_pred1} was already understood.

\underline{\emph{log}$^1(U)$ \emph{sector}}. There are $\lfloor\frac{p}{2}\rfloor$ 
degenerate long double-particle operators of twist $p+2$ in the $su(4)$ rep $[0,p-2,0]$. 
The space is spanned by $\big\{ {\cal O}_2\partial^l{\cal O}_p,{\cal O}_3\partial^l{\cal O}_{p-1},\ldots \big\}$. 
Since we know the operators, the prediction we are looking for can be obtained by 
gathering together data from all four-point correlators where these operators are exchanged.
Generalizing the argument explained around \eqref{rewrite_pred5}, we obtain the formula
\begin{equation}\label{situationlog1U2dpij}
 \sum_{ R_{p+2,[0,p-2,0]} }\!\!\!\!\!{\cal C}^{(\frac{1}{2})}_{[2^2]p; {\cal D}_{}} \, \eta_{{\cal D}_{}} \, {\cal C}^{(0)}_{2p ;{\cal D}_{} } = 
  \sum_{ij}\frac{ \left( {\cal G}_{[2^2]pij}^{(\frac{1}{2})} \Big|_{[0,p-2,0],\tau=p+2,l} \right) \times \left(  {\cal H}^{(1)}_{ij2p} \Big|_{\log(U),[0,p-2,0],\tau=p+2,l} \right) }{ 
\left(  {\cal G}^{(0)} _{ijij} \Big|_{[0,p-2,0],\tau=p+2,l} \right)}\;,
\end{equation}
where the sum goes over the set $ij=\{(2,p),(3,p-1),\ldots\}$ and corresponds to the rectangle explained in section \ref{REVdP}.   

At this point, note that the only non-vanishing disconnected contribution 
${\cal G}_{[2^2]pij}^{(\frac{1}{2})}$ is when $i,j=2,p$, namely, ${\cal G}_{[2^2]p2p}^{(\frac{1}{2})}$. 
In particular, the only disconnected diagram in $\langle {\cal O}_2^2{\cal O}_p{\cal O}_2{\cal O}_p \rangle$ is 
\begin{equation}\label{fig2pptimes22}
\begin{array}{c}
\begin{tikzpicture}[scale=1.2]  
\def\shift{.35}

\def\latoxuno{-.35}
\def\latoxdue{-.38+0.8}
\def\latoyuno{.3}
\def\latoydue{-.25}

\def\latoxuno{-.35}
\def\latoxdue{-.38+1.5}
\def\latoyuno{.7}
\def\latoydue{-.25}

\draw[fill=red!60] 
(\latoxuno+.05, \latoydue) -- (\latoxdue, \latoyuno-.05) --  
(\latoxdue-.05, \latoyuno) -- (\latoxuno, \latoydue+.05) -- cycle;

\draw (\latoxuno-.1, \latoyuno) --  (\latoxuno-.1, \latoydue+.1);
\draw (\latoxuno-.1, \latoyuno) --  (\latoxdue, \latoyuno);

\draw (\latoxdue-.05, \latoydue) --  (\latoxuno, \latoyuno+.2-.05);
\draw (\latoxdue, \latoydue+.05) --  (\latoxuno+.05, \latoyuno+.2);

\draw[fill=white!60,draw=white] (\latoxuno-.2, \latoyuno-.1) rectangle (\latoxuno, \latoyuno+.1);
\draw[fill=white!60,draw=white] (\latoxdue-.1, \latoydue-.05) rectangle (\latoxdue+.15, \latoydue+.1);

    \draw[fill=white!60,draw=white] (\latoxdue+.02, \latoyuno+.02) circle (2.5pt);
    \draw[fill=white!60,draw=white] (\latoxuno-.02, \latoydue+.02) circle (2.5pt);

\draw(\latoxuno-.1, \latoyuno+.2+.08) node[scale=.8] {${\cal O}_2$};
   \draw(\latoxuno-.2, \latoyuno+.08) node[scale=.8] {${\cal O}_2$};
   \draw(\latoxuno-.04, \latoydue-.08) node[scale=.8] {${\cal O}_p$};
   \draw(\latoxdue+.14, \latoyuno+.03) node[scale=.8] {${\cal O}_p$};
   \draw(\latoxdue+.1, \latoydue-.05) node[scale=.8] {${\cal O}_2$};

\end{tikzpicture} 
\end{array}
 \end{equation}
which at order $O(\frac{1}{N})$ disconnects into the product of a two 
point function $\langle {\cal O}_2{\cal O}_2\rangle$ and a NE three-point function 
$\langle {\cal O}_p{\cal O}_p{\cal O}_2\rangle$. 
It follows that the sum over $ij$ truncates to $i,j=2,p$, thus
\begin{equation}\label{prediction_p_plus_2}
  \sum_{ R_{p+2,[0,p-2,0]}  }\!\!\!\!\!\!\!{\cal C}^{(\frac{1}{2})}_{[2^2]p; {\cal D}_{}}\eta_{{\cal D}_{}} \, {\cal C}^{(0)}_{2p ;{\cal D}_{} } = 
\frac{ \left( {\cal G}_{[2^2]p2p}^{(\frac{1}{2})} \Big|_{[0,p-2,0],\tau=p+2,l} \right) \times \left(  {\cal H}^{(1)}_{2p2p} \Big|_{\log(U),[0,p-2,0],\tau=p+2,l} \right) }{ 
\left(  {\cal G}^{(0)} _{2p2p} \Big|_{[0,p-2,0],\tau=p+2,l} \right)}\;.
\end{equation}
The relevant CFT data is obtained as follows.  From the disconnected free theory, we find
 \begin{equation}\label{prediction_p_plus_2_disc}
{\cal G}^{(0)} _{2p2p} \Bigg|_{[0,p-2,0],\tau=p+2,l}\!\!\!=\, M^{2p2p}_{0,p+2,[2+l,2]} 
= \frac{p^2(p-1)}{(p+2)!}\frac{(l+1)(l+p+1)!(l+p+4)!}{(l+p+2)(l+p+3)(2l+p+4)!}\;.
\end{equation}
From tree level supergravity we find\footnote{Recall that ${\cal H}^{(1)}_{2p2p}=-{2p}/{(p-2)!} U^2 \overline{D}_{2,p+2,2,p}$} 
 \begin{align}
 &
 \!\!\!\!\!\!{\cal H}^{(1)}_{2p2p} \Bigg|_{\log(U),[0,p-2,0],\tau=p+2,l} \!\!\!=   
 \frac{(l+3)!(l+p+1)!}{(2l+p+4)!} \left\{\begin{array}{ll}  
(-1)^p c_{p}(-l-p-5) - {4(p-1)p^2}{}\frac{1+(-1)^p}{2}   & l\ {\rm even}\\[.2cm] 
 c_p(l)&  l\ {\rm odd}\\
 \end{array}\right.
\end{align}
where the polynomials $c_p(l)$ can be found case-by-case, for example 
\begin{align}
c_3(l)&=-12(l+1)\;,\\
c_4(l)&=-8(l+1)(l+8)\notag\;,\\
c_5(l)&=-\frac{10}{3}(l+1)(60+14l+l^2)\notag\;,\\
c_6(l)&=-(l+1)(l+10)(48+11l+l^2)\notag\;,\\
\vdots &\rule{1cm}{0pt}    \vdots\;  \notag 
\end{align}
Finally, from connected free theory, we find
\begin{align}\label{prediction_p_plus_2_conn}
 {\cal G}_{[2^2]p2p}^{(\frac{1}{2})} \Bigg|_{[0,p-2,0],\tau=p+2,l}& = 
 (2p)\times M^{4,p,2,p}_{0,p+2,[2+l,2]} 
 =  \frac{ 2p^2 (p-1)^2 (p-2) }{ (p+2)!} \frac{(l+1)(l+p+4)(l+p+1)!^2}{(l+p+1)(2l+p+4)!} \;.
 \end{align}
In the above formula, the value $\frac{2p}{N}$ is the color factor corresponding to the propagator 
structure drawn in \eqref{fig2pptimes22}. It can be understood by looking at~\eqref{fig2pptimes22}: 
at leading order in $N$, the double contraction between ${\cal O}_2^2$ and ${\cal O}_2$ yields 
a factor of $2|{\cal O}_2|^2$, while the remaining three point function yieds $2p |{\cal O}_p|^2$. Thus, 
their product is $2|{\cal O}_2|^2 2p |{\cal O}_p|^2\simeq \left( |{\cal O}_p|^2 |{\cal O}_2||{\cal O}_2^2|\right) 
\left(\frac{2p}{N}\right)$, where in the last step we isolated the prefactor for ${\cal P}_{[2^2]p2p}$ 
  in~\eqref{prefactorswith_22_NNE} and used~\eqref{eq:O2nor} at large $N$.
  
We now have all the data that we need to assemble the prediction in \eqref{prediction_p_plus_2}. 
Let us note again that the free theory parts of the prediction in \eqref{prediction_p_plus_2} 
are almost trivial to find since, as in the examples before, one just needs 
to decompose the tree-level supergravity result by following the approach of~\cite{Aprile:2025nta}.
Nicely enough, the form of the prediction given as a one-variable sum over spin is very simple, 
\begin{equation}\label{pred_2dp2plog}
  {  {\cal H}_{2^2p2p}\Bigg|_{\frac{\log(U)}{N^3}} =  -4p\times (p-2)(p-1)^2p\Bigg[ U^2\sum_{\lm} \frac{(\lm+1)(\lm+2)}{(\lm+p+1)_3} x_1^\lm + \ldots\Bigg]}\;.
\end{equation}
The sum over $l$ gives
a Gauss hypergeometric function that can be written explicitly by introducing the Lerch function $\Phi (z,s,a)$, 
  \begin{equation}
    \label{eq:Lerch}
    \Phi (z,s,a)  =\sum_{k=0}^\infty \frac{z^k}{(k+a)^s}\;,
  \end{equation}
e.g.~one can verify the identity
  \begin{equation}
    \label{eq:sdf}
    \sum_{l}\frac{(\lm+1)(\lm+2)}{(\lm+p+1)_3} x_1^\lm = \frac{p-1}{2 x_1}-\frac{p+2}{2x_1^2} + \left[\frac{(p+1) (p+2)}{2 x_1^2}-\frac{(p+1) p}{x_1}+\frac{(p-1) p}{2} \right] \Phi (x_1,1,p+1) \,,
 \end{equation}
 where  $\Phi (x_1,1,p+1)=\frac{1}{x_1^p}( -\frac{\log(1-x_1)}{x_1}-\sum_{n=0}^{p-1} \frac{x_1^n}{n+1})$.

With formula \eqref{eq:sdf} is then immediate to check that the prediction at $p=3$ coincides 
with \eqref{pred_warmup_logU}, and one can compute the predictions for an arbitrary integer $p$. 
The structure of $\Phi (x_1,1,p+1)$ gives a $\log(1-x_1)$ term and a rational function,
thus the prediction is compatible with a space of correlators written in terms of D-functions.

\underline{\emph{log}$^0(U)$ \emph{sector}}. The unitarity bound in the $[0,p-2,0]$ 
rep is $\tau(=2a+b+2)=p$. The relevant operators are protected double particle operators. For $p=4,5,$ 
we are still in a situation of non-degenerate operators, namely ${\cal O}_2\partial^l{\cal O}_2$ and 
${\cal O}_2\partial^l{\cal O}_3$. However, for $p\ge 6$ we find degenerate operators, for example, 
${\cal O}_2\partial^l {\cal O}_4$ and ${\cal O}_3 \partial^l {\cal O}_3$ at  twist $6$ double, 
and so on so forth. The crucial observation is again this: 
at order $\frac{1}{N^3}$, we can restrict attention to double-particle operators and
compute the relevant three-point couplings by constructing a Gram matrix whose determinant vanishes.  
This Gram matrix  generalises the one considered in \eqref{020_Gram_M},  and will have the form
\begin{equation}\label{situationlog0U2dp2p}
\ \ \ \ \ \ \left(\begin{array}{cc} 
		\langle {\cal O}_{i} {\cal O}_{j} {\cal O}_{a}{\cal O}_{b} \rangle  & V^T_{2p} \\[.2cm] 
				V_{[2^2]p} & \langle {\cal O}_2^2{\cal O}_p{\cal O}_2{\cal O}_p\rangle   \end{array}\right)\;,
\end{equation}
where the vector $V_{2p}$ is 
\begin{equation}\label{situationlog0U2d2p_vec2p}
V_{2p}=\Big( \langle {\cal O}_{2} {\cal O}_{p-2} {\cal O}_2{\cal O}_p \rangle ,  \langle {\cal O}_{3} {\cal O}_{p-3} {\cal O}_2{\cal O}_p \rangle ,\ldots \Big)\,,
\end{equation}
thus it includes all correlators of the type $\langle {\cal O}_{a} {\cal O}_{b}{\cal O}_2{\cal O}_p \rangle$ such that $a+b=p-2$, while the vector $V_{[2^2]p}$ is
\begin{equation}
V_{[2^2]p}=\Big( \langle {\cal O}^2_{2} {\cal O}_{p} {\cal O}_2{\cal O}_{p-2} \rangle ,  \langle  {\cal O}^2_2{\cal O}_p  {\cal O}_{3} {\cal O}_{p-3}\rangle ,\ldots \Big)\,.
\end{equation}
 Crucially, all the correlators listed in $V_{2p}$ and  $V_{[2^2]p}$ vanish for external single particle operators!  
 The list in $V_{2p}$ corresponds to four-point correlators that are near extremal. The list in $V_{[2^2]p}$ can 
 be understood as five-point correlators that are near extremal. As proven in \cite{Aprile:2020uxk}, they all vanish 
 for external single particle operators. 
Thus 
\begin{equation}
{\cal S}_{[2^2]p2p}=0\;.
\end{equation}
The one variable spin prediction at twist $p$ is simply, 
\begin{equation}\label{pred_2dp2pnonlog}
{\cal H}_{[2^2]p2p}\Bigg|_{\frac{ \log^0(U)}{N^3} }=  2p\times {(p-2)(p-1)} \Bigg[- U \sum_{\lm=0}^\infty \frac{(\lm+1)(\lm+2p)}{(\lm+p)(\lm+p+1)} x_1^\lm    + \ldots\Bigg]\;,
\end{equation}
and with similar technique as in \eqref{eq:sdf} we find
\begin{equation}
  \label{eq:sum2}
   \sum_{\lm=0}^\infty \frac{(\lm+1)(\lm+2p)}{(\lm+p)(\lm+p+1)} x_1^\lm = -\frac{(p-1) p (x_1-1)}{x_1} \Phi (x_1,1,p)-\frac{1+p(x_1-1)}{(x_1-1) x_1}\;.
\end{equation}
As before, this shows explicitly the compatibility with a correlator made by D-functions.

{\bf OPE predictions for }${\cal H}_{[2^2]2pp}$. 
The relevant exchanged operators belong to the $su(4)$ rep $[0,2,0]$.  Since the 
threshold for exchanging triple-particle operators in this orientation is twist six, independent of $p$, 
there is nothing we can say about $\tau\ge 6$, and we can only make a prediction 
for the exchange of $\tau<6$ operators.  Then, the relevant operators are again the $\tau=4$ 
protected semishort double particle operators at the unitarity bound, ${\cal O}_2\partial^l {\cal O}_2$. 
Therefore, the logic of our next discussion will be similar to the one for 
${\cal H}_{[2^2]233}$, with only minor modifications.

Let ${\cal S}^{[2^2]2pp}$ be the prediction for the exchange of the semishort operators. 
Repeating the discussion of the Gram matrix, as in \eqref{020_Gram_M}, we find, 
\begin{equation}\label{prediction_Stwist4_p}
{\cal S}_{[2^2]2pp,[020],\tau=4,l}\Bigg|_{\frac{1}{N^3}}=
\frac{ \left( {\cal G}^{(\frac{1}{2})}_{[2^2]222} \Big|_{[0,2,0],\tau=4,l} \right) \times\left( 
{\cal G}^{(1)}_{22pp} \Big|_{[0,2,0],\tau=4,l} \right)  }{  \left( {\cal G}^{(0)}_{2222} \Big|_{[0,2,0],\tau=4,l}\right) }\;.
\end{equation}
More explicitly from~\eqref{eq:sfr2}
\begin{equation}
{{\cal S}_{[2^2]2pp,[020],\tau=4,l}\Bigg|_{\frac{1}{N^3}}\!\!\!\!\!\!=
\frac{ 4 \left[ M^{4222}_{0,4,[l+2]} + M^{4222}_{1,4,[l+2]} \right] \times\left[ 2p(p-1) M^{22pp}_{1,4,[l+2]} \right]  }{  
\left( M^{2222}_{0,4,[l+2]} + M^{2222}_{2,4,[l+2]}\right) } = \frac{ 16p(p-1) (l+2)!(l+3)! }{(2l+6)!}\frac{1+(-1)^l}{2}\,.}
\end{equation}
Considering now that the free theory term is
\begin{equation}
A_{[2^2]2pp,[020],{4,l}}\Bigg|_{\frac{1}{N^3}}\equiv 
4p (p-1)\big( M^{42pp}_{0,4,[l+2]}+M^{42pp}_{1,4,[l+2]}\big)=\frac{4p\times 2(p-1) (l+3)!(l+4)!}{(2l+6)!}\frac{1+(-1)^l}{2}\;,
\end{equation}
we obtain the prediction
\begin{equation}
{\cal H}_{[2^2]2pp}\Bigg|_{ \frac{\log^0(U)}{N^3} } = \sum_{l} \Big( - A_{[2^2]2pp}+{\cal S}_{[2^2]2pp} \Big) H_{[020],4,l}  +\ldots \;.
\end{equation}
The spin dependence is the same for all $p$, thus we can use the result for $p=3$ 
and multiply by an overall prefactor. As a result,
\begin{align}\label{pred_2d2ppnonlog}
{\cal H}_{[2^2]2pp}\Bigg|_{ \frac{\log^0(U)}{N^3} }&
 =  -x_2 \Bigg[  \frac{ 4p(p-1) (2-x_1) }{ (1-x_1)x_1 } + \frac{ 8p(p-1) \log(1-x_1)}{x_1^2} +\ldots \Bigg]\;.
\end{align}
Again, the result for all $p$ is compatible with a correlator described by D-functions.

{\bf Bootstrap algorithm: from position space to Mellin space.} 
We are ready to implement the position space bootstrap for ${\cal H}_{[2^2]2pp}$, 
as we did for the case $p=3$. Since the correlators are N$^2$E, the ansatz 
has a similar structure, 
\begin{equation}
{\cal H}^{\emph ansatz}_{[2^2]p2p} = \frac{ N_2\, {\cal P}^{(1)}(x_1,x_2) }{(x_1-x_2)^{d-1} } 
+ \frac{ N_{1,U}\, \log(U)  + N_{1,V}\, \log(V) }{ V(x_1-x_2)^{d-2} }  + \frac{ N_{0} }{ V(x_1-x_2)^{d-2} } \;,
\end{equation}
where $N_2$, $N_{1,U}$, $N_{1,V}$ and $N_0$ are polynomials in $x_1,x_2$ with 
arbitrary coefficients. We only need a guess for the exponent $d$.
Our initial guess is to take the same exponent as for the single-particle correlators, 
$d=(2+2)+2+p+p$, namely $d=2p+6$. We will now proceed in position space, 
on a case-by-case basis, i.e.~by fixing $p$. Then, we will go to Mellin space to 
find the dependence on $p$.

The totality of available constraints is:
\begin{itemize}
\item[$\bullet$]  ${\cal H}^{\emph ansatz}_{[2^2]p2p} |_{\frac{\log(U)}{N^3} }$ starts at $\tau=p+2$,  
${\cal H}^{\emph ansatz}_{[2^2]p2p} |_{\frac{\log^0(U)}{N^3} }$ starts at $\tau=p$,  and
${\cal H}^{\emph ansatz}_{[2^2]2pp}|_{\frac{\log^0(U)}{N^3} }$ starts at $\tau=4$. 
The corresponding predictions are\footnote{
Crossing here implies the following relation:
${\cal H}_{[2^2]2pp}(x_1,x_2)={\cal H}_{[2^2]p2p}(\frac{1}{x_1},\frac{1}{x_2})$. }
\begin{equation}
{\cal H}^{\emph ansatz}_{[2^2]p2p} \Big|_{\frac{\log^1(U)}{N^3} } = \eqref{pred_2dp2plog} 
\qquad;\qquad 
{\cal H}^{\emph ansatz}_{[2^2]p2p} \Big|_{\frac{\log^0(U)}{N^3} }=\eqref{pred_2dp2pnonlog} 
\qquad;\qquad  
{\cal H}^{\emph ansatz}_{[2^2]2pp} \Big|_{\frac{\log^0(U)}{N^3}}=\eqref{pred_2d2ppnonlog}\,. \notag
\end{equation}

\item[$\bullet$] We impose that there are no spurious poles at $x_1=x_2$.  
\item[$\bullet$] We impose crossing symmetry, 
\begin{equation}
\frac{U^2}{V^2} {\cal H}^{\emph ansatz}_{[2^2]p2p}(1-x_1,1-x_2)=  {\cal H}^{\emph ansatz}_{[2^2]p2p}(x_1,x_2) \,.
\end{equation}
\end{itemize}

For any $p=4,5,6,\ldots$  we find a unique function. Moreover, we find that 
the output value of $d$ is actually $d=2p+4$. In other words, our initial guess is 
again reduced by $-2$. 

We are interested in a formula where the $p$ dependence is explicit, however, 
the polynomials $N_{2}$, $N_{1}$ and $N_{0}$ become complicated since their degree 
grows as a function of $p$, as it grows the exponent of $(x_1-x_2)$ in the common 
denominator. The situation can be dramatically improved by going to Mellin space where 
D-functions are mapped to simple rational functions. The reason is that, in this map,
part of the complexity of the position space result is absorbed into a convenient choice of Gamma functions, 
and the rest is a rational function. Thus, by performing the Mellin transform 
on the first few cases, $p=4,5,6$, it is simple to see a pattern and make an 
ansatz for the correlator directly in Mellin space, such that it depends only on a
few free coefficients. At this point, we impose the OPE predictions and fix the ansatz. 

{The result, which generalises~\eqref{eq:ms2q233}, is:}
\begin{equation}\label{solu2d2pp}
{\cal H}_{[2^2]2pp}\Big|_{\frac{1}{N^3}}= 
4p\frac{(p-1)}{(p-3)!}\oiint_{-i\infty}^{+\infty}\!\!\frac{dsdt}{(2\pi i)^2} U^{s+2} V^t
\frac{ \Gamma[-s+p-3]^2\Gamma[-t]^2 \Gamma[-u]^2}{(-s)_{p-3}}  \frac{\frac{1}{2}(3+s)}{(1+s)(1+t)(1+u)}\,,
\end{equation}
where $u=-s-t-4$. Nicely, the factor $3+s$ is the same as for $p=3$ correlator. The result 
for the overall factor has the following interpretation: The factor $4p$ 
is just the two-point function normalization, and even though we started from 
$p=3$, the denominator $(p-3)!$ imposes kinematically 
that the reduced correlator vanishes for $p=2$.\footnote{This $1/(p-3)!$ is present also in 
the tree-level single-particle ${\cal H}_{42pp}$, confirming that is a consequence of kinematics.}

A posteriori, we can write the result also in position space, {generalizing~\eqref{eq:cs2q233}}
\begin{equation}
{{\cal H}_{[2^2]2pp}\Big|_{\frac{1}{N^3}}= 2p\frac{(p-1)}{(p-3)!} U^{p-1} \partial_U^{p-3} ( -3  - U \partial_U) \overline{D}_{2422}\;.}
\end{equation}

{\bf Comments.} Repeating the position space bootstrap by assuming $d=2p+4$ from 
the beginning, as for $p=3$, we discover that the totality of constraints is over-constraining 
with respect to the parameters of the ansatz. In particular, the $\log^1(U)$ prediction 
comes out for free once the $\log^0(U)$ predictions and the rest of the constraints are imposed.  

In the orientation ${\cal H}_{[2^2]2pp}$, triple-particle operators appear 
starting from $\tau\ge 6$, and the double-particle bootstrap only makes a prediction at $\tau=4$. 
(We made a similar observation for ${\cal H}_{[2^2]p2p}$ starting from twist $p+4$.)
However, since the final result is uniquely determined, we find that \eqref{solu2d2pp} 
has more information than what we input, 
and in particular, it contains average data for three-point couplings and anomalous 
dimensions with triple particle operators with $\tau\ge6$. It is a good point to review what these data is.

Consider again ${\cal H}_{[2^2]2pp}$. The general explanation that we gave 
in section \ref{overV} here gives the following picture:
Let ${\cal T}_{\vec{\tau}}$ be a triple-particle operator. The three-point coupling 
${\cal C}_{[2^2]2{\cal T}}$ is $O(1)$ if $\tau\ge 6$. Similarly,  ${\cal C}_{pp{\cal T}}=O(\frac{1}{N})$ 
if $\tau\ge 2p$, otherwise, then ${\cal C}_{pp{\cal T}}=O(\frac{1}{N^3})$. 
Now, let ${\cal D}_{\vec{\tau}}$ be a double-particle operator. 
The three-point couplings ${\cal C}_{[2^2]2{\cal D}}$ is $O(\frac{1}{N})$ if $\tau\ge 6$. Similarly, 
${\cal C}_{pp{\cal D}}=O(1)$ if $\tau\ge 2p$, otherwise, ${\cal C}_{pp{\cal D}}=O(\frac{1}{N^2})$.   
The block expansion of \eqref{solu2d2pp} for $\tau\ge6$  contains the CFT data of these operators. 
The conversion from $s$ to twist is $\tau=2+2(s+2)$. Thus, the simple poles $s=0,1,\ldots,p-4$,  
correspond to $\tau=6,\ldots,2p-2$, and contain a combination of three-point couplings, 
\begin{equation}
\sum_{\cal T} C^{(0)}_{[2^2]2{\cal T}}\times C^{(\frac{3}{2})}_{pp{\cal T}}  + 
\sum_{\cal D} C^{(\frac{1}{2})}_{[2^2]2{\cal D}}\times  C^{(1)}_{pp{\cal D}}\,.
\end{equation}
The double poles $s\ge p-3$ correspond to $\tau\ge 2p$, and contain a 
combination of three-point couplings and anomalous dimensions 
\begin{equation}
\sum_{\cal T} C^{(0)}_{[2^2]2{\cal T}}\, \eta_{\cal T} \,C^{(\frac{1}{2})}_{pp{\cal T}}  
+ \sum_{\cal D} C^{(\frac{1}{2})}_{[2^2]2 {\cal D}}\, \eta_{\cal D} \,C^{(0)}_{pp{\cal D}}\,.
\end{equation}
The value of $\eta_{\cal D}$, ${\cal C}^{(0)}_{pp{\cal D}}$ and ${\cal C}^{(1)}_{pp{\cal D}}$ 
is known in general from the study of the tree-level four-point single particle correlators \cite{Aprile:2018efk}, 
but the rest of the data, involving the external double-particle operator or the exchanged 
triple-particle operator, is not known. For general spin of the exchanged triple-particle 
operators, the mixing of double- and triple-particle operators is a hard problem. 
However, for low values of the spin, say $l=0,1$, it is possible to discuss 
the unmixing equations, by including also four-point correlators with single-particle 
operators and correlators with two double- and two single-particle operators.

\subsection{More bootstrap results}\label{more_correlators_sec}

In this section, we present results for other families of correlators. 
As in the previous sections, the bootstrap algorithm starts by 
making a list of predictions for all orientations of the external operators and 
all $su(4)$ channels.  The aforementioned predictions are then computed
by gathering together CFT data from the free theory and from 
the tree-level single-particle correlators. As we showed in the test cases, 
this data is naturally arranged into matrices. For ${\cal H}_{[2^2]2pp}$, we could 
reduce the problem to one-dimensional matrix multiplication, but for 
${\cal H}_{[r_1r_2]spq}$ we generically deal with matrices. 
As a representative of more general families of reduced correlators, 
we will consider here, 
\begin{equation}
{\cal H}_{[2^2]sp(p+s-2)}
\qquad;\qquad 
{\cal H}_{[3^2]2pp}\,,\,{\cal H}_{[24]2pp}
\qquad;\qquad 
{\cal H}_{[r^2]2pp}\;.
\end{equation}
As the degree of extremality of the correlator increases, and similarly, 
as the charge of the double-particle increases, 
some steps of the bootstrap become more involved.   
Nevertheless, it is always possible to bootstrap a given 
correlator of interest, as we shall demonstrate. 

In all cases that we studied, we have found that ${\cal H}_{[r_1r_2]spq}$ 
in position space comes with a denominator  $(x_1-x_2)^{d-1}$ where
\begin{equation}
d=(r_1+r_2)+s+p+q-2\;,
\end{equation}
and, as in our test cases, it is possible to overconstraint the initial ansatz in position space. 
Note that tree-level single particle correlators at four points have a similar property, and 
by analogy with what happens there, we are led to think that also ${\cal H}_{[r_1r_2]spq}$ might be determined 
by imposing symmetries on a Witten diagram expansion, as in the derivation of 
Rastelli-Zhou \cite{Rastelli:2016nze,Rastelli:2017udc} of the four-point single-particle 
correlators.\footnote{We thank Xinan Zhou for a conversation about this point. In fact, this computation 
is closely related to the five-point computation depicted in Figure \ref{fig:Witten_diagrams_5pts}, 
since all Witten diagrams corresponding to the exchange of bulk fields can be enumerated.}

\subsubsection{Higher extremality with ${\cal O}_2^2$}\label{sec_22_NNNE}

We are interested in the reduced correlators of the form ${\cal H}_{[2^2]p_2p_3p_4}$ with ${\cal O}_2^2$ fixed. 
Depending on the charge of the single-particle operators, $p_i$, these correlators
can be N$^2$E, or N$^3$E, or N$^4$E. In the previous section, we studied a 
family of N$^2$E correlators. Here, we will study correlators  
${\cal H}_{[2^2]p_2p_3p_4}$ that are N$^3$E.

Let us consider the family ${\cal H}_{[2^2]p3(p+1)}$. We shall assume that $p\ge3$, otherwise for $p=2$ 
we go back to the case of N$^2$E correlator that we studied already.
Given these external charges, there are three independent orientations to study. Let us begin from 
the orientation ${\cal H}_{[2^2]p3p+1}$. In this case, there are three long 
$su(4)$ rep: $[0,p-2,0]$, $[1,p-2,1]$ and $[0,p,0]$. The threshold for the exchange 
of triple-particle operators is the same for ${\cal O}_{2}^2{\cal O}_p$ 
and ${\cal O}_3{\cal O}_p$, namely $p+4$, thus we can only make predictions for a value 
of the twist strictly below this threshold, but greater or equal than the unitary bound of each rep. 
In particular, in this orientation, we can only make predictions for the non-log$(U)$ sector. 
A similar reasoning will be repeated in the other two orientations of the correlator.

The tables below summarise what predictions the double-particle bootstrap computes 
in the two orientations, ${\cal H}_{[2^2]p3(p+1)}$ and ${\cal H}_{[2^2]3p(p+1)}$. 
We grouped these two together because they have a similar structure:
\begin{equation}\label{pred2dnontrivialN3E}
\!\!\!\!\!\begin{array}{ccc}
\begin{tikzpicture}
\draw (0,0) node[scale=.8] {$
\begin{array}{|c|c|c|c|}
 \hline  & & &\\
 \ \log^0(U)\ & \rule{0pt}{.6cm}\tau=p+2  & \begin{array}{c} \left[0,p,0 \right] \\ \left[1,p-2,1\right] \\ \left[0,p-2,0\right] \end{array}  &
  \emph{non\,trivial}   \ \  \\[5ex]
 \hline  \ \log^0(U)\ & \rule{0pt}{.6cm}\tau=p  &   \left[0,p-2,0 \right]  & \ \ - A_{[2^2]p3(p+1)} \ \  \\[2ex]
 \hline
 \multicolumn{4}{c}{\rule{0pt}{.8cm}\textrm{In the orientation}\ {[2^2]p3(p+1)}}  
 \end{array}$};
 \end{tikzpicture} 
 & \rule{.5cm}{0pt} &
 \begin{tikzpicture}
\draw (0,0) node[scale=.8] {$
\begin{array}{|c|c|c|c|}
 \hline  & & &\\
 \ \log^0(U)\ & \rule{0pt}{.6cm}\tau=5  & \begin{array}{c} \left[0,3,0 \right] \\ \left[1,1,1\right] \\ \left[0,1,0\right] \end{array}  &
  \emph{non\,trivial}   \ \  \\[5ex]
 \hline  \ \log^0(U)\ & \rule{0pt}{.6cm}\tau=3  &   \left[0,1,0 \right]  & \ \ - A_{[2^2]3p(p+1)} \ \  \\[2ex]
 \hline 
 \multicolumn{4}{c}{\rule{0pt}{.8cm}\textrm{In the orientation}\ {[2^2]3p(p+1)}} 
 \end{array}$};
 \end{tikzpicture} 
 \end{array}
\end{equation}
The predictions labelled as \emph{non trivial} are explained in Appendix \ref{appendix_more}. 
Here we will give some more details on
the predictions in the orientation ${\cal H}_{[2^2](p+1)3p}$, which are the following
\begin{equation}\label{pred2dpp13p}
\begin{array}{c}
\begin{tikzpicture}
\draw (0,0) node[scale=.8] {$
\begin{array}{|c|c|c|c|}
 \hline  & & &\\
\ \log^1(U)\ & \rule{0pt}{.6cm} \ \tau=p+3 \ &  \begin{array}{c} \left[0,p-3,0 \right] \\ \left[1,p-3,1\right] \\ \left[0,p-1,0\right] \end{array}   &  
\begin{array}{c}  0 \\ 0  \\ {\cal G}^{(\frac{1}{2})}_{[2^2](p+1)2p+1}{\cal H}^{(1)}_{2(p+1)3p}/ {{\cal G}^{(0)}_{2(p+1)2(p+1)}}\end{array} \\[5ex]
 \hline & &  &\\
 \ \log^0(U)\ & \rule{0pt}{.6cm}\tau=p+1  &  \begin{array}{c} \left[0,p-3,0 \right] \\ \left[1,p-3,1\right] \\ \left[0,p-1,0\right] \end{array} &
 \ \ - A_{[2^2](p+1)3p} \ \  \\[5ex]
 \hline  \ \log^0(U)\ & \rule{0pt}{.6cm}\tau=p-1  &   \left[0,p-3,0 \right]  & \ \ 0 \ \  \\[2ex]
 \hline 
 \multicolumn{4}{c}{\rule{0pt}{.8cm} \textrm{In the orientation}\ {[2^2](p+1)3p}} 
 \end{array}$};
 \end{tikzpicture}
 \end{array}
\end{equation}
Let us begin by noting that the orientation ${\cal H}_{[2^2](p+1)3p+1}$ is 
the only one in which we can predict a piece of data in the $\log(U)$ sector. 
However, out of the three possible reps, only $[0,p-1,0]$ has a 
non-trivial contribution, and the other vanish:
In the case of $[1,p-3,1]$ the predictions vanish because among the correlators 
obtained from $R_{\tau=p+1,[1,p-3,1]}$ and ${\cal O}_2^2{\cal O}_{p+1}$, 
none has a disconnected contribution, $O(\frac{1}{N})$.
In the case of $[0,p-3,0]$, a similar consideration applies to the correlators obtained from $R_{\tau=p+1,[0,p-3,0]}$,\footnote{ 
Correlators obtained from $R_{p+1,[0,p-3,0]}$, such as ${\cal G}_{[2^2](p+1)3(p-2)},
{\cal G}_{[2^2](p+1)3p}$ etc\ldots, do not have a $O(\frac{1}{N})$ contribution.}  
but for the correlators ${\cal G}_{[2^2](p+1)2,p-1}$. However, 
even so, the prediction vanishes because the latter is a near extremal correlator that vanishes.
Moving on to the predictions at the unitarity bound, $\tau=p+1$,
the correlators that would be forming the Gram matrix vanish and therefore 
the prediction only depends on the free theory part of the correlator itself,  
${\cal G}_{[2^2](p+1)3p+1}$. We will give the latter directly in the large $N$ expansion, 
since full dependence on $N$ is not relevant for our purpose here, and would clutter 
notation. With the prefactor 
\begin{equation}
{\cal P}_{[2^2](p+1)3p}=\sqrt{24p(p+1)} N^{4+p} g_{12}^4 g_{24}^{p-3}g_{34}^3
\end{equation}
the relevant free theory is
\begin{equation}
{\cal G}^{(\frac{3}{2})}_{[2^2](p+1)3p}=\sqrt{24p(p+1)}\Bigg[  U^2 \hat{\sigma}^2+2(p-1)\frac{
U^2\hat\sigma\hat\tau}{V} + \frac{(p-1)(p-2)}{2}\frac{U^2\hat{\tau}^2}{V^2}+ 
\frac{2U^3\hat{\sigma}^2\hat{\tau}}{V} + (p-1)\frac{\hat\sigma\hat{\tau}^2 }{V^2} \Bigg]\,.
\end{equation}
Note that there are no contributions at twist $\tau=p-1$, rather the free theory starts at twist 
$\tau=p+1$ with $O(U^2)$ terms, and this is why the predictions in table \eqref{pred2dpp13p} 
at twist $\tau=p-1$ vanish.

In order to bootstrap the correlator, we need to consider the totality of the 
predictions, and crossing symmetry, which implies that
\begin{align}\label{crossing223ppp1}
{\cal H}_{[2^2]3p(p+1)}(U,V,\hat{\sigma},\hat{\tau}) &= U\hat{\sigma}\, {\cal H}_{[2^2]p3(p+1)}\!\!\left(\frac{1}{U},\frac{V}{U},\frac{1}{\hat{\sigma}},\frac{\hat{\tau}}{\hat{\sigma}}\right)\;,\\
{\cal H}_{[2^2](p+1)3p}(U,V,\hat{\sigma},\hat{\tau}) &= \frac{U^3}{V^3}\hat{\tau} \,{\cal H}_{[2^2]p3(p+1)}\!\!\left(V,U,\frac{\hat{\sigma}}{\hat{\tau}},\frac{1}{\hat{\tau}}\right)\;.
\end{align}

Collecting all the information,
we find a unique solution. Below we shall give the result in the orientation ${\cal H}_{[2^2]p3(p+1)}$.\footnote{
Results in other orientations can be obtained from \eqref{crossing223ppp1}.} As before, we will 
work directly in the large $N$ expansion.  
Then,
\begin{equation}
{\cal P}_{[2^2]p3p+1}= \sqrt{ 24 p(p+1) }N^{p+4} g_{12}^3 g_{14} g_{24}^{p-3} g_{34}^3\;.
\end{equation}
and our result for the tree-level reduced correlator is
\begin{align}\label{MellinN3E}
{\cal H}_{[2^2]p3(p+1)}\Bigg|_{\frac{1}{N^3}} = &~ \sqrt{ 24 p(p+1)}\oiint\!\!\frac{dsdt}{(2\pi i)^2} U^{s+3} V^t \frac{ \Gamma[-s]^2\Gamma[-t]^2 \Gamma[-u+p-3]^2}{ (-u)_{p-3}} \frac{ 1}{(1+s)(1+t)(1+u)} \\ \times
& ~ \Bigg[\frac{1}{2(p-3)!} \frac{(p-1)t + p (s+2)}{(2+s)} +\frac{\hat\sigma}{2(p-2)!} \frac{2t+p(u+2)}{(2+u)} -\frac{\hat\tau}{2(p-3)!}\Bigg]\,, \notag
\end{align}
where $u=-s-t-5$.

A consistency check of this result is that for 
$p=2$ we have to re-obtain ${\cal H}_{[2^2]233}$. In fact, when $p=2$, the
contributions from $1$ and $\hat{\tau}$ are trivially (kinematically) vanishing, and we 
recover the expected result from looking at the rational function multiplying 
$U\hat{\sigma}$.\footnote{The following identity is 
needed $(U\hat{\sigma})\,{\cal P}_{[2^2]p3p+1}|_{p=2}={\cal P}_{[2^2]233}$. 
For the RHS see \eqref{prefactorswith_22_NNE}, and for the cross ratios 
see \eqref{def_cross_ratiosFA}.}

An important observation about ${\cal H}_{[2^2]p3p+1}$ in \eqref{MellinN3E} is that,  differently 
from ${\cal H}_{[2^2]2pp}$ in \eqref{solu2d2pp}, the rational part  now has an explicit dependence  
on the charge of the external single-particle operators, through the value of $p$. 
This implies that in general we should expect ${\cal H}_{[r_1r_2]spq}$ 
at tree level to depend on both the double-particle operators inserted \emph{and} the three 
single-particle operators. We will come back to this observation later on
when we compare with the case of the tree-level single particle correlators.

{\bf More correlators.} The same steps as above can be repeated for all ${\cal H}_{[2^2]p_2p_3p_4}$ 
that are N$^3$E, thus $p_4=p_3+p_2-2$. When the result is expressed in Mellin space, its expression
in terms of $p_2,p_3$ is a very mild generalization of the result in \eqref{MellinN3E}.
Then, we also bootstrapped a handful of correlators ${\cal H}_{[2^2]p_2p_3p_4}$ that 
are N$^4$E.\footnote{For illustration we give the details and the result for ${\cal H}_{[2^2]444}$ 
in Appendix \ref{appendix_more}.} We do not find it convenient to list more results and give 
more details here, since in section \ref{sec:[22]pqr} we will give a complete formula for all 
${\cal H}_{[2^2]p_2p_3p_4}$ that encompasses all the results obtained so far.

\subsubsection{Generic double-particle operators }\label{generic_double_particle_sec}

We consider now correlators with external double-particle operators of higher charge.
The case of ${\cal O}_2^2$ at twist four was unique, but for higher charges, there is a 
space of degenerate half-BPS operators that we can explore, and therefore different 
instances of composite operators with the same total charge. 

\begin{center}
\underline{\bf $\langle$[3$^2$]$,$2$,$p$,$p$\rangle$ and $\langle$[24]$,$2$,$p$,$p$\rangle$}
\end{center}

Of the same form as ${\cal O}_2^2$ we have ${\cal O}_3^2$ at twist six. However, 
this is not the only double-particle operator at twist six. There is another operator: ${\cal O}_2{\cal O}_4$. 
Thus, we have the two possibilities: 
\begin{equation}
{\cal H}_{[3^2]2pp}\qquad;\qquad {\cal H}_{[24]2pp}\;.
\end{equation}
The computations that follow will show how the double-particle bootstrap distinguishes these two.

We shall assume that $p\ge 4$ because otherwise the correlator is protected or vanishes. 
Let us begin with the orientation ${\cal H}_{{\bf 6}p2p}$ where ${\bf 6}$ stands for one of 
the two possible double-particle operators. The $su(4)$ rep that is relevant here is $[0,p-2,0]$ 
and since the threshold for triple-particle exchange is $\tau=p+6$, we can make predictions for
$\tau=p,p+2,p+4$. 

For $\tau=p$ we are in a situation similar to the one encountered in \eqref{situationlog0U2dp2p}. 
The prediction is
\begin{equation}
{\cal H}_{{\bf 6}p2p}\Bigg|_{ \frac{\log^0(U)}{N^3} } = \sum_{l} \Big( {\cal S}^{(\frac{3}{2})}_{{\bf 6}p2p} - A^{(\frac{3}{2})}_{{\bf 6}p2p} \Big) H_{[0,p-2,0],\tau=p,l} +\ldots \;,
\end{equation}
where $A_{{\bf 6}p2p}$ comes from free theory and ${\cal S}_{{\bf 6}p2p}$ is computed from 
a Gram matrix. The exchanged operators are the same as in \eqref{situationlog0U2dp2p}, 
in particular the matrix $\langle {\cal O}_i {\cal O}_j {\cal O}_a {\cal O}_b\rangle$, and 
the vector $V_{2p}$ in \eqref{situationlog0U2d2p_vec2p}, are the same. 
The vector $V_{{\bf 6}p}$ will depend on whether ${\bf 6}=[3^2],[24]$. However,
since the vector $V_{2p}$ vanishes, the Gram determinant gives ${\cal S}_{{\bf 6}p2p}=0$, 
and the prediction at twist $\tau=p$ reduces to 
\begin{equation}
{\cal H}_{{\bf 6}p2p}\Bigg|_{ \frac{\log^0(U)}{N^3} } = \sum_{l} \Big( - A^{(\frac{3}{2})}_{{\bf 6}p2p} \Big) H_{[0,p-2,0],\tau=p,l} +\ldots \;.
\end{equation}
The free theory diagrams relevant to compute $A^{(\frac{3}{2})}_{{\bf 6}p2p}$ are\footnote{Upon 
extracting $g_{12}^4 g_{14}^2 g_{24}^{p-4}g_{34}$, the diagram on the LHS 
correspond to $U\hat{\sigma}$ and the one on the RHS corresponds to $U\hat{\tau}/V$. 
}
\begin{equation}
\begin{array}{ccc}
\begin{tikzpicture}[scale=1.2]  
\def\shift{.35}

\def\latoxuno{-.35}
\def\latoxdue{-.38+0.8}
\def\latoyuno{.3}
\def\latoydue{-.25}

\def\latoxuno{-.35}
\def\latoxdue{-.38+1.5}
\def\latoyuno{.7}
\def\latoydue{-.25}

\draw[fill=red!60] 
(\latoxuno+.05, \latoydue) -- (\latoxdue, \latoyuno-.05) --  
(\latoxdue-.05, \latoyuno) -- (\latoxuno, \latoydue+.05) -- cycle;

\draw (\latoxuno-.2, \latoyuno) --  (\latoxuno-.2, \latoydue+.1);
\draw (\latoxuno-.15, \latoyuno) --  (\latoxuno-.15, \latoydue+.1);
\draw (\latoxuno-.1, \latoyuno) --  (\latoxuno-.1, \latoydue+.1);

\draw (\latoxuno-.1, \latoyuno-.05) --  (\latoxdue, \latoyuno-.05);
\draw (\latoxuno-.1, \latoyuno) --  (\latoxdue, \latoyuno);

\draw (\latoxdue-.05, \latoydue) --  (\latoxuno, \latoyuno-.05);
\draw (\latoxdue, \latoydue) --  (\latoxdue, \latoyuno);

\draw[fill=white!60,draw=white] (\latoxuno-.25, \latoyuno-.1) rectangle (\latoxuno, \latoyuno+.1);
\draw[fill=white!60,draw=white] (\latoxdue-.1, \latoydue-.05) rectangle (\latoxdue+.15, \latoydue+.1);

    \draw[fill=white!60,draw=white] (\latoxdue+.02, \latoyuno+.02) circle (2.5pt);
    \draw[fill=white!60,draw=white] (\latoxuno-.02, \latoydue+.02) circle (2.5pt);

   \draw(\latoxuno-.15, \latoyuno+.08) node[scale=.8] {${\bf 6}$};
   \draw(\latoxuno-.11, \latoydue-.08) node[scale=.8] {${\cal O}_p$};
   \draw(\latoxdue+.14, \latoyuno+.03) node[scale=.8] {${\cal O}_p$};
   \draw(\latoxdue+.1, \latoydue-.05) node[scale=.8] {${\cal O}_2$};

\end{tikzpicture} 
&\qquad\qquad &
\begin{tikzpicture}[scale=1.2]  
\def\shift{.35}

\def\latoxuno{-.35}
\def\latoxdue{-.38+0.8}
\def\latoyuno{.3}
\def\latoydue{-.25}

\def\latoxuno{-.35}
\def\latoxdue{-.38+1.5}
\def\latoyuno{.7}
\def\latoydue{-.25}

\draw[fill=red!60] 
(\latoxuno+.05, \latoydue) -- (\latoxdue, \latoyuno-.05) --  
(\latoxdue-.05, \latoyuno) -- (\latoxuno, \latoydue+.05) -- cycle;

\draw (\latoxuno-.2, \latoyuno) --  (\latoxuno-.2, \latoydue+.1);
\draw (\latoxuno-.15, \latoyuno) --  (\latoxuno-.15, \latoydue+.1);
\draw (\latoxuno-.1, \latoyuno) --  (\latoxuno-.1, \latoydue+.1);

\draw (\latoxuno-.1, \latoyuno-.1) --  (\latoxdue, \latoyuno-.1);
\draw (\latoxuno-.1, \latoyuno-.05) --  (\latoxdue, \latoyuno-.05);
\draw (\latoxuno-.1, \latoyuno) --  (\latoxdue, \latoyuno);

\draw (\latoxdue-.05, \latoydue) --  (\latoxuno, \latoydue);
\draw (\latoxdue, \latoydue) --  (\latoxdue, \latoyuno);

\draw[fill=white!60,draw=white] (\latoxuno-.25, \latoyuno-.1) rectangle (\latoxuno, \latoyuno+.1);
\draw[fill=white!60,draw=white] (\latoxdue-.1, \latoydue-.05) rectangle (\latoxdue+.15, \latoydue+.1);

    \draw[fill=white!60,draw=white] (\latoxdue+.02, \latoyuno+.02) circle (2.5pt);
    \draw[fill=white!60,draw=white] (\latoxuno-.02, \latoydue+.02) circle (2.5pt);

   \draw(\latoxuno-.15, \latoyuno+.08) node[scale=.8] {${\bf 6}$};
   \draw(\latoxuno-.11, \latoydue-.08) node[scale=.8] {${\cal O}_p$};
   \draw(\latoxdue+.14, \latoyuno+.03) node[scale=.8] {${\cal O}_p$};
   \draw(\latoxdue+.1, \latoydue-.05) node[scale=.8] {${\cal O}_2$};

\end{tikzpicture} 
\end{array}
 \end{equation}
and the color factors of these diagrams at order $O(\frac{1}{N^3})$ coincide 
up to the two-point normalizations. This implies the following relations between 
block coefficients,
\begin{equation}
 \frac{ A_{[3^2]p2p}}{ \sqrt{ (18)(p)(2)(p)}}\Bigg|_{\frac{1}{N^3}}=\frac{ A_{[24]p2p}}{\sqrt{(8)(p)(2)(p)} }\Bigg|_{\frac{1}{N^3}}=  3 (p-2) M^{6p2p}_{0,p,[l+2]} + (p-2)(p-3) M^{6p2p}_{1,p,[l+2]}\;,
\end{equation}
where again we used the used the formula ${M}^{\vec{p}}_{k,\gamma}$~\cite{Aprile:2025nta}.
By accident, this prediction does not distinguish the two correlators.   

For $\tau=p+2$, instead, we are in a situation similar to the one encountered 
in \eqref{situationlog1U2dpij}, namely 
\begin{equation}\label{situationlog1U2dpij_6bold}
 \sum_{ {R}_{p+2,[0,p-2,0]}  }\!\!\!\!\!\!{\cal C}^{(\frac{1}{2})}_{{\bf 6}p; {\cal D}_{}}\, \eta_{{\cal D}_{}} \, {\cal C}^{(0)}_{2p ;{\cal D}_{} } = 
  \sum_{ij}\frac{ \left( {\cal G}_{{\bf 6}pij}^{(\frac{1}{2})} \Big|_{[0,p-2,0],\tau=p+2,l} \right) \times \left(  {\cal H}^{(1)}_{i,j,2,p} \Big|_{[0,p-2,0],\tau=p+2,l} \right) }{ 
\left(  {\cal G}^{(0)} _{i,j,i,j} \Big|_{[0,p-2,0],\tau=p+2,l} \right)}\;,
\end{equation}
where the sum goes over the set $ij=\{(2,p),(3,p-1),\ldots\}$. 
This is the rectangle we explained in section \ref{REVdP}. In particular, 
there are at most $\lfloor\frac{p}{2}\rfloor$ elements in this set corresponding to 
the degeneracy of $[0,p-2,0]$ and the fact that we are looking at the lowest possible 
twist for double-particle operators in that rep.

On the one hand, if ${\bf 6}=[3^2]$, then 
only ${\cal G}_{[3^2]p3p-1}^{(\frac{1}{2})} \neq 0$, and therefore the sum localizes to 
$ij=\{(3,p-1)\}$:
\begin{equation}
\begin{tikzpicture}
\draw (0,0) node[scale=.9] {$\displaystyle
 \sum_{ {R}_{p+2,[0,p-2,0]}  }\!\!\!\!\!\!{\cal C}^{(\frac{1}{2})}_{[3^2]p; {\cal D}_{}}\, \eta_{{\cal D}_{}} \, {\cal C}^{(0)}_{2p ;{\cal D}_{} } = 
\frac{ \left( {\cal G}_{[3^2]p3p-1}^{(\frac{1}{2})} \Big|_{[0,p-2,0],\tau=p+2,l} \right) \times \left(  {\cal H}^{(1)}_{3,p-1,2,p} \Big|_{[0,p-2,0],\tau=p+2,l} \right) }{ 
\left(  {\cal G}^{(0)} _{3,p-1,3,p-1} \Big|_{[0,p-2,0],\tau=p+2,l} \right)}\;.$};
\end{tikzpicture}
\end{equation}
To compute the prediction, we have to perform the block decomposition of 
the $\log(U)$ projection of ${\cal H}^{(1)}_{3,p-1,2,p}$. 

On the other hand, if ${\bf 6}=[24]$, then 
only ${\cal G}_{[24]p2p}^{(\frac{1}{2})} \neq 0$ and therefore the sum localizes to $ij=\{(2,p)\}$:
\begin{equation}
\begin{tikzpicture}
\draw (0,0) node[scale=.9] {$\displaystyle
 \sum_{ {R}_{p+2,[0,p-2,0]}  }\!\!\!\!\!\!{\cal C}^{(\frac{1}{2})}_{[24]p; {\cal D}_{}}\, \eta_{{\cal D}_{}} \, {\cal C}^{(0)}_{2p ;{\cal D}_{} } = 
\frac{ \left( {\cal G}_{[24]p2p}^{(\frac{1}{2})} \Big|_{[0,p-2,0],\tau=p+2,l} \right) \times \left(  {\cal H}^{(1)}_{2,p,2,p} \Big|_{[0,p-2,0],\tau=p+2,l} \right) }{ 
\left(  {\cal G}^{(0)} _{2p2p} \Big|_{[0,p-2,0],\tau=p+2,l} \right)}\;.$};
\end{tikzpicture}
\end{equation}
In this case, we have to compute the block decomposition of the $\log(U)$ projection of  
${\cal H}^{(1)}_{2p2p}$. 

Conclusion: The predictions at twist $\tau=p+2$ for ${\bf 6}=[3^2]$ and ${\bf 6}=[24]$ 
are different and thus the double-particle bootstrap does distinguish the two correlators.

The reasoning for $\tau=p+4$ 
is similar. We have to consider 
\begin{equation}\label{situationlog1U2dpij_6boldmore}
 \sum_{ {R}_{p+4,[0,p-2,0]}  }\!\!\!\!\!\!{\cal C}^{(\frac{1}{2})}_{{\bf 6}p; {\cal D}_{}}\, \eta_{{\cal D}_{}} \, {\cal C}^{(0)}_{2p ;{\cal D}_{} } = 
  \sum_{ij}\frac{ \left( {\cal G}_{{\bf 6}pij}^{(\frac{1}{2})} \Big|_{[0,p-2,0],\tau=p+4,l} \right) \times \left(  {\cal H}^{(1)}_{ij2p} \Big|_{[0,p-2,0],\tau=p+4,l} \right) }{ 
\left(  {\cal G}^{(0)} _{i,j,i,j} \Big|_{[0,p-2,0],\tau=p+2,l} \right)}\,.
\end{equation}
In this case, there are twice the number of degenerate operators in $R_{p+4}$, 
compared to $R_{p+2}$, because we are looking at the next-to-lowest twist. 
In particular, the sum $\sum_{ij}$ goes over the set  
$ij=\{\{(2,p),(3,p+1),\ldots\},\{(3,p-1),(4,p),\ldots\}\}$. Again, each subset contains 
at most $\lfloor\frac{p}{2}\rfloor$ elements corresponding to the degeneracy of the rep $[0,p-2,0]$. 
Then, when ${\bf 6}=[3^2]$ the sum will localize on $ij=\{(3,p+1),(3,p-1)\}$, 
and when ${\bf 6}=[24]$ the sum will localize on 
$ij=\{(2,p),(4,p)\}$.

To complete the OPE predictions we have to consider the crossed orientation ${\cal H}_{{\bf 6}2pp}$. 
The relevant $su(4)$ rep is $[0,4,0]$, whose unitarity bound is $\tau=6$. Since the threshold 
for triple-particles is ${\bf 6}+2=8$, we can only predict the contribution at twist $6$. There 
are two such twist six operators ${\cal O}_2{\cal O}_4$ and ${\cal O}_3{\cal O}_3$ and 
one has to arrange a Gram matrix in order to compute the prediction. This prediction is 
again sensitive to the double-particle operator ${\bf 6}$ being $[3^2]$ or $[24]$. The 
computation here is not different from what we described already in previous sections, 
therefore, we will not give further details. 

Collecting all the predictions and implementing our bootstrap algorithm, 
we find a unique solution. Upon
extracting the following (large $N$) prefactors, 
$$
{\cal P}_{[3^2]2pp}= \sqrt{ 36p^2 N^{2p+8} }g_{12}^2 g_{13}^2 g_{14}^2 g_{34}^{p-2} \qquad;
\qquad {\cal P}_{[24]2pp}= \sqrt{ 16p^2 N^{2p+8} } g_{12}^2 g_{13}^2 g_{14}^2 g_{34}^{p-2}
$$
the final result is
\begin{align}
\!\!\!\!{\cal H}_{[24]2pp}\Bigg|_{\frac{1}{N^3}}&=
(4p)\frac{(p-2)}{(p-4)!}\oiint_{-i\infty}^{+\infty}\!\!\frac{dsdt}{(2\pi i)^2} U^{s+2} V^t
\frac{ \Gamma[-s+p-4]^2\Gamma[-t]^2 \Gamma[-u]^2 }{ (-s)_{p-4}} \frac{\frac{1}{6}(4+s)(2 tu + s+3)}{(s+1)(t+1)(u+1)} 
\label{M242pp}\\ 
{\cal H}_{[3^2]2pp}\Bigg|_{\frac{1}{N^3}}&=
(6p)\frac{(p-2)}{(p-4)!}\oiint_{-i\infty}^{+\infty}\!\!\frac{dsdt}{(2\pi i)^2} U^{s+2} V^t
\frac{ \Gamma[-s+p-4]^2\Gamma[-t]^2 \Gamma[-u]^2 }{ (-s)_{p-4}} \frac{\frac{1}{6}( 2(s+4)^2+(s+1)(tu+1)) }{(s+1)(t+1)(u+1)}
\label{M332pp}
\end{align}
where $u=-s-t-4$.

\begin{center}
\underline{\bf A family of double-particle operators with arbitrary charge: $\langle$[r$^2$]$,$2$,$p$,$p$\rangle$}
\end{center}

We now move on to the study of ${\cal H}_{[r^2]2pp}$. In this case, the correlator 
is again N$^2$E, so the reduced correlator does not depend on $\hat{\sigma},\hat{\tau}$. 
We shall assume that $p\ge r+1$, otherwise the correlator (when it exists) is protected. 

Recall that ${\cal O}_r^2$ is just an operator in the space of degenerate 
operators with charge $\Delta=2r$, so the scope here is to make explicit the dependence on $r$. 
To this end, we first bootstrap correlators by keeping fixed the value of $r=3,4,\ldots$.
This allows us to find the correlators ${\cal H}_{[r^2]2pp}$ in Mellin space
as a function of $p$. Then we vary the value of $r$ and we study the dependence on $r$. 

Upon extracting the prefactor
\begin{equation}
{\cal P}_{[r^2]2pp}=(2rp) N^{r+p+1}\, g^2_{12}g_{13}^{r-1}g_{14}^{r-1}g_{34}^{1+p-r}
\end{equation}
we can write the final result in the following form
\begin{equation}\label{dr2pp}
{\cal H}_{[r^2]2pp}\Bigg|_{\frac{1}{N^3}}= 
(2rp)\frac{ (p+1-r) }{(p-1-r)!}\oiint_{-i\infty}^{+\infty}\!\!\frac{dsdt}{(2\pi i)^2} 
U^{s+2} V^t\Gamma[-s]\Gamma[-s+p-r-1] \Gamma[-t]^2 \Gamma[-u]^2  \frac{N_r(s,t)}{(s+1)(t+1)(u+1)}
\end{equation}
where $u=-s-t-4$,  and where $N_r(s,t)$ is a polynomial in $s,t$. We will shortly 
give the general expression.  In preparation for that, we list the first few of them, 
\begin{align}
N_2 & =(3+s)\left( \frac{1}{2}\right)\\
N_3 &= (3+s) \left( \frac{s}{3}+\frac{11}{6} \right) +\frac{tu}{6} (s+1)\notag\\
N_4 &= (3+s) \left( \frac{s^2}{8}+\frac{35s}{24}+\frac{13}{3} \right) + \left[  \frac{t^2u^2}{40}+ tu\left(\frac{s}{10}+\frac{71}{120}\right) \right](s+1)\notag\\
N_5 &=(3+s) \left( \frac{s^2}{30}+\frac{5s^2}{8}+\frac{157s}{40}+\frac{25}{3} \right) +\left[  \frac{t^3u^3}{504} 
+ t^2u^2\left( \frac{11}{126}+\frac{s}{72} \right) 
+t u \left(\frac{43s^2}{1260} + \frac{547s}{168}+\frac{233}{168}\right)
\right] (s+1) \notag\\
\ldots \notag
\end{align}
Results up to $N_{10}(s,t)$ are given in an ancillary file. 
It is straightforward to check that $N_2$ and $N_3$ give back the results quoted in \eqref{solu2d2pp} and \eqref{M332pp}.

Firstly, as for the other N$^2$E correlators that we studied 
in previous sections, the only $p$ dependence in \eqref{dr2pp} comes from ${ (p+1-r) }/{(p-1-r)!}$ 
and the two-point function normalization $(2rp)$. The factorial $(p-1-r)!$ is a result of our computation 
and a posteriori it ensures kinematically that for $p<r+1$ the dynamical function vanishes. 

Secondly, the $N_r(s,t)$ appear to be complicated at first, but
the idea is that the degree in $s,t$ grows because there are hidden sums that 
involve shifted Gamma functions compared to the ones we extracted in \eqref{dr2pp}. 
The intuition for this comes from the OPE predictions: Differently from $[2^2]$, 
when we consider $[r^2]$ the OPE predictions are obtained by summing over 
many different correlators. This is a consequence of the degeneracy of 
double-particle operators. We already saw this phenomenon for the case $r=3$ 
in \eqref{situationlog1U2dpij_6bold} or \eqref{situationlog1U2dpij_6boldmore}.
Now, the predictions for ${\cal H}_{[r^2]p2p}$ at the top of the window, $\tau=p+2r-2$, 
will involve at least $r-1$ terms in the sum.\footnote{Recall that 
$p\ge r+1$, otherwise the correlator is protected or vanishes. At the lowest twist 
$\tau=p+2$ the rectangle $R_{p+2}$ contains 
all pairs from $(2,p)$ to $(\frac{p}{2},\frac{p}{2})$, therefore it
will certainly contain a pair $(i,j)$ for which either $i$ or $j$ are equal to $r$.} 
Therefore, it seems plausible that by re-absording $N_r(s,t)$ into the Gamma functions 
we could obtain a simpler expression. 

Let us then consider the splitting 
\begin{equation}\label{first_rewrtNr}
N_r=\frac{r(r-1)}{2}+(s+1)n_r(s,t)
\end{equation}
and rewrite the $n_r(s,t)$ as an expansion in Pochhammer symbols of the form $(-1-t)_n$, 
$(-1-u)_n$ and $(-s)_m$. The first few such polynomials are
\begin{align}
n_3 &= \ \frac{1}{6}(-1-t)(-1-u)\ + \ \ \Bigg[ 3 - \frac{1}{2}(-s)\,\Bigg] \\
n_4 &= \frac{1}{40}(-1-t)_2(-1-u)_2+ \Bigg[ \frac{2}{3} - \frac{1}{8}(-s)\Bigg] (-1-t)(-1-u)+ \Bigg[ 9- 3(-s)+\frac{1}{4}(-s)_2\Bigg]\\
\vdots& \notag
\end{align}
With the $n_r$ written in this way, we can absorb the various terms of $n_r/(t+1)/(u+1)$ 
into the $\Gamma[-s]\Gamma[-t]\Gamma[-u]$ functions. 
It turns out that the numerical coefficients in this rewriting 
are given by a simple formula, 
\begin{equation}\label{formula_nr}
n_r=\sum_{I=0}^{r-2}\sum_{J=0}^{I} {\frac{r(r-1)}{2}}
\frac{   (-1)^{I+J} \frac{\Gamma[2+I]}{\Gamma[1+J]\Gamma[r-1-I]\Gamma[r+1-J]} }{ ((r-1-J)+(r-2-I))\Gamma[1+I-J]} (-s)_{I-J} (-1-t)_{r-2-I}(-1-u)_{r-2-I}
\end{equation}
We could leave the summation indices unbounded since they are cut off 
by the Gamma functions in the denominator: $\Gamma[r-1-I]\Gamma[r+1-J]$ 
and $\Gamma[1+I-J]$. 

Note that we could to add to \eqref{formula_nr}  the term $\frac{r(r-1)}{2}$ 
of $N_r$ in \eqref{first_rewrtNr}. It can be done by using the prescription that 
$I=r-2$ first and then $J\rightarrow I+1$ (since the numerical coefficient is not continuous there).  

It is evident from \eqref{formula_nr} that the dependence on the double-particle 
operator ${\cal O}_r^2$ is entering in two ways: 1) there is now a sum of terms that 
grows with $r^2$ and 2) there is explicit dependence on $r$ in the numerical coefficients.  
We will come back to this formula in the next section.

\section{Mellin amplitudes with one double-particle operator} \label{sec_MellinAmps}

So far we used the double-particle bootstrap to obtain 
explicit results for a number of tree-level dynamical correlators with one double- and three 
single-particle operators, ${\cal H}_{[r_1r_2]spq}$. In particular, we have established that 
the correlators ${\cal H}_{[r_1r_2]spq}$ are  given by a
sum of four-point contact diagrams, namely D-functions, 
and that the outcome of the double-particle bootstrap is unique.
Moreover, by taking advantage of the simplicity of D-functions in Mellin space, 
we have been able to obtain formulae with explicit dependence on the external 
charges, for some families of correlators. Building on these results, in this section, 
we would like to take one step further and discuss what is \emph{the} Mellin amplitude 
${\cal M}_{[r_1r_2]spq}$ that corresponds to ${\cal H}_{[r_1r_2]spq}$?


In order to explore this question, we will be inspired by the AdS$\times$S 
formalism \cite{Aprile:2020luw} for the single-particle amplitudes, 
and subsequently, by the flat-space limit of Penedones  \cite{Penedones:2010ue}.  In the 
case of four single-particle operators, ${\cal H}_{rspq}$, it is well known how to define 
\emph{the} Mellin amplitude ${\cal M}_{rspq}$. Let us start by reviewing how this works \cite{Aprile:2020luw}:
Introduce continuous Mellin variables for AdS, denoted by $\delta_{ij}$, and non-negative integer 
variables for S, denoted by $d_{ij}$, where $i,j=1,\ldots 4$.
These variables are further constrained to be: symmetric, $\delta_{ij}=\delta_{ji}$, $d_{ij}=d_{ji}$, 
off-diagonal, $\delta_{ii}=d_{ii}=0$, and subject to the on-shell relations,\footnote{For example, 
if $p_i=2$ then $\sum d_{ij}=0$, thus $d_{ij}=0$, and there is no dependence on the sphere variables.} 
\begin{equation}\label{constraintMellin4p}
\sum_{j=1}^{4} \delta_{ij}= p_i+2\qquad;\qquad \sum_{j=1}^4 d_{ij}= p_i-2\,.
\end{equation}
There are two independent $\delta_{ij}$'s and two $d_{ij}$'s; in the following, we shall call them $s,t$ and $\tilde{s},\tilde{t}$, respectively. Introduce 
now the AdS$\times$S ``measure''
\begin{equation}\label{measureAdSSfor4p}
\intsumfour \equiv \sum_{[\tilde{s},\tilde{t}]} \prod_{i<j}\frac{\vec{Y}_{ij}^{2d_{ij} }}{\Gamma[d_{ij}+1]} 
\varoiint_{-i\infty}^{i\infty}  \frac{\!\!dsdt }{ (2\pi i)^2 } \prod_{i<j}    \frac{  \Gamma[\delta_{ij}]}{ \vec{X}_{ij}^{2\delta_{ij} }  }\,.
\end{equation}
Note that, for given external charges, $p_i$, $\displaystyle\intsumfour$ comprises arbitrary free theory superconformal structures. 

Let ${\cal O}_{p_i}$ be any half-BPS single-particle operator;  by combining the implications of the 
partial non-renormalization theorem \cite{Eden:2000bk,Heslop:2002hp}, 
and the use of AdS$\times$S measure, we obtain the following general expression 
for the four-point correlator,
\begin{equation}\label{AdSSfor4pt}
\langle {\cal O}_{p_1}(z_1) {\cal O}_{p_2}(z_2) {\cal O}_{p_3}(z_3) {\cal O}_{p_4}(z_4)\rangle 
= \emph{free}+ {\cal R}_{1234}\, 
\Bigg(  N^{-2+{\frac{1}{2}}\sum_i p_i} {p_1p_2p_3p_4} 
\intsumfour\, {\cal M}_{\vec{p}}(\delta_{ij},d_{ij})
\Bigg)\,,
\end{equation}
where 
\begin{equation}\label{eq:Rfactor}
{\cal R}_{1234}= \vec{X}^{\,4}_{13} \vec{X}^{\,4}_{24} \vec{Y}^{\,4}_{13} \vec{Y}^{\,4}_{24}\, \prod_{i,j=1,2}(x_i-y_j)\,.
\end{equation}
The Mellin amplitude  ${\cal M}_{\vec{p}}(\delta_{ij},d_{ij})$ 
depends implicitly on the central charge and the t' Hooft coupling. 
It is a non-trivial function of the Mellin variables, and the shift by $\pm2$ in \eqref{constraintMellin4p} 
was introduced to take into account the reduced degree
of extremality of the dynamical correlator due to ${\cal R}_{1234}$. 
However, note that in total ${\cal M}_{\vec{p}}(\delta_{ij},d_{ij})$  is  a function of \emph{8} independent variables. 
A choice of these variables would be to take the four external charges $\vec{p}$ and the independent Mellin variables of 
each type, $s,t$ and $\tilde s,\tilde t$.

In order to match the AdS$\times$S representation with the one we have been using for 
the bootstrap, see \eqref{basic_c_1}, we should consider 
again the definition in \eqref{notation4pt_intro_symcr2} and translate the factor ${\cal R}_{1234}$ into (using $\prod_{i,j} (x_i-y_j)={\cal I}(U,V,\hat{\sigma},\hat{\tau})/(\hat\sigma)^{2}$) 
\begin{equation}
{{\cal R}_{1234}= \vec{X}^{\,4}_{13} \vec{X}^{\,4}_{24} \vec{Y}^{\,4}_{12} \vec{Y}^{\,4}_{34}\,{\cal I}(U,V,\hat{\sigma},\hat{\tau})\,.}
\end{equation}
Then, upon extracting the prefactor ${\cal P}_{\vec{p}}[\, g_{ij} ]$, consisting of propagators 
and two-point function normalizations, the ``measure'' in \eqref{measureAdSSfor4p} can be 
written in terms of cross ratios, $U,V,\hat{\sigma},\hat{\tau}$ introduced in  \eqref{notation4pt_intro_symcr1}. 
For fixed external charges $\vec{p}$, the integer variables $\tilde s,\tilde t$ span a finite set, associated to finitely many monomials 
in $\hat{\sigma}$ and $\hat{\tau}$. In fact, thanks to the $\Gamma[d_{ij}+1]$ factors in the 
denominators, the sum over $\tilde s,\tilde t$ automatically truncates to the correct number 
of monomials, that equals $\frac{1}{2}(\kappa+2)(\kappa+1)$ where $\kappa$ is the 
degree of extremality in \eqref{degree_extre}.  Along with $\delta_{ij}$ and $d_{ij}$, we also introduce 
the bold-face variables defined via
\begin{equation}\label{rho4ptvar}
\pmb{\rho}_{ij}=\delta_{ij}-d_{ij}\qquad;\qquad \sum_{i} \pmb{\rho}_{ij}=+4\,.
\end{equation}
The on-shell constraints on $\pmb{\rho}_{ij}$ follow from those of $\delta_{ij}$ and $d_{ij}$, but crucially, the $\pmb{\rho}_{ij}$ do not depend on the KK levels, $p_i$, and we have the identifications:
\begin{equation}\label{boldFA}
-{\bold s}\equiv \pmb{\rho}_{12}=\pmb{\rho}_{34}\qquad;\qquad 
-{\bold t}\equiv \pmb{\rho}_{14}=\pmb{\rho}_{23}\qquad;\qquad
-{\bold u}\equiv \pmb{\rho}_{13}=\pmb{\rho}_{24}\,.
\end{equation}

As shown in \cite{Aprile:2020luw}, 
in the case of four single-particle operators, {the Mellin amplitude ${\cal M}_{\vec{p}}(\delta_{ij},d_{ij})$ that appears in \eqref{AdSSfor4pt} has the form}
\begin{equation}
{\cal M}_{\vec{p}}(\delta_{ij},d_{ij})= {\cal M}^{\emph{tree-level}}_{\vec{p}}(\delta_{ij},d_{ij})+\frac{1}{N^2}{\cal M}^{\emph{1-loop}}_{\vec{p}}(\delta_{ij},d_{ij}) +\ldots
\end{equation}
and the leading order tree level term is quite special:  
\begin{equation}\label{singlep_treelevel_AdSxS}
 {\cal M}^{\emph{tree-level}}_{\vec{p}}(\delta_{ij},d_{ij})= \frac{1}{({\bold s}+1) ({\bold t}+1) ({\bold u}+1) }\,.
\end{equation}
It only depends on the bold variables,
and has no explicit dependence on the external KK levels.
This is the manifestation of the accidental 10d \emph{conformal} symmetry 
discovered in \cite{Caron-Huot:2018kta}, which  implies that all tree-level single-particle 
four-point correlators are uniquely determined by the four-point correlator of the lowest 
KK mode, ${\cal O}_2$; hence, the amplitude ${\cal M}_{\vec{p}}^{\emph{tree-level}}$ 
can be obtained by covariantizing ${\cal M}^{\emph{tree-level}}_{2222}(s,t)$, namely by 
replacing the $s,t,u$ variables of the latter with  the bold variables ${\bold s},{\bold t},{\bold u}$ 
\cite{Aprile:2020luw}.

\subsection{All tree-level  amplitudes $\langle {\cal O}_2^2\,{\cal O}_{p_2}{\cal O}_{p_3}{\cal O}_{p_4}\rangle$}\label{sec:[22]pqr}

The superconformal Ward identities and the partial non-renormalization theorem imply that a general 
expression of the form of \eqref{AdSSfor4pt} applies to four-point correlators of \emph{any} half-BPS 
operators, single or multi-particle ones. When some of the operators are multi-particles, two complications 
might arise: 1) in general, a multi-particle operator is not uniquely determined by its charge, $p_i$, 
and hence the amplitude ${\cal M}_{\vec{p}}(\delta_{ij},d_{ij})$ might depend on additional parameters 
needed to specify the operator; 2) for generic correlators, the amplitude might have a complicated dependence 
on all the relevant parameters, and hence it is not apriori obvious how to organize it in a simple expression.  

In this section, we consider all four-point correlators with three half-BPS single-particles and the 
double-particle operator ${\cal O}_2^2$. For this case, problem 1) does not apply and we will show that the 
AdS$\times$S amplitude  ${\cal M}_{[2^2]p_2p_3p_4}$ is a fairly mild generalization of the simple result 
for single-particles in \eqref{singlep_treelevel_AdSxS}, in the following sense:
${\cal M}_{[2^2]p_2p_3p_4}$ does not only depend solely on the bold variables, however, the 
dependence on $d_{ij}$ and $p_i$ is polynomial and quite concise. Hence, the AdS$\times$S 
formalism introduced in the previous section is an efficient way to represent correlators in this class. 
Analogously to \eqref{AdSSfor4pt}, we parametrize the correlators as 
\begin{equation}\label{eq:22p2p3p4}
\langle {\cal O}^2_{2}(z_1) {\cal O}_{p_2}(z_2) {\cal O}_{p_3}(z_3) {\cal O}_{p_4}(z_4)\rangle 
= \emph{free}+
{\cal R}_{1234}\, 
\Bigg( N^{-1+\frac{1}{2}\sum_{i=2}^4 p_i} \,(4)p_2p_3p_4 
\intsumfour\, {\cal M}_{[2^2]p_2p_3p_4}(\delta_{ij},d_{ij})
\Bigg)\,.
\end{equation}
To deduce ${\cal M}_{[2^2]p_2p_3p_4}(\delta_{ij},d_{ij})$ for generic $p_2$, $p_3$, $p_4$ we 
first compute ${\cal M}_{[2^2]p_2,p_3,p_2+p_3-2}$ by rewriting our previous result for 
${\cal H}_{[2^2]p_2,p_3,p_2+p_3-2}$ in the form of \eqref{eq:22p2p3p4}, and then conjecture 
a natural generalization valid for arbitrary values of the single-particle charges. We thus produce 
a general expression for all the correlators $\langle {\cal O}_2^2\,{\cal O}_{p_2}{\cal O}_{p_3}{\cal O}_{p_4}\rangle$, 
including the N$^2$E, N$^3$E cases already contained in the previous sections and the N$^4$E 
cases that we predict here.

As a preliminary step, we want to understand the pole structure of the Mellin amplitude ${\cal M}$ that 
corresponds to the N$^3$E correlators ${\cal H}_{[2^2]p_2 p_3 p_2+p_3-2}$. With the bootstrap method of the previous section, 
we have computed the three propagator structures, $1,\hat{\sigma},\hat{\tau}$, for this class of correlators. Now, 
we want to rewrite them by factoring out the AdS$\times$S measure and, by doing so, reconstruct the 
pole structure in bold variables ${\bold s},{\bold t},{\bold u}$. Note that the bold variables are 
given unambiguously by the expressions in \eqref{rho4ptvar}-\eqref{boldFA}. If we denote by ${\cal M}^{1}$, 
${\cal M}^{\hat{\sigma}}$, ${\cal M}^{\hat{\tau}}$ the function ${\cal M}(\delta_{ij},d_{ij})$ in \eqref{eq:22p2p3p4} 
evaluated at the values 
of $d_{ij}$ corresponding to the three structures $1,\hat{\sigma},\hat{\tau}$, the result of this exercise gives,\footnote{More explicitly, 
one has ${\cal M}^{1}\equiv {\cal M}(d_{12}=1,d_{13}=0)$, 
${\cal M}^{\hat{\sigma}}\equiv {\cal M}(d_{12}=0,d_{13}=1)$, ${\cal M}^{\hat{\tau}}\equiv {\cal M}(d_{12}=0,d_{13}=0)$,
and respectively, 
$\prod_{i<j} \Gamma[d_{ij}+1]$ evaluates to  $(p_2-3)!(p_3-2)!$, $(p_2-2)!(p_3-3)!$, $2(p_2-3)!(p_3-3)!$.}
\begin{align}
{\cal M}^{1}_{[2^2] p_2 p_3(p_2+p_3-2)} =&\ {+\frac{p_2-1}{(1+{\bf s})(1+{\bf t})(1+{\bf u})}
+\frac{p_3-2}{({\bf s})(1+{\bf t})(1+{\bf u})}+\frac{3-p_2-p_3}{({\bf s})({\bf t})(1+{\bf u})}}\,,\\
\label{MN3extra2}
{\cal M}^{\hat{\tau}}_{[2^2] p_2 p_3(p_2+p_3-2)} =&\ {-\frac{2}{(1+{\bf s})({\bf t})(1+{\bf u})}+\frac{2}{(1+{\bf s})(-1+{\bf t})(1+{\bf u})}} 
\end{align}
and ${\cal M}^{\hat{\sigma}}_{[2^2] p_2 p_3(p_2+p_3-2)} =  {\cal M}^{1}_{[2^2] p_3 p_2(p_2+p_3-2)}( {\bf s}\leftrightarrow {\bf u})$.

From the above rewriting, we recognize the crossing symmetric poles, $(1+{\bf s})(1+{\bf t})(1+{\bf u})$, that also 
appear in the single-particle correlator \eqref{singlep_treelevel_AdSxS},  
and three novel types of shifted poles: $(1+{\bf s})({\bf t})(1+{\bf u})$, $(1+{\bf s})(-1+{\bf t})(1+{\bf u})$, 
$({\bf s})(1+{\bf t})({\bf u})$ plus their crossing. We see that this structure of poles persists by 
sampling some N$^4$E cases, such as ${\cal H}_{[2^2]444}$ that is given in appendix \ref{appendix_more}, 
and this motivates us to make a general ansatz of the form
\begin{equation}\label{ansatzM2p2p3p4}
{\cal M}_{[2^2]p_2p_3p_4}= \bigg({\cal M}_{[2^2]su}+ \emph{crossing}\bigg)+ {\cal M}_{[2^2]stu} \,,
\end{equation}
where 
\begin{equation}
\begin{tikzpicture}
\draw (0,0) node[scale=.95]  {$\displaystyle
{\cal M}_{[2^2]su}=
\frac{p_{1,0,1}(d_{ij},\vec{p}) }{(1+{\bf s})({\bf t})(1+{\bf u})} 
+ \frac{p_{1,-1,1}(d_{ij},\vec{p})}{ (1+{\bf s})(-1+{\bf t})(1+{\bf u}) } 
+  \frac{p_{0,1,0}(d_{ij},\vec{p})}{ ({\bf s})(1+{\bf t})({\bf u})},\qquad {\cal M}_{[2^2]stu}= \frac{p_{1,1,1}(d_{ij},\vec{p})}{(1+{\bf s})(1+{\bf t})(1+{\bf u})}\,.$};
\end{tikzpicture}
\end{equation}
The polynomials $p_{abc}(d_{ij},p_2,p_3,p_4)$ should be constructed so as to reproduce all the 
results we know from the bootstrap. One way to find them is to repeat the construction
in \cite[(2.10)-(2.12)]{Aprile:2020mus} which consists of making an ansatz that span over 
a basis of polynomials in the variables $d_{ij},p_2,p_3,p_4,$ that are symmetric under 
exchange of points $2\leftrightarrow 3$, $3\leftrightarrow4$, $2\leftrightarrow 4$.\footnote{In our problem 
here we have a reduced set of crossing equations because $p_1=4$ is fixed, 
three out six compared to \cite{Aprile:2020mus}. 
The other modification compared to \cite{Aprile:2020mus} is to eliminate $\Sigma$, since there are 
really three-independent charges here: $p_2,p_3,p_4$.}
After doing so, the only freedom is the choice of degree for the polynomials and the numerical 
constants to be fitted against the data. We append this basis to an ancillary file. 

It turns out that there is a unique solution such that the $p_{abc}(d_{ij},p_2,p_3,p_4)$ have total
degree three with degree at most two in the $d_{ij}$. To write it down in an even simpler form, 
we introduce the variables
\begin{equation}\label{def_ijmn}
f_{ij;mn}=\frac{p_i+p_j-p_m-p_n}{2}+d_{ij}+d_{mn},\qquad\qquad 
\begin{array}{l}
f_{s}:=f_{12;34}\\[.2cm]
f_{t}\,:=f_{14;23}\,.\\[.2cm]
f_{u}:=f_{13,24}
\end{array}
\end{equation}
The final result reads as follows, 
\begin{equation}\label{eq:Msu}
\begin{tikzpicture}
\draw (0,0) node[scale=1] {$\displaystyle
{\cal M}_{[2^2]su}=
\frac{ \frac{1}{4} \big( (p_4-p_2-p_3) (f_{t}-3)+(f_{t}-4)\big)f_{t}}{(1+{\bf s})({\bf t})(1+{\bf u})} 
+ \frac{ \frac{1}{8}(p_2+p_3-p_4) (f_{t}-2) f_{t}}{ (1+{\bf s})(-1+{\bf t})(1+{\bf u}) } 
+  \frac{ \frac{1}{8}(4-p_2-p_3-p_4) f_{s} f_{u}  }{ ({\bf s})(1+{\bf t})({\bf u})}$ 
};
\end{tikzpicture}
\end{equation}
and
\begin{equation}\label{eq:Mstu} 
\!\!\!\!\!\begin{tikzpicture}
\draw (0,0) node[scale=.95] {$\displaystyle
{\cal M}_{[2^2]stu}=\frac{4-\frac{1}{4} ( f_{s}^2 + f_{t}^2 + f^2_{u} ){-} \frac{1}{8} \big( 
(4-p_2+p_3-p_4) f_{s} f_{t}  + 
(4-p_2-p_3+p_4) f_{s} f_{u} +  
(4+p_2-p_3-p_4) f_{t} f_{u} \big)  }{(1+{\bf s})(1+{\bf t})(1+{\bf u})}\,.$
};
\end{tikzpicture}
\end{equation}
To reconstruct the full amplitude in \eqref{ansatzM2p2p3p4}, one needs ${\cal M}_{[2^2]stu}$, 
${\cal M}_{[2^2]su}$ and its crossings. 

It is interesting to analyse the formula we have deduced for  ${\cal M}_{[2^2]p_2p_3p_4}$ from the 
point of view of the flat space limit of Penedones \cite{Penedones:2010ue}. 
Mathematically, this is the limit  in which we scale the variables 
$\delta_{ij}\rightarrow \Lambda \delta_{ij}$ with $\Lambda\rightarrow \infty$, while keeping the 
other variables fixed. Even though we will leave the physical interpretation of the 
double-particle operator in the flat space limit for  future studies, we will show below that the limit 
$\Lambda\rightarrow \infty$ of ${\cal M}_{[2^2]p_2p_3p_4}(\Lambda \delta_{ij},d_{ij},p_i)$ is well 
defined and consistent across all values of $p_2,p_3,p_4$.

The first observation we want to make is that the flat space limit of ${\cal M}_{[2^2]p_2p_3p_4}$ is 
subleading compared to the flat space limit of the tree-level single-particle amplitude ${\cal M}_{p_1p_2p_3p_4}$, 
see \eqref{singlep_treelevel_AdSxS}. Note that since ${\cal M}_{[2^2]p_2p_3p_4}$ is written as a sum over
shifted poles of the form $(a+{\bf s})(b+{\bf t})(c+{\bf u})$, the leading term would be naively of the 
same order as the one for the tree-level single-particle amplitude. However, there will be a 
non-trivial cancellation among the terms. 

To clarify what happens in the flat space limit, let us analyze first the case of ${\cal M}_{[2^2]su}$. 
In the limit $\Lambda\rightarrow \infty$, the RHS of the on-shell relations for the $\delta_{ij}$, 
see \eqref{constraintMellin4p}, goes to zero, and one can solve the  constraints in terms of 
variables $\underline{s},\underline{t},\underline{u}$, where $\underline{s}+\underline{t}+\underline{u}=0$. 
Then, the leading term in the flat space limit reads,
\begin{equation}
 {\cal M}_{[2^2]su}(\Lambda \delta_{ij},d_{ij},p_i)\Bigg|_{\frac{1}{\Lambda^3}}= \frac{1}{{\underline{s}}{\underline{t}}{\underline{u}} }
 \Big(  \tfrac{1}{8}(2+p_4-p_2-p_3) f_t^2+   \tfrac{1}{8} (4-p_2-p_3-p_4) f_s f_u - \tfrac{1}{2} (2+p_4-p_2-p_3) f_t\Big)\,,
\end{equation}
and it is non-trivial. When we add together this contribution, its images under crossing, and the limit of ${\cal M}_{[2^2]stu}$,
we can rewrite the final result as follows, 
\begin{equation}\label{vanishing_of_flatspace_limit}
\!\!\!\!\!\!\!{\cal M}_{[2^2]p_2p_3p_4}(\Lambda \delta_{ij},d_{ij},p_i)\Bigg|_{\frac{1}{\Lambda^3}}= \frac{1}{{\underline{s}}{\underline{t}}{\underline{u}} }
\left(f_s+f_t+f_u-4\right)\Big( -\tfrac{1}{3} + \tfrac{1}{8}(p_2-p_3-p_4) f_s+\emph{crossing}\Big) =0
\end{equation}
which vanishes because $f_s+f_t+f_u=p_1=4$ in our computation. 

A posteriori, the cancellation of the $O(\frac{1}{\Lambda^3})$ terms could be used as a bootstrap constraint, 
additional to the double-particle bootstrap.  This would parallel the bootstrap approach for the four-point single-particle correlators,
where basic knowledge about the flat space limit is used as a constraint. 

Coming back to ${\cal M}_{[2^2]p_2p_3p_4}$, the flat space limit  is thus sub-leading compared to the case of 
tree-level single-particle operators at four points. An explicit computation gives the result
\begin{align}
\lim_{\Lambda\rightarrow\infty} {\cal M}_{[2^2]p_2p_3p_4}(\Lambda \delta_{ij},d_{ij},p_i)\Bigg|_{\frac{1}{\Lambda^4}} & = 
\frac{  \frac{1}{8}(p_2+p_3 +p_4-4) \Big(- \underline{s}^2(4-6 f_s +f_s^2 ) + 2 (-f_s+f_t f_u)\underline{t} \underline{u} + \emph{crossing} \Big)}{  \underline{s}^2 \underline{t}^2 \underline{u}^2} + \notag\\
&\ \ \ \  - \frac{ \Big( \underline{s}^2(-2+f_s) + \frac{1}{2} f_s(d_{12}+d_{34}+2-\frac{1}{2}f_s) \underline{t}\underline{u} +\emph{crossing} \Big)}{  \underline{s}^2 \underline{t}^2 \underline{u}^2 }\,.
\end{align}
By construction, this expression 
is still a crossing symmetric function in the variables 
$\underline{s},\underline{t},\underline{u}$, $d_{ij}$ and $p_2,p_3,p_4$.

There is another limit that we could study, in which we rescale by $\Lambda$ all the Mellin variables $\delta_{ij},d_{ij}$ and 
also the three charges $p_2,p_3,p_4$ and then we send $\Lambda\rightarrow\infty$. Since both $\delta_{ij}$ and $d_{ij}$ are rescaled, 
this limit preserves the bold variables. 
In particular, the constraint \eqref{rho4ptvar} is replaced by ${\bf s}+{\bf t}+{\bf u}\rightarrow 0$, 
and at the same time $f_s+f_t+f_u\rightarrow 0$, 
which follows from the identity $f_s+f_t+f_u=4$ and the fact that $p_1=4$ 
is fixed while $f_s$, $f_t$, $f_u$ are sent to infinity. 
To compute the limit 
we introduce the {rescaled} variables of each type, 
such that $\underline{\bf s}+\underline{\bf t}+\underline{\bf u}=0$ and 
${f}_{\underline{s}}+{f}_{\underline{t}}+{f}_{\underline{u}}=0$. 
The result is again non-trivial: The naive leading order term, this time of order $O(1)$, cancels due 
to the same mechanism explained in \eqref{vanishing_of_flatspace_limit},
and the first non-trivial contribution is remarkably simple:
\begin{align}
{\cal M}_{[2^2]p_2p_3p_4}(\Lambda \delta_{ij},\Lambda d_{ij},\Lambda p_i) \approx&\ \ 
 \underbrace{({f}_{\underline{s}}+{f}_{\underline{t}}+{f}_{\underline{u}})}_{=\,0}\big(\ldots\big)\ +\, \frac{1}{\Lambda} \left( 
-\frac{(p_2+p_3 +p_4)}{8}\frac{ (\, \underline{\bf s}\, {f}_{\underline{t}} - \underline{\bf t}\, {f}_{\underline{s}}\, )^2 }{
(\underline{\bf s} )^2( \underline{\bf t} )^2( \underline{\bf u} )^2} \right) + O\!\left(\frac{1}{\Lambda^{2}}\right)\,,
\end{align}
Let us note that now the result is only crossing symmetry 
under ${\bold s}\leftrightarrow {\bold t}$ and $f_s\leftrightarrow f_t$.

\subsection{All tree-level amplitudes $\langle {\cal O}_r^2{\cal O}_{2}{\cal O}_{p}{\cal O}_{p}\rangle$}

It is interesting to ask if the structure seen in the previous section generalises to correlators containing 
a double-particle operator other than $\mathcal{O}_2^2$. In this section, we will derive the AdS$\times$S Mellin 
amplitude of ${\cal H}_{[r^2],2,p,p}$, which form a family of  N$^2$E correlators with ${\cal O}_r^2$ 
that we studied in section  \ref{generic_double_particle_sec}. We will see that for generic $r$ 
these Mellin amplitudes have a more complicated structure of poles than the one displayed 
in \eqref{eq:Msu}-\eqref{eq:Mstu}, and the number of poles grows quadratically with $r$. 

As in the previous section, we want to rewrite the correlator following the convention 
of the AdS$\times$S Mellin space, thus
\begin{equation}
\langle {\cal O}^2_{r}(z_1) {\cal O}_{2}(z_2) {\cal O}_{p}(z_3) {\cal O}_{p}(z_4)\rangle 
= \emph{free}+
{\cal R}_{1234}\,
\Bigg( N^{-2+r+p} \,r^2 (2)p^2 
\intsumfour\, {\cal M}_{[r^2]2pp}(\delta_{ij},d_{ij})
\Bigg)\,.
\end{equation}
Some preliminary comments first: Recall that for tree-level single-particle operators all N$^2$E 
are given by a single D-function. In particular,  we could consider ${\cal H}_{2r,2, p, p}$ 
which would have the same external charges as ${\cal H}_{[r^2] 2 p p}$, 
and it would be given by a single D-function. Now, when the single-particle operator 
is replaced by the double-particle operator, the Mellin amplitude  that we deduce 
from  \eqref{formula_nr} reads
\begin{equation}\label{numericalMIJ0}
\!\!\!\begin{tikzpicture}
\draw (0,0) node[scale=.93] {$\displaystyle
{\cal M}_{[r^2]2pp}= 
\!\!\sum_{I,J\,\ge 0}\!\frac{\Gamma[\delta_{12}+I-J]\Gamma[\delta_{14}-(2+I)]\Gamma[\delta_{13}-(2+I)]}{\Gamma[\delta_{12}]\Gamma[\delta_{14}]\Gamma[\delta_{13}]} 
\frac{\Gamma[1+d_{14}]\Gamma[1+d_{13}]}{\Gamma[1+d_{14}-(1+I)]\Gamma[1+d_{13}-(1+I)] } 
{\cal M}^{(I,J)}_{[r^2]2pp} $};
\end{tikzpicture}
\end{equation}
where the numerical coefficient ${\cal M}^{(I,J)}_{[r^2]2pp}$ is
\begin{equation}\label{numericalMIJ1}
{\cal M}^{(I,J)}_{[r^2]2pp}={{(p+1-r)r(r-1)}{}}
\frac{   (-1)^{I+J} \frac{\Gamma[2+I]\Gamma[r-1-I]}{\Gamma[1+J]\Gamma[r+1-J]} }{ ((r-1-J)+(r-2-I))\Gamma[1+I-J]}\,.
\end{equation}
Let us note that the sum over $I,J$ in \eqref{numericalMIJ0} truncates automatically, because 
of $\Gamma[r-1-I]\Gamma[r+1-J]$ and $\Gamma[1+I-J]$ in \eqref{numericalMIJ1}.
This is reminiscent of the mechanism underlying the summation over the sphere variables $d_{ij}$, 
with the difference that here $d_{ij}$ are already fixed (in fact the correlator is even N$^2$E) while $I,J$ 
would be additional variables that appear because of the double-particle operator. 
Similarly to \eqref{formula_nr}, the number of non-zero terms in the sum is 
$\frac{r(r-1)}{2}+1$.\footnote{The extra term that we count as the $+1$ is the one for 
which $I=r-2$, $J=I+1$. It can be included in the $\sum_{I,J}$ 
by the prescription that $I$ is evaluated first and then $J\rightarrow r-1$.} 

In the way ${\cal M}_{[r^2]2pp}$ is written in \eqref{numericalMIJ0},  each term of the sum
manifestly corresponds to a D-function. To see this note that the factor of $\Gamma[\delta_{12}]
\Gamma[\delta_{14}]\Gamma[\delta_{13}]$ simplifies against the $\prod_{i<j}
\Gamma[\delta_{ij}]$, coming from the AdS$\times$S measure, and gets replaced by 
$\Gamma[\delta_{12}+I-J]\Gamma[\delta_{14}-(2+I)]\Gamma[\delta_{13}-(2+I)]$. 
In doing so, ${\cal M}^{(I,J)}$ is a numerical coefficient and plays no role.
Therefore ${\cal M}_{[r^2]2pp}$ consists of a sum of D-functions, $\frac{r(r-1)}{2}+1$ 
of them with arguments depending on $I$ and $J$. This result has quite more structure compared 
to the result for single-particle operators, say ${\cal M}_{(2r)2pp}$.
As we remarked above, all such N$^2$E correlators are given by a single D-function. 

The way we wrote ${\cal M}_{[r^2]2pp}$ in \eqref{numericalMIJ0} shows that the leading 
term in the flat space limit, $\delta_{ij}\rightarrow \Lambda \delta_{ij}$ and $\Lambda\rightarrow \infty$, 
is the one for which $I=J=0$. With the same conventions on the rescaled Mellin variables 
$\underline{s},\underline{t},\underline{u}$ as in the previous section, 
we obtain the result
\begin{equation}
{\cal M}_{[r^2]2pp}(\Lambda \delta_{ij},d_{ij},p_i)\Bigg|_{\frac{1}{\Lambda^3}}=0,
\qquad\qquad {\cal M}_{[r^2]2pp}(\Lambda \delta_{ij},d_{ij},p_i)\Bigg|_{\frac{1}{\Lambda^4}}=\frac{1}{(2r-3)(r-2)!^2} \frac{1} {\underline{t}^2 \underline{u}^2}
\end{equation}
where $\underline{s}+\underline{t}+\underline{u}=0$. Again we find that the flat space limit is 
subleading compared to the case of single-particle operators which would be $O(\frac{1}{\Lambda^3})$. 
Moreover, we find that the result is consistent across different values of the external 
charges and different compositions of the double-particle operator, since the limit is order 
$O(\frac{1}{\Lambda^4})$ exactly as in the case of ${\cal M}_{[2^2]p_2p_3p_4}$. The resulting expression, 
as function of $\underline{s},\underline{t}$ and the relevant variables,  
depends of course on the details of the external operators. 

At this point, we would like to rewrite ${\cal M}_{[r^2]2pp}$ in \eqref{numericalMIJ0} by 
displaying the bold variables in this case. To do so, we have to rearrange the sum over $I$, $J$. 
It is simple to understand why by comparing with the expression for ${\cal M}_{[2^2]p_2p_3p_4}$ 
when $p_2=2$ and $p_3=p_4=p$. In the latter we only have $(a+{\bf s})(b+{\bf t})(c+{\bf u})$ 
poles, while in \eqref{numericalMIJ0} each contribution labelled by $I$ and $J$ gives 
a term proportional to 
\begin{equation}
\frac{ (-s)_{I-J} }{ (1+s) (r-I-3-t)_{I+2}(r-I-3-u)_{I+2} }\,.
\end{equation}
The denominator of the latter contains more than three factors of the form $(s+1)(t+b)(u+c)$, 
therefore, in order to proceed, we have to rewrite it. We seek a rewriting consisting of a sum 
of terms with denominators of the form $(s+1)(t+b)(u+c)$ and varying values of $b,c$. 
From studying the first few cases, it is not difficult to guess the algebraic identity that is needed, 
and write down the result, which of course can be checked a posteriori for arbitrary 
values of $p$ and $r$. Finally, we obtain
\begin{equation}\label{ultimated2pp}
\!\!\!{\cal M}_{[r^2]2pp}= \frac{ {(p+1-r)(r-1)} }{(1+{\bold s})(1+{\bold t})(1+{\bold u})} + \sum_{m,n=0}^{r-2} \frac{(-1)^{m+n+1} (p+1-r)(r-1)!^2 }{
 (m+n+1)_2  (r-2-m)!(r-2-n)!m! n!}\frac{2}{(1+{\bold s})(-m+{\bold t})(-n+\bold{u})}
\end{equation}
In the case $r=2$ we find perfect agreement with the way we wrote ${\cal M}_{[2^2]2pp}$ 
in \eqref{ansatzM2p2p3p4}. Note that also here the sum over $m,n$ can be left unbounded 
because of the factorials $(r-2-m)!(r-2-n)!$. 

Let us comment that in formula \eqref{ultimated2pp} the flat space limit is not manifest (as it 
was not manifest in ${\cal M}_{[2^2]p_2p_3p_4}$) since naively all terms contribute with 
a $1/\underline{s}\underline{t}\underline{u}$ term, as in the case of single-particle correlators. 
But of course formula \eqref{ultimated2pp} was derived by  \eqref{numericalMIJ0}, so indeed 
one can check that the naive leading term vanishes:
\begin{equation}
\!\!\!\!\!\!\!{\cal M}_{[r^2]2pp}(\Lambda \delta_{ij})\Bigg|_{\frac{1}{\Lambda^3}}= \frac{{(p+1-r)(r-1)}}{\underline{s}\underline{t}\underline{u} }+
\frac{2}{\underline{s}\underline{t}\underline{u}}\sum_{m,n} \frac{(-1)^{m+n+1} (p+1-r)(r-1)!^2 }{
 (m+n+1)_2  (r-2-m)!(r-2-n)!m! n!}=0
\end{equation}
where the sum $\sum_{m,n}$ equals $-\frac{{(p+1-r)(r-1)}}{2}$. Again, this cancellation 
could be used as a bootstrap constraint.

\section{The double-particle limit of the 5pt correlators}\label{doublelimitsection}

We will now consider the recently constructed five-point correlators 
\cite{Fernandes:2025eqe}. When the R-symmetry polarizations, $\vec{Y}$, and the spacetime 
insertion points, $\vec{X}$, of two of the external operators are taken to be equal, these five-point amplitudes 
should reduce to the four-point correlators with one double-particle operator that we considered in 
the previous sections. Thus, one expects that 
\begin{align}\label{OPE_5pt}
\!\!\!\!\!\!\!\!\!\!\!\!\!\!\!\!\!\lim_{z_0\rightarrow z_1} 
\langle {\cal O}_{p_0}(z_0) {\cal O}_{p_1}(z_1)  {\cal O}_{p_2}(z_2) & {\cal O}_{p_3}(z_3) {\cal O}_{p_4}(z_4)\rangle 
=  
\langle [{\cal O}_{p_0}{\cal O}_{p_1}]\!(z_1)  {\cal O}_{p_2}(z_2) {\cal O}_{p_3}(z_3) {\cal O}_{p_4}(z_4)\rangle\,.
\end{align}
It will be the purpose of this section to explain how to check this identity, and 
in particular, how to take the limit on the five-point correlator. We have checked \eqref{OPE_5pt} for
all charges $\vec{p}=\{p_0,p_1,p_2,p_3,p_4\}$ that give the four-point correlators 
of the form  $\langle [{\cal O}_{r_1}{\cal O}_{r_2}] {\cal O}_s 
{\cal O}_p {\cal O}_q\rangle$ bootstrapped in the previous sections and we have found a perfect match. 
Our result provides a highly non-trivial test of the formula conjectured in 
\cite{Fernandes:2025eqe}.

To clarify the double-particle limit, we examine in more detail the OPE of the operators 
${\cal O}_{p_0}(z_0)$ and  ${\cal O}_{p_1}(z_1)$ that appear in the LHS of \eqref{OPE_5pt}. 
Using the notation of \cite{Doobary:2015gia}, we can write
\begin{equation}\label{OPEdplimit}
{\cal O}_{p_0}(\vec{X}_0,\vec{Y}_0) {\cal O}_{p_1}(\vec{X}_1,\vec{Y}_1) = 
\sum_{\!\!\!\gamma=|p_0-p_1|}^{\,\,\,\,\,p_0+p_1}\!\! \sum_{\,\,O_{\gamma,\underline{\lambda}}}
 \, g_{01}^{ \frac{p_0+p_1}{2}-\frac{\gamma}{2} }\ C_{p_0p_1}^{\ \ \,O_{\gamma,\underline{\lambda}}}\, 
 \mathbb{D}^{ \gamma,\underline{\lambda} }(\vec{X}_{01},\partial_1)O_{\gamma,\underline{\lambda}}(\vec{X}_1)
\end{equation}
where the operator $\mathbb{D}^{ \gamma,\underline{\lambda} }$ generates descendant 
of the exchanged primary operators $O_{\gamma,\underline{\lambda}}$, and
$C_{p_0p_1}^{\ \ \,O_{\gamma,\underline{\lambda}}}$ is the (un-normalized) three-point point coupling between 
$O_{\gamma,\underline{\lambda}}$ and the external operators ${\cal O}_{p_0}$, ${\cal O}_{p_1}$.
In particular, the exchanged superprimary operators ${\cal O}_{\gamma,\underline{\lambda}}$ 
can be schematically written as follows,  
\begin{equation}\label{operator_Paul}
O_{\gamma,\underline{\lambda}}\sim {\rm span}\Bigg[ {\phi}^{ \frac{p_0-p_1+\gamma}{2} } (\partial^{|\underline{\lambda}|}) \phi^{ \frac{p_1-p_0+\gamma}{2}  }+\ldots \Bigg]
\end{equation}
where $\phi$ is the elementary scalar defined in \eqref{elementaryscalar}, and for given $\gamma$, 
the operators $O_{\gamma,\underline{\lambda}}$ are obtained by taking all possible Wick 
contractions. Note that this accounts for trace structures and/or degenerate operators. 
In \eqref{operator_Paul}, the  Young diagram $\underline{\lambda}$ gives the symmetrisation 
of the indices of the derivatives, and the ellipses denote terms with derivatives redistributed 
in such a way that $O_{\gamma,\underline{\lambda}}$ is a primary.

The double-particle limit that we are considering works as follows: 
\begin{equation}\label{eq:doubleplimit}
\rule{.4cm}{0pt}\lim_{z_0\rightarrow z_1}\langle{\cal O}_{p_0}(z_0) {\cal O}_{p_1}(z_1)  {\cal O}_{p_2}(z_2)  {\cal O}_{p_3}(z_3) {\cal O}_{p_4}(z_4)\rangle\ \equiv \lim_{\vec{X}_0\rightarrow \vec{X}_1} \Bigg( \lim_{\vec{Y}_0\rightarrow \vec{Y}_1} \langle{\cal O}_{p_0}(z_0) {\cal O}_{p_1}(z_1)  {\cal O}_{p_2}(z_2)  {\cal O}_{p_3}(z_3) {\cal O}_{p_4}(z_4)\rangle \Bigg)
\end{equation}
where we first take the limit on the polarizations, $\vec{Y}_0\rightarrow \vec{Y}_1$, and then 
we take the limit on the spacetime points, $\vec{X}_0\rightarrow \vec{X}_1$.

The first limit $\vec{Y}_0\rightarrow \vec{Y}_1$ projects the OPE in \eqref{OPEdplimit} onto the 
$\gamma=p_0+p_1$ sector, because otherwise the null constraint on the polarization vectors, 
$\vec{Y}_i\cdot\vec{Y}_i=0$, would  give a vanishing result, since in the limit $g_{01}=\vec{Y}_{0}
\cdot\vec{Y}_1/\vec{X}^{2}_{01}\rightarrow 0$. This guarantees that the second limit 
$\vec{X}_0\rightarrow \vec{X}_1$ is regular. Then, the leading term in the limit 
$\vec{X}_0\rightarrow \vec{X}_1$ projects on the half-BPS primary corresponding to 
empty Young diagram, $\underline{\lambda}=[\varnothing]$  and dimension 
$\Delta=\gamma=p_0+p_1$. Finally, for these quantum numbers, we have to look 
at $C_{p_0p_1}^{\ \ \,O_{\gamma,\underline{\lambda}}}$. In general, there is a space 
of half-BPS operators exchanged with the same value of $\gamma$, thus the proper 
definition is
\begin{equation}
C_{p_0p_1}^{\ \ \,O_{\gamma,\underline{\lambda}}}= 
\sum_{\underline{t}\,\vdash \gamma} |\langle {\cal O}_{p_0} {\cal O}_{p_1} O_{\underline{t}}\rangle| 
\cdot \big( |\langle O_{\underline{t}} O_{\gamma,\underline{\lambda}}\rangle| \big)^{\!-1}
\end{equation}
where $\underline{t}$ is a partition of $\gamma$, $|\langle {\cal O}_{p_0} {\cal O}_{p_1} 
O_{\underline{t}}\rangle|$ is the three-point coupling between the three operators 
and  $|\langle O_{\underline{t}} O_{\gamma,\underline{\lambda}}\rangle|$ refers to  
the two-point function metric.
In our case, since we have already established that $\Delta=\gamma=p_0+p_1$
we find that, at leading order in the $1/N$ expansion, 
the only exchanged operator is the double-particle, $O_{\gamma,\underline{\lambda}}
=O_{\underline{t}}=[{\cal O}_{p_0}{\cal O}_{p_1}]$ and the corresponding three-point coupling, 
$|\langle {\cal O}_{p_0} {\cal O}_{p_1} [{\cal O}_{p_0}{\cal O}_{p_1}]\rangle|$, just cancels the 
two-point function $|\langle O_{\underline{t}} O_{\gamma,\underline{\lambda}}\rangle|$. Beyond the 
leading order, there could be mixing with other half-BPS multi-particle operators corresponding to 
other partitions, $\underline{t}$, of $p_0+p_1$. However, the three-point couplings 
$|\langle {\cal O}_{p_0} {\cal O}_{p_1} O_{\underline{t}}\rangle|$ vanishes because there 
cannot be contractions between the multi-particle constituents in $O_{\underline{t}}$ and, by displacing 
them on fictitious points, the resulting 
diagram is near-extremal and thus it vanishes by orthogonality of the single-particle operators \cite{Aprile:2020uxk}.

Let us see more concretely how the double-particle limit works on the five-point functions. 
Following \cite{Goncalves:2019znr,Goncalves:2023oyx,Fernandes:2025eqe}, we represent the 
five-point function as an expansion over {\it all} free theory propagator structures. 
The AdS$\times$S Mellin space formalism, already introduced at four points in 
Section~\ref{sec_MellinAmps}, is again very useful. 
The relevant five-point generalization is the following,
\begin{equation}\label{fivep_Joao}
\langle {\cal O}_{p_0}(z_0) {\cal O}_{p_1}(z_1)  {\cal O}_{p_2}(z_2)  {\cal O}_{p_3}(z_3) {\cal O}_{p_4}(z_4)\rangle =\emph{free} + N^{{-3+\frac{1}{2}\sum_i p_i}} \intsumfive
\widetilde{\cal M}^{(5)}_{\vec{p}}(\tilde{\delta}_{ij},\tilde{d}_{ij})
\end{equation}
where we introduced the measure
\begin{equation}\label{fivep_measure}
\intsumfive \equiv \sum_{[\tilde{d}\,]} \prod_{i<j}\frac{\vec{Y}_{ij}^{2\tilde d_{ij} }}{\Gamma[\tilde d_{ij}+1]} 
\oint_{-i\infty}^{i\infty}  \frac{\!\![d\tilde\delta] }{ (2\pi i)^5 } \prod_{i<j}    \frac{  \Gamma[\tilde\delta_{ij}]}{ \vec{X}_{ij}^{2\tilde\delta_{ij} }  }\,
\end{equation}
and we are using the
``tilde" variables $\tilde{d}_{ij}=\tilde{d}_{ij}$ 
and $\tilde{\delta}_{ij}=\tilde{\delta}_{ji}$ with $i,j=0,1,\ldots, 4$ 
such that $\tilde{d}_{ii}=\tilde{\delta}_{ii}=0$, and 
\begin{equation}\label{on_shell_constraints_5}
\sum_{j=0}^{4} \tilde{\delta}_{ij} = p_i \qquad;\qquad \sum_{j=0}^{4} \tilde{d}_{ij}= p_i\,.
\end{equation}
This leaves five independent $\tilde{d}$'s and five independent $\tilde{\delta}$'s 
and the sum and the integral in \eqref{fivep_measure} are understood 
to be carried over these independent variables. The 
free theory contribution to the correlator goes on its own since it has 
vanishing Mellin transform. Thus, the AdS$\times$S Mellin amplitude 
$\widetilde{\cal M}^{(5)}_{\vec{p}}(\tilde{\delta}_{ij},\tilde{d}_{ij})$ is, in general, 
a function of \emph{15} independent variables, including also the five external weights $p_i$. 
The crucial difference between four and five-points is the absence of a uniquely 
defined ``reduced correlator" obtained by factoring out a factor, $\mathcal{R}\sim \mathcal{I}$, 
analogous to the one in \eqref{eq:Rfactor}. Therefore, we should compare the five-point 
AdS$\times$S Mellin space formalism \eqref{fivep_Joao} with the four-point version of it in which 
${\cal I}(U,V,\hat{\sigma},\hat{\tau})$ has been expanded out and has been absorbed 
into a ``bigger" AdS$\times$S Mellin amplitude. 
The factor ${\cal I}(U,V,\hat{\sigma},\hat{\tau})$ is also responsible for the $-2$ 
difference in the tilde-type on-shell constraints \eqref{on_shell_constraints_5} 
versus \eqref{constraintMellin4p}. It is to make clear this 
distinction that we are using here the ``tilde" variables.

At this point, it is important to emphasize that the constraints used by \cite{Goncalves:2019znr,Goncalves:2023oyx,Fernandes:2025eqe} 
to determine the five-point correlators at tree level
do not use $\langle [{\cal O}_{r_1}{\cal O}_{r_2}] {\cal O}_s{\cal O}_p{\cal O}_q\rangle$ 
as input. (We will review what they are in section \ref{bootstrap_Joao}.) Therefore, the 
double-particle limit \eqref{OPE_5pt} is non-trivial in two ways:
We must recover  both the universal factor ${\cal I}$ in \eqref{intriligator}  
and the dynamical function ${\cal H}_{[p_0p_1]p_2p_3p_4}$  obtained from the  
double-particle bootstrap. For this reason, the double-particle limit is a highly stringent 
check on the five-point correlators given in \cite{Fernandes:2025eqe}.

Coming back to the operations to be done in order to take the double-particle limit 
\eqref{eq:doubleplimit}, the one on the R-symmetry variables, $\vec{Y}_0\rightarrow \vec{Y}_1$, 
is immediate to implement. As for the four-point correlators, the sum over $\tilde{d}_{ij}$ in \eqref{fivep_measure} 
gives rise to a finite number of terms, each one corresponding to a propagator structure. The limit 
$\vec{Y}_0\rightarrow \vec{Y}_1$ can be done on the individual propagator structures, setting 
to zero all the structures containing $\vec{Y}_{01}^{2}$. It is then simple to check that all the 
propagator structures of the four-point function are recovered. In doing so, each propagator structure 
comes multiplied by a function of $\vec{X}_{i=0,1,2,3,4}$ that is still a five-point function:
\begin{align}
\lim_{\vec{Y}_0\rightarrow \vec{Y}_1} 
\langle {\cal O}_{p_0}(z_0)  {\cal O}_{p_1}(z_1)  &{\cal O}_{p_2}(z_2) {\cal O}_{p_3}(z_3) {\cal O}_{p_4}(z_4)\rangle=  \notag\\
&
\!\!\!{\cal P}_{[p_0p_1]p_2p_3p_4}\big[{g_{ij}}\big]\, \sum_{\gamma=\gamma_{\min}}^{\gamma_{\max}}\!\!\left(
\sum_{i+j=\frac{1}{2}(\gamma-\gamma_{\min})}\!\!\!\!\!\!(\hat{\sigma})^i (\hat{\tau})^j \ \mathcal{C}_{i,j}(\vec{X}_0,\vec{X}_1,\vec{X}_2,\vec{X}_3,\vec{X}_4)\right)\,,
\label{OPE_5pt_ony}
\end{align}
where $\gamma_{\min}=\max(|p_0+p_1-p_2|,|p_3-p_4|)$, 
$\gamma_{\max}=\min(p_0+p_1+p_2,p_3+p_4)$ and $\gamma=\gamma_{\min},\gamma_{\min}+2,\ldots,\gamma_{\max}$.

Now, $\mathcal{C}_{i,j}$ is a function of the five points and the limit 
$\vec{X}_0\rightarrow \vec{X}_1$ is therefore non-trivial. 
As we will prove later, ${\cal C}_{i,j}$ can be written as a sum of five-point D-functions. So, 
if one had a systematic algorithm to perform this rewriting, the limit $\vec{X}_0\rightarrow \vec{X}_1$
on each individual D-function would be immediate to carry out because the following identity holds 
\begin{equation}\label{resultDfunction}
\lim_{\vec{X}_0\rightarrow \vec{X}_1} D_{\Delta_0,\Delta_1,\Delta_2,\Delta_3,\Delta_4}(\vec{X}_0,\vec{X}_1,\vec{X}_2,\vec{X}_3,\vec{X}_4) = 
D_{\Delta_0+\Delta_1,\Delta_2,\Delta_3,\Delta_4}(\vec{X}_1,\vec{X}_2,\vec{X}_3,\vec{X}_4)\,.
\end{equation}
The proof of this identity is straightforward from the position space representation of the 
D-functions. We refer to appendix \ref{appendix_toolkit} for details, see~\eqref{eq:Dfuncdef} 
and~\eqref{eq:OPEinDfuncs}. On the other hand, the proof of this result in Mellin space is 
found by integrating over three (out of the five independent) Mellin variables. One integration 
is straightforward and corresponds to localize on the \emph{simple} pole of $\Gamma[\tilde{\delta}_{01}]$ 
at  $\tilde{\delta}_{01}=0$, corresponding to $\vec{X}_{01}^{2}=0$. The other two integrations 
can be done by using the first Barnes' Lemma~\eqref{eq:Barneslemma}.

In practise, the general five-point correlators are expressed through the AdS$\times$S Mellin 
space formalism, \eqref{fivep_Joao}, and the $\vec{X}_0\rightarrow \vec{X}_1$ limit has 
to be carried out directly in Mellin space. In the following section, we will consider 
a pragmatic approach similar to that of section \ref{section_analitic_boot}. 
First, we will provide some additional context to  understand important aspects of the
five-point Mellin amplitude, and thus of the ${\cal C}_{i,j}$ amplitudes introduced above, and then we will explain
the steps needed to take the limit $\vec{X}_0\rightarrow \vec{X}_1$, and ultimately prove the equality \eqref{OPE_5pt}, for the test 
case of ${\cal H}_{[2^2]2pp}$.
We treated all other cases in the same way by implementing the routine with a
computer program.

\subsection{Comments on the tree-level five-point Mellin amplitude of \cite{Fernandes:2025eqe}}
\label{bootstrap_Joao}

In this section, we review some salient features 
of the five-point Mellin amplitude presented in \cite{Fernandes:2025eqe}. 
Our discussion should clarify that the approach of  \cite{Fernandes:2025eqe} does not use the double-particle limit as an input. 

The solution technique of \cite{Fernandes:2025eqe} 
builds on the tree-level Witten diagram expansion
but avoids completely the diagrammatic expansion in AdS$_5$ by imposing a series of 
general consistency requirements that we discuss in the following.  
A crucial piece of information comes from the properties of Mellin amplitudes under factorization and our first task here 
will be to review how factorization formulae \cite{Fitzpatrick:2011ia,Goncalves:2014rfa} of tree-level Mellin 
amplitudes, first in AdS  \cite{Goncalves:2019znr,Goncalves:2023oyx} 
and then in AdS$\times$S \cite{Fernandes:2025eqe}, imply the following structure 
\begin{equation}
\label{eq: 5pts_gen_func}
{\widetilde{\cal M}^{(5)}_{\vec{p}}(\tilde{\delta}_{ij},\tilde{d}_{ij})} = \sum_{k_1,k_2=0}^{2} \frac{P_{k_1,k_2}(\tilde\delta_{ij},\tilde d_{mn},\vec{p}\,)}{(\pmb{\tilde\rho}_{12}+k_1)(\pmb{\tilde\rho}_{45}+k_2) } + \sum_{k_3=0}^{2} \frac{P_{k_3}(\tilde\delta_{ij},\tilde d_{mn},\vec{p}\,)}{(\pmb{\tilde\rho}_{12}+k_3)} + P(\tilde\delta_{ij},\tilde d_{mn},\vec{p}\,)+ \emph{(perms)}
\end{equation}
where $\pmb{\tilde\rho}_{ij}=\dlta_{ij}-\tilde d_{ij}$ and  $P_{k_1,k_2}$, $P_{k_3}$, $P$ are polynomial functions.

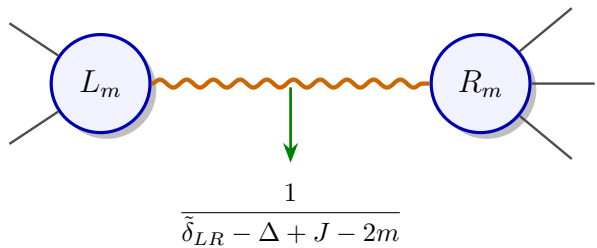
\begin{figure}
\centering
\begin{tikzpicture}[>=Stealth, every node/.style={font=\small}]

\tikzset{
  blob/.style={
    draw=blue!70!black,
    fill=blue!5,
    very thick,
    circle, minimum size=1.3cm, inner sep=4pt, font=\large,
    drop shadow
  },
  propagator/.style={
    decorate, decoration={snake, amplitude=1.5pt, segment length=10pt},
    draw=orange!80!black, line width=1.6pt
  },
  external/.style={draw=black!70, line width=0.9pt},
  arrowlab/.style={draw=green!50!black, -{Stealth[length=6pt]}, line width=1.0pt}
}

\node[blob] (L) at (0,0) {${L}_m$}; 
\node[blob] (R) at (5,0) {${R}_m$};

\draw[propagator] (L) -- (R);

\draw[external] (L) -- ++(-1.2,0.8);
\draw[external] (L) -- ++(-1.2,-0.8);

\draw[external] (R) -- ++(1.2,1.0);
\draw[external] (R) -- ++(1.2,-1.0);
\draw[external] (R) -- ++(1.5,0);

\draw[arrowlab,->] (2.5,-0.05) -- ++(0,-1.0);
\node[below, align=center] at (2.5,-1.2) 
  {$\displaystyle \frac{1}{\tilde{\delta}_{LR} - \Delta+J - 2m}$};
\end{tikzpicture}
\caption{Factorization of a five-point Mellin amplitude with $k=2$. The amplitude develops a 
simple pole at the twist $\Delta-J$ of a single-trace operator exchanged in the OPE 
and its conformal descendants. The residues are determined in terms of the 
lower-point amplitudes $M_L$ and $M_R$.
\label{fig:Mellin_factorization}}
\end{figure}

Let us begin with the properties of conventional AdS Mellin amplitude under factorization. 
It is well known \cite{Mack:2009mi,Penedones:2010ue,Fitzpatrick:2011ia,Goncalves:2014rfa}
that the Mellin amplitude of a $n$-point exchange Witten diagram, 
as depicted in Figure~\ref{fig:Mellin_factorization}, that corresponds
to the exchange of a bulk field in AdS dual to a single-trace operator 
(at large $N$), of dimension $\Delta$ and spin $J$,
has an infinite sequence of poles,
\begin{equation}\label{Mwithpole}
\widetilde{\cal M}(\,\dlta_{ij})
\approx\ \sum_{m} 
\frac{{\cal Q}_m(\,\tilde{\delta}_{ij})}{\tilde{\delta}_{LR}-(\Delta-J)-2 m} 
\end{equation}
where
$\tilde{\delta}_{LR}$ is the variable 
\begin{equation}
\label{eq: pole_var}
\tilde{\delta}_{LR}=\sum_{a=1}^{k}\sum_{b=k+1}^{n}\tilde{\delta}_{ab}\,.
\end{equation}
The definition of $\tilde{\delta}_{LR}$
corresponds to splitting the $n$-point diagram into two lower-point functions of $k+1$ and $n-k+1$ points.\footnote{For example for an $s$-channel Witten diagram at four-points: $n=k=2\, , \,\,\tilde{\delta}_{LR}=\tilde{\delta}_{13}+\tilde{\delta}_{14}+\tilde{\delta}_{23}+\tilde{\delta}_{24}=p_1+p_2-2\tilde{\delta}_{12}$.}
The value of $m=0,1,2\ldots$ corresponds to a descendant of the operator exchanged.
The residue functions  ${\cal Q}_m$ are non-trivial functions of the Mellin variables and are determined by the lower-point 
Mellin amplitudes according to factorization formulas studied in~\cite{Goncalves:2014rfa}. In the simplest case of a scalar 
exchanged operator 
\begin{equation}
{\cal Q}_m=\frac{-2\Gamma(\Delta)m!}{(\Delta-1)_m} {L}_m {R}_m 
\end{equation}
where $L_0=\widetilde{\cal M}_L$ ($R_0=\widetilde{\cal M}_R$) is the Mellin amplitude for the left (right) sub-diagram, 
and the others, $L_m(R_m)$, are determined through a recursion. 
This and analogous formulae for the exchange of a spin $J$ single-trace operator are given in~\cite{Goncalves:2014rfa}.

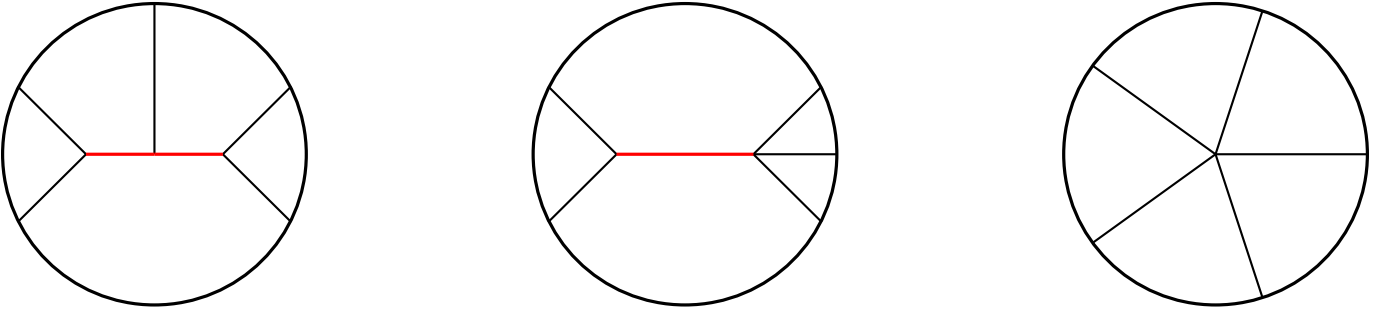
\begin{figure}[h!]
\centering
\begin{tikzpicture}[scale=1.8]
    \draw[very thick] (0,0) circle (1.11);

    \coordinate (top) at (0,1.11);
    \coordinate (upper-left) at (-1.,0.5);
    \coordinate (lower-left) at (-1,-0.5);
    \coordinate (lower-right) at (1,-0.5);
    \coordinate (upper-right) at (1,0.5);

    \coordinate (left-vertex) at (-0.5,0);
    \coordinate (top-vertex) at (0,0);
    \coordinate (right-vertex) at (0.5,0);

    \draw[thick] (upper-left) -- (left-vertex);
    \draw[thick] (lower-left) -- (left-vertex);

    \draw[thick] (top) -- (top-vertex);

    \draw[thick] (upper-right) -- (right-vertex);
    \draw[thick] (lower-right) -- (right-vertex);

    \draw[very thick,red] 
        (left-vertex) -- (top-vertex);
    \draw[very thick,red] 
        (right-vertex)--(top-vertex) ;
    
\end{tikzpicture}
\hfill
\begin{tikzpicture}[scale=1.8]
    \draw[very thick] (0,0) circle (1.11);

    \coordinate (upper-left) at (-1.,0.5);
    \coordinate (lower-left) at (-1,-0.5);
    \coordinate (lower-right) at (1,-0.5);
    \coordinate (upper-right) at (1,0.5);
     \coordinate (center-left) at (1.11,0);
    
    \coordinate (left-vertex) at (-0.5,0);
    \coordinate (right-vertex) at (0.5,0);

    \draw[thick] (upper-left) -- (left-vertex);
    \draw[thick] (lower-left) -- (left-vertex);

    \draw[thick] (upper-right) -- (right-vertex);
    \draw[thick] (center-left) -- (right-vertex);
    \draw[thick] (lower-right) -- (right-vertex);

    \draw[very thick, red] 
        (left-vertex) -- (right-vertex);
    
\end{tikzpicture}
\hfill
\begin{tikzpicture}[scale=1.8]
    \draw[very thick] (0,0) circle (1.11);

    \coordinate (center) at (0,0);

    \draw[thick] (center) -- (1.11,0);

    \draw[thick] (center) -- +(72:1.11);
    \draw[thick] (center) -- +(144:1.11);
    \draw[thick] (center) -- +(216:1.11);
    \draw[thick] (center) -- +(288:1.11);
    
\end{tikzpicture}
\caption{Tree-level Witten diagrams contributing to a five-point correlator. The red lines correspond to the exchange of single-trace operators in the OPE of the two external operators. 
\label{fig:Witten_diagrams_5pts}}
\end{figure}

For a five-point correlator in a given channel (we can think about the (12) channel for concreteness),  
the relevant process is the factorization into a three-point Mellin amplitude on one side and a four-point Mellin amplitude on the other. 
 The four-point Mellin amplitude  itself can be an exchange diagram or a contact diagram. 
The exchange diagram contributes to terms with two poles, 
e.g.~$(\tilde\delta_{12}+\k_1)(\tilde\delta_{45}+\k_2)$, the contact diagram 
contributes to terms with a single pole
e.g.~$(\tilde\delta_{12}+\k_3)$. 
Isolating the AdS variables $\tilde \delta_{ij}$ from the bold combination,
one finds precisely these two types of poles in the first two terms of \eqref{eq: 5pts_gen_func}. 
Finally, in the AdS$_5$ effective action, there are genuine five-point contact terms which determine the last polynomial
$P$ in \eqref{eq: 5pts_gen_func}.
Diagrammatically, these considerations can be understood in terms 
of the Witten diagrams drawn in  Figure~\ref{fig:Witten_diagrams_5pts}.

The exchanged field in a factorized channel are dual
to any field of the AdS$_5$ effective action with allowed quantum numbers. 
These are the half-BPS protected single-particle operators and their $su(4)$ bosonic descendant, as 
summarized by the following table 
\begin{equation}\label{tab: spectrum}
\begin{array}{c}
\begin{tikzpicture}
\draw (0,0) node {$
\begin{tabular}{|c|c|c|c|c|c|c|}
\hline
 & $\mathcal{O}_p$ & $\mathcal{J}_p$ & $\mathcal{T}_p$ & $\mathcal{A}_p$ & $\mathcal{C}_p$ & $\mathcal{F}_p$ \\
\hline
$(\Delta, J)$ & $(p,0)$ & $(p+1,1)$ & $(p+2,2)$ & $(p+2,0)$ & $(p+3,1)$ & $(p+4,0)$ \\
\hline
${[a,b,a]}$ & $[0,p,0]$ & $[1,p-2,1]$ & $[0,p-2,0]$ & $[2,p-4,2]$ & $[1,p-4,1]$ & $[0,p-4,0]$ \\
\hline
\end{tabular}$};
\end{tikzpicture}
\end{array}
\end{equation}
As we now review from \cite{Fernandes:2025eqe}, the properties of AdS Mellin amplitudes under factorization uplift 
to the $\text{AdS}\times${S} formalism, which therefore proves to be very advantageous 
for the bootstrap, since one needs to take into account factorization formulae for all the fields in \eqref{tab: spectrum} for all $p$, 
in order to bootstrap ${\widetilde{\cal M}^{(5)}_{\vec{p}}(\tilde{\delta}_{ij},\tilde{d}_{ij})}$. 

Let us focus on the exchange of the scalar fields ${\cal O}_p$,
to start with. Rather than looking at them individually, we would like to consider the 
exchange of the master field
\begin{equation}\label{masterscalar}
{\cal O}_M(z)=\sum_{p}  {\cal O}_{p}(z)\,,
\end{equation}
which includes all the $S^5$ KK modes.
This can be achieved by combining factorization formulas for both AdS and S, which share 
similar forms as suggested by noticing that the conformal $SO(4,2)$ Casimir 
and the R-symmetry $SO(6)$ Casimir are related by a Wick rotation.
In the case of the scalar operator $\mathcal{O}_M$, the generalization of the known formula in \eqref{Mwithpole} reads 
\begin{align}
\label{eq: AdsxS_factorization}
 \widetilde{\mathcal{M}}_{\vec{p}}(\tilde \delta_{ij}, \tilde d_{ij}) &\approx 
{\sum_q}\sum_{m+r=q} 
 A_{(\Delta, m \,|\, [a,b,a], r)} \,
\frac{\widetilde{\mathcal{M}}_{L,m,r} \times \widetilde{\mathcal{M}}_{R,m,r}}
{\widetilde{\pmb{\rho}}_{LR} - (\Delta-J-b-2a) - 2q}
\end{align}
where $\widetilde{\pmb{\rho}}_{LR}$ is defined as in~\eqref{eq: pole_var}, namely, 
\begin{equation}
\widetilde{\pmb{\rho}}_{LR}=\sum_{a=1}^{k}\sum_{b=k+1}^{n}\widetilde{\pmb{\rho}}_{ab}\,.
\end{equation}
At $m=r=0$ the left and right amplitudes, $\widetilde{\cal M}_{L}$ and $\widetilde{\cal M}_{R}$, are the factorized 
lower-point $\text{AdS}\times${S} amplitudes. As before, $m$ labels the conformal 
descendants, and $r$ the R-symmetry, or $SO(6)$, "descendants". Following 
 \cite{Fernandes:2025eqe}, the amplitudes for the descendants, $\widetilde{\mathcal{M}}_{L,m,r}$ and $\widetilde{\mathcal{M}}_{R,m,r}$
are obtained from those of the primaries through a recursion. 
For the left amplitudes, we find
\begin{align}
\widetilde{\mathcal{M}}_{L,m,r}&= 
\frac{(\hat{\delta}_{L})^m}{m!}
\frac{(\hat{d}_{L})^r}{r!}
\circ \widetilde{\mathcal{M}}_{L}(\tilde\delta_{ab},\tilde d_{ab})\,
\end{align}
where the shift operators are defined by
\begin{equation}
\hat{\delta}_{L} \circ \widetilde{\cal M}_{}
\equiv \sum_{a,b\in L, a<b}\tilde\delta_{ab}\, \widetilde{\cal M}_{}(\tilde\delta_{ab}+1, \tilde d_{ab})\qquad;\qquad
\hat{d}_{L}\circ \widetilde{\cal M}
\equiv \sum_{a,b\in L, a<b}\tilde d_{ab}\, \widetilde{\cal M}(\tilde\delta_{ab}, \tilde d_{ab}-1)\,,
\end{equation}
and where $L$ denotes the set of external points attached to the left part of the factorized amplitude in Figure~\ref{fig:Mellin_factorization}.
The same formula with the replacement $L\rightarrow R$ gives $\widetilde{\mathcal{M}}_{R,m,r}$ in terms of $ \widetilde{\mathcal{M}}_{R}$.

As a concrete example of \eqref{eq: AdsxS_factorization}, we give the form of $A$ for scalar operators
\begin{equation}
\label{eq: AdsxS_factorization_2}
A_{(\Delta, m\,|\,[a,b,a], r)} = 
\frac{1}{|{O}|^2}
\frac{-2\Gamma(\Delta)\, m!}{(\Delta-1)_m \,}
\frac{(-1)^r \, (b+2a+1-r)! r!}{(b+2a+1)!(b+2a)! }\,,
\end{equation}
where we introduced the two-point function normalization. 
The generalization of this formula for AdS$\times$S spinning operators, thus for operators like
${\cal J}_p,{\cal T}_p,{\cal C}_p$,
can be found in~\cite{Fernandes:2025eqe}.

Now, the importance of the AdS$\times$S factorization formulae is the following: It is easy to 
check from the table in~\eqref{tab: spectrum} that the value of $\Delta-J-b-2a$ for the exchanged 
operators ${\cal O}_p$ is an integer independent of the KK level $p$. Hence, if we 
consider $O_M$ to start with, the AdS$\times$S factorization formula in \eqref{eq: AdsxS_factorization} 
will have a single pole for the whole tower of KK modes, ${\cal O}_p$. 
This property generalizes for each one of the operators in the spectrum 
of table~\eqref{tab: spectrum}, ${\cal J}_p,{\cal T}_p,{\cal A}_p,{\cal C}_p,{\cal F}_p$
and allowed the authors of~\cite{Fernandes:2025eqe} to probe the factorization 
properties of the Mellin amplitude $\widetilde{\cal M}^{(5)}$ for all KK modes in one go, 
just by considering the contribution of the six master operators: ${\cal O}_M$ 
in \eqref{masterscalar} and ${\cal J}_M,{\cal T}_M,{\cal A}_M,{\cal C}_M,{\cal F}_M$ 
defined similarly. The AdS$\times$S factorization formulae also manifest the fact that the variables that describe the poles in \eqref{eq: 5pts_gen_func} are the bold variables $\widetilde{\pmb{\rho}}_{ij}$.  
 
Using the AdS$\times$S factorization formulae and imposing that only poles in $\widetilde{\pmb{\rho}}_{ij}$ (and no other spurious poles induced by the factorization formulas) can exist in the Mellin amplitude, the authors of~\cite{Fernandes:2025eqe} fixed the singular part of the five-point Mellin amplitude and in this way also determined the form of the lower order amplitudes, in particular the four-point amplitudes with 
one field in table \eqref{tab: spectrum} and three-single particle operators ${\cal O}_{p_i}$. This is indeed one of the possible ways to use factorization of Mellin amplitudes. 
If one knows the form of the necessary lower-point amplitudes explicitly, it is possible to construct the RHS of \eqref{eq: AdsxS_factorization} and constrain 
${\widetilde{\cal M}^{(5)}_{\vec{p}}}$ on the LHS, but the reverse is also true. 
This means that using the factorization formula from left to right, one finds constraints for the polynomials $P_{k_1k_2},P_{k_3}$ in \eqref{eq: 5pts_gen_func},
while using it in the other direction, one finds constraints for the lower-point amplitudes.
Note, however, that factorization properties 
alone cannot fix the contact terms in $\widetilde{\cal M}^{(5)}_{\vec{p}}$, i.e.~the $P$ polynomial, 
but this can be constrained by implementing the Drukker-Plefka twist~\cite{Drukker:2009sf} which 
makes all the $P_{k_1k_2},P_{k_3},P$ talk to each other. 
These constraints should clearly highlight that the approach of \cite{Fernandes:2025eqe}
is alternative to the double-particle limit.

As already anticipated, in~\cite{Fernandes:2025eqe} the authors make an ansatz for ${\widetilde{\cal M}^{(5)}_{\vec{p}}}$ and 
thus for the polynomials $P_{k_1k_2},P_{k_3}, P$ in \eqref{eq: 5pts_gen_func}.
This also includes setting a bound for the sums. 
At this point, it is useful to have information about the degree of these polynomial w.r.t.~their variables.
From the AdS$\times$S factorization 
formulae it is possible to argue that $\widetilde {\cal M}^{(5)}$ scales at least 
as $\Lambda^7$ in the large $p$ limit, i.e. when $\tilde\delta\rightarrow \Lambda \tilde\delta$, 
$\tilde d\rightarrow \Lambda \tilde d$, $p_i\rightarrow \Lambda p_i$ and $\Lambda\to\infty$. 
Then, it is also possible to fix the degree in the $\tilde\delta_{ij}$ by 
considering the previous studies of 
$\langle{\cal O}_2{\cal O}_2{\cal O}_2{\cal O}_p {\cal O}_p\rangle$ in~\cite{Goncalves:2019znr,Goncalves:2023oyx} 
using the flat space limit in which only the $\tilde\delta\rightarrow \Lambda \tilde\delta$ are rescaled, 
while the other variables are fixed. This gives that  the $P_{k_1,k_2}$, $P_{k_3}$ and $P$ are
expected to be of degree at most 2, 1 and 0 in $\tilde\delta_{ij}$, so that the scaling 
of $\widetilde{\cal M}^{(5)}$ in the limit of large $\tilde\delta_{ij}$ is $(\tilde\delta_{ij})^0$. 

Assuming all the above inputs, \cite{Fernandes:2025eqe} gave a solution for the polynomials 
$P_{k_1,k_2}$, $P_{k_3}$ and $P$. As already stated, in fact  \cite{Fernandes:2025eqe} does simultaneously an ansatz for $\widetilde{\cal M}^{(5)}$ and for the four-point amplitudes relevant to factorization, 
and remarkably the bootstrap returns both the five- and the four-point correlators. 
A non-trivial outcome of the bootstrap is the fact  
that the sum over the poles $\pmb{\tilde\rho}_{ij}$ actually truncates.\footnote{The reason for this truncation is yet to be understood. A similar truncation was also previously observed in \cite{Goncalves:2023oyx}.}

For completeness, we point out that the results of  \cite{Fernandes:2025eqe}, thus the polynomials  
$P_{k_1,k_2}$, $P_{k_3}$ and $P$, are provided by using a different choice of independent 
variables compared to the ones we have been mentioning above. The \emph{15} independent 
variables of \cite{Fernandes:2025eqe} are taken to be the \emph{10}+\emph{10} Mellin variables, 
which below we rename as $\tilde \delta_{ij}\rightarrow \gamma_{ij} $ and $\tilde d_{ij}\rightarrow n_{ij}$, subject to 
the \emph{5} constraints
\begin{equation}\label{onshelljoao}
\sum_{j\neq i} {\gamma}_{ij} = \sum_{j\neq i} n_{ij}\,.
\end{equation}
The choice of  \cite{Fernandes:2025eqe}  is to trade the explicit dependence on $p_i$ 
by considering \emph{5} extra ${n}_{ij}$. But in total there are still \emph{15} independent variables. 
With this choice, the amplitude presented  in \cite{Fernandes:2025eqe} is written as
\begin{equation}
\label{eq_joao_5pts_gen_func}
\widetilde {\cal M}_5 = 
\sum_{k_1,k_2=0}^{2} \frac{P_{k_1,k_2}(\gamma_{ij},n_{ij} )}{(\pmb{\tilde\rho}_{12}+k_1)(\pmb{\tilde\rho}_{45}+k_2)} + 
\sum_{k_3=0}^{2} \frac{P_{k_3}(\gamma_{ij},n_{ij})}{(\pmb{\tilde\rho}_{12}+k_3)} + P(\gamma_{ij},n_{ij})+ \emph{(perms)}\,,
\end{equation}
and it can be read off from the ancillary file attached to \cite{Fernandes:2025eqe}.

\subsection{The double-particle limit algorithmically}

In this section, we come back to the double-particle limit of tree-level five-point correlators and explain 
how to take the limiting procedure directly from their AdS$\times$S Mellin space representation:
\begin{equation}
\lim_{\vec{X}_0\rightarrow \vec{X}_1}\left[ \lim_{\vec{Y}_0\rightarrow \vec{Y}_1} \intsumfive {\widetilde{\cal M}^{(5)}_{\vec{p}}(\tilde{\delta}_{ij},\tilde{d}_{ij})} \right]\,.
\end{equation}
Our procedure will work for \emph{all} correlators, and in particular for those studied in 
section \ref{more_correlators_sec}. Using this algorithm, we checked extensively the 
agreement between our bootstrap results at four points and the double-particle limit 
taken on $\widetilde {\cal M}^{(5)}$ \cite{Fernandes:2025eqe}. However, for the purpose of this section, 
we will illustrate the double-particle limit for the simple test cases, such as the correlators $\widetilde{\cal M}^{(5)}_{222pp}$. 

For the particular case of $\widetilde{\cal M}^{(5)}_{222pp}$ we (re)checked that the
result of \cite{Fernandes:2025eqe}, quoted in \eqref{eq_joao_5pts_gen_func}, equals the one 
previously presented in \cite{Goncalves:2023oyx}, up to an overall $1/(p-2)!$ that comes from the 
$\Gamma$'s in the sphere part of the measure \eqref{fivep_measure}, that was not included in the formalism of \cite{Goncalves:2023oyx}, 
and up to a factor of $8 \sqrt{2}$ that is due to a different normalization of \cite{Goncalves:2023oyx}. 
In this paper, we do yet another choice of the normalization of single-particle operators that introduces in this case of interest an extra global constant $2p$ in comparison to these works.~\footnote{More generally, a five-point correlator computed from the generating-function of~\cite{Fernandes:2025eqe} is given in our conventions by $\widetilde{M}^{(5), \text{there}}_{\vec{p}}=\sqrt{2}/{\sqrt{p_1 p_2\dots p_5}}\, \widetilde{M}^{(5),\text{here}}_{\vec{p}}$.}

Compared to the ancillary file of \cite{Fernandes:2025eqe}, the rational functions 
in \cite{Goncalves:2023oyx} have the advantage of being concise, and for this reason 
we shall use them as we go through our exercise.

After taking the limit $\vec{Y}_0\rightarrow \vec{Y}_1$ on $\widetilde{\cal M}^{(5)}_{222pp}$, 
we find the following expected R-symmetry structure
\begin{equation}\label{expression5pt}
\lim_{\vec{Y}_0\rightarrow \vec{Y}_1} \intsumfive \widetilde{\cal M}^{(5)}_{222pp} = 
\vec{Y}_{12}^4 \vec{Y}_{14}^2 \vec{Y}_{13}^2 \vec{Y}_{34}^{2p-2} 
\sum_{0\leq m+n\leq 2}\!\!\!\!\!\!(\hat\sigma)^m (\hat\tau)^n 
 \oint_{-i\infty}^{+i\infty}\frac{ [d\tilde{\delta}] }{ (2\pi i)^5 }  \prod_{{i <  j}}  
 \frac{  \Gamma[\dlta_{ij}]}{ X_{ij}^{2\dlta_{ij} }  }\, \widetilde{\cal M}_{\hat \sigma^m\hat \tau^n}(\dlta_{ij})\,,
\end{equation}
where the sum spans over the six propagator structures $\{1,\hat\sigma,\hat\tau,\hat\sigma^2,\hat\sigma\hat\tau,\hat\tau^2\}$. 
At this point, it  is still  convenient to write $\widetilde{\cal M}_{\hat\sigma^m\hat\tau^n}$ by using 
the \emph{unconstrained} $\dlta_{ij}$. We find
\begin{align}\label{expression5pttau2}
\!\!\!\!\!\!\!\!\!\!\!\!\!\widetilde{\cal M}_{\hat\tau^2}&=\frac{2p}{(p-3)!}\Bigg[ 
\frac{\dlta_{34}(-1-2p+p(\dlta_{02}+\dlta_{04}-\dlta_{13}+\dlta_{23}) -\dlta_{34})}{(-1+\dlta_{04})(-1+\dlta_{23})}
-\frac{p\,\dlta_{24}\dlta_{34}}{(-1+\dlta_{03})(-1+\dlta_{14})} + (0\leftrightarrow 1)\Bigg]\,, \\
\!\!\!\!\!\!\!\!\!\!\!\!\!\widetilde{\cal M}_{\hat\sigma^2}&= \widetilde{\cal M}_{\tau^2}\Big|_{ 3 \leftrightarrow 4} \,,
\end{align}
and 
\begin{align}
\!\!\!\widetilde{\cal M}_{\hat\sigma\hat\tau}=\frac{2p}{(p-3)!}
\Bigg[ 
& \frac{ \dlta_{34}(-1-2p+p^2 +p(\dlta_{01}+\dlta_{04}+\dlta_{13})-\dlta_{34})}{(-1+\dlta_{04})(-1+\dlta_{13})} +\Bigg[ \frac{p(p-\dlta_{13})\dlta_{34}}{(-1+\dlta_{03})(-1+\dlta_{24})} + (3\leftrightarrow 4)\Bigg]+ (0\leftrightarrow 1) \Bigg]\notag\\
& \!\!\!\!+  \frac{2p}{(p-3)!}\Bigg[ \frac{2p \dlta_{34}}{(1-\dlta_{03})} + \frac{2p \dlta_{34}}{(1-\dlta_{04})} + (0\leftrightarrow 1) + (0\leftrightarrow 2) \Bigg] \,.
\end{align}
Expressions for the other amplitudes are lengthier. They take the form of
\begin{equation}
\left\{\begin{array}{c} {\widetilde{\cal M}}_{\hat\tau}= \frac{(\ldots)+(p-2)(\ldots) }{(p-2)!} \\[.5cm]
{\widetilde{\cal M}}_{\hat\sigma}={\cal M}_{\hat\tau}\Big|_{3\leftrightarrow 4}
\end{array}\right.
\qquad ;\qquad 
{\widetilde{\cal M}}_1= \frac{(\ldots)+(p-2)(\ldots) }{(p-2)!}
\end{equation}
and we will not write the explicit expressions in the parentheses here. We refer to appendix \ref{appendix_toolkit} for more details. It is easy to check that the terms of the five-point Mellin amplitude with poles in $\dlta_{01}$ are killed by the limit $\vec{Y}_0\rightarrow \vec{Y}_1$.

Note that for $p=2$ the expressions of $\widetilde{\cal M}_{\hat{\sigma}^2}, \widetilde{\cal M}_{\hat{\tau}^2}$, 
$\widetilde{\cal M}_{\hat{\sigma}\hat{\tau}}$ vanish trivially, but $\widetilde{\cal M}_{\hat{\sigma}},
\widetilde{\cal M}_{\hat{\tau}},\widetilde{\cal M}_{1}$ do not. 
This is interesting because the correlator  ${\cal C}_{2^2,2,2,2}$ is protected and therefore it coincides 
with its free theory value, therefore it must  happen that the limit $\vec{X}_0\rightarrow \vec{X}_1$ kills the contributions corresponding to 
$\widetilde{\cal M}_{\hat{\sigma},\hat\tau,1}$, up to free theory terms. In Mellin space, this means that after performing 
the integration, the result must vanish. Thus, even for ${\cal C}_{2,2,2,2,2}$, the double-particle limit is non-trivial.

Now, in order to reduce \eqref{expression5pt} to a four-point function (with two independent Mellin variables) we have to 
perform three out of five integrations in $[d\dlta]$. The first one is obvious: We 
choose $\dlta_{01}$ to be one of the independent variables, and then we pick the 
residue at $\dlta_{01}=0$ coming from $\Gamma[\dlta_{01}]$.
To perform the other two integrations, we need to first massage the expressions for ${\cal M}_{\hat{\sigma}^m\hat{\tau}^n}$. 
This can be done at the level of unconstrained variables because the rewriting we are looking for is algebraic. 
The idea is to ``absorb" the Mellin amplitude into the $\Gamma$ functions by rewriting it as 
a sum over $\Gamma$ functions with shifted arguments. The interpretation as a sum over 
D-functions is then manifest because a single string of 10 such $\Gamma$ functions is by definition a D-function,
as can be understood from relation~\eqref{eq:DfuncstoMellin}.\footnote{Note that the gamma functions $\Gamma(\tilde{\delta}_{ij})$ 
in \eqref{eq:DfuncstoMellin} cancel against the gamma factors in the definition of the Mellin amplitude in~\eqref{expression5pt}.}

To achieve our rewriting, there are three-types of identities that we shall use:
First, if a variable $\dlta_{ij}$ appears 
both in the numerator and the denominator, for example $\dlta_{04}$ 
or $\dlta_{23}$ in $\widetilde{\cal M}_{\hat{\tau}^2}$, see \eqref{expression5pttau2},
we have to split it as follows, 
\begin{equation}
\frac{\dlta}{-k+\dlta}= 1+\frac{k}{-k+\dlta}\,.
\end{equation}
This step is crucial to avoid nested integration at later steps. 
Then we shall use the identities quoted below

{\bf Identity on denominators:}
\begin{equation}
\label{eq:denomtogammas}
\frac{1}{-k+\dlta} = \sum_{j=1}^{k} \frac{ \Gamma[k] }{\Gamma[k-j+1]} \frac{\Gamma[\dlta-j]}{\Gamma[\dlta]}\,.
\end{equation}

{\bf Identity on numerators:}
\begin{equation}
\label{eq:numtogammas}
\dlta^k = \sum_{j=0}^{{k}} (-1)^{k-j} S_{k,j} \frac{\Gamma[\dlta+j]}{\Gamma[\dlta]} \,,
\end{equation}
where $S_{k,j}=\sum_{i=0}^j \frac{ (-1)^{j-i} i^k}{i!(j-i)!}$ are known as second Stirling numbers.

For illustration, let us quote a piece of the result for $\widetilde{\cal M}_{\hat\tau^2}$ 
\begin{equation}\label{sampleMtau2Mellin}
\frac{(p-3)!}{2p}\prod_{i<j}\Gamma[ \dlta_{ij} ]\widetilde{\cal M}_{\tau^2}=  
p \Gamma[ \dlta_{01}] \Gamma[ \dlta_{12}]\Gamma[ \dlta_{03} ]\Gamma[1+\dlta_{02}]\Gamma[-1+\dlta_{04}]\Gamma[\dlta_{14}]\Gamma[\dlta_{13}]\Gamma[-1+\dlta_{23}]\Gamma[\dlta_{24}]\Gamma[1+\dlta_{34}] + \ldots
\end{equation}
The full result can be found in appendix \ref{appendix_toolkit}, see formula \eqref{appMtoDfunc_testcase}. 
Each string of 10 $\Gamma$ functions is simply a five-point D-function in Mellin space. In total, there are 12 D-functions in \eqref{sampleMtau2Mellin}.

At this point, we proceed by solving the on-shell constraints in terms 
of $\dlta_{01},\dlta_{12},\dlta_{14}$ and $\dlta_{34},\dlta_{23}$, and 
we shall integrate over $\dlta_{01},\dlta_{12},\dlta_{14}$. We will first take 
the residue at $\dlta_{01}=0$ and then integrate in $\dlta_{12},\dlta_{14}$ 
using twice the first Barnes' Lemma. From the four-point correlator 
we know that the result \emph{must} reproduce 
\begin{equation}\label{sampeldbarnescalMtau2}
 \oint_{-i\infty}^{+i\infty}\frac{ [d\dlta] }{ (2\pi i)^5 } \prod_{i< j} 
 \frac{  \Gamma[\dlta_{ij}]}{ X_{ij}^{2\dlta_{ij} }  }\, \widetilde{\cal M}_{\hat{\tau}^2}(\dlta_{ij})= \frac{1}{X_{12}^4 X_{14}^2 X_{13}^2 X_{34}^{2p-2}}\, U\, {\cal H}_{[2^2]2pp}\,,
\end{equation}
and in fact, by redefining $\dlta_{23}=-t$ and $\dlta_{34}=-s-4+p$ 
in \eqref{sampeldbarnescalMtau2} we reobtain the known result~\eqref{solu2d2pp}
\begin{equation}
\label{eq:Hfrom5pts}
{\cal H}_{[2^2]2pp}= 2p\frac{(p-1)}{(p-3)!}\varoiint_{-i\infty}^{+i\infty}\!\!\frac{dsdt}{(2\pi i)^2} U^{s+2} V^t\Gamma[-s]\Gamma[-s+p-3]\Gamma[-t]^2 \Gamma[-u]^2  \frac{3+s}{(s+1)(t+1)(u+1)}\,.
\end{equation}
Similarly to the case of $\widetilde{\cal M}_{\tau^2}$, one can readily check all other contributions, ${\cal M}_{\hat{\sigma}^2, \hat{\sigma} \hat{\tau}, \hat{\tau},\hat{\sigma},1}$, by using the same procedure. 
Collecting all the contributions we reconstruct the polynomial ${\cal I}(U,V,\sigma,\tau)$ and arrive at the expected result, 
\begin{equation}
\lim_{z_0\rightarrow z_1} 
\langle {\cal O}_{2}(z_0) {\cal O}_{2}(z_1)  {\cal O}_{2}(z_2)  {\cal O}_{p}(z_3) {\cal O}_{p}(z_4)\rangle \Big|_{\frac{N^{\frac{1}{2}\sum p_i }}{N^3}}= \emph{free}+{\cal P}_{[2^2]2pp}\,{\cal I}(U,V,\sigma,\tau)\, {\cal H}_{[2^2]2pp}(U,V)\,.
\end{equation}
In particular, for $p=2$ we obtain ${\cal H}_{[2^2]222}=0$ after the last integration with Barnes' lemma.

For $\widetilde{\cal M}^{(5)}_{222pp}$ we can also check that the double-particle limit $z_3\to z_4$ 
gives ${\cal H}_{222[p^2]}=0$. However, one can see that this result is due to kinematics. First let us show that 
\begin{equation}\label{otherlimit222p2}
{\lim_{\vec{Y}_3\rightarrow \vec{Y}_4} \langle{\cal O}_{2}(z_0) {\cal O}_{2}(z_1)  {\cal O}_{2}(z_2)  {\cal O}_{p}(z_3) {\cal O}_{p}(z_4)\rangle = 0} \qquad \mathrm{for}\qquad p\ge 4\,.
\end{equation}
This follows from the condition on the degree of extremality of the full correlator ${\cal C}_{2,2,2,[p^2]}$. This 
extremality, $\tilde \kappa$, is related to the extremality, $\kappa$, 
of the reduced correlator defined in \eqref{degree_extre}, by  $\tilde \kappa =\kappa+2$, 
where the $+2$ shift come from adding $\mathcal{I}$. We use $\tilde \kappa$, rather than $\kappa$, 
because we are taking the limit of the five-point correlator, for which there is no 
reduced correlator to start with. For ${\cal C}_{2,2,2,[p^2]}$ we obtain
$\tilde \kappa=3-p$, and thus the vanishing of the limit in \eqref{otherlimit222p2} is 
automatically implemented in the skeleton expansion of $\langle{\cal O}_{2} {\cal O}_{2}  
{\cal O}_{2}  {\cal O}_{p} {\cal O}_{p}\rangle$ for $p\ge 4$. The special case $p=2$ has been  
discussed above. For $p=3$ we have $\tilde \kappa=0$ and thus we only 
get $\widetilde{\cal M}_1$, whose vanishing is again due to kinematics. Indeed, one 
has $\widetilde{\cal M}_1\propto \dlta_{34}(\ldots)$, and thus the Mellin integral vanishes 
when one takes the double-particle limit at the pole $\dlta_{34}=0$.

\section{Conclusions and discussion}\label{concl_discuss}

In this paper we focused on ${\cal N}=4$ SYM in the supergravity regime and showed 
that it is possible to bootstrap a new class of four-point correlators, i.e.~those involving 
one double- and three single-particle operators:
$\langle [{\cal O}_{r_1}{\cal O}_{r_2}]{\cal O}_s {\cal O}_p {\cal O}_q\rangle$.
Our approach only uses the consistency of the Operator Product Expansion 
and knowledge about the spectrum of double-particle operators 
in the supergravity regime \cite{Aprile:2017xsp,Aprile:2018efk}.
The inputs are the free theory and the tree-level single-particle 
correlators computed in \cite{Rastelli:2016nze}. The algorithm builds on the one previously 
developed for one-loop amplitudes \cite{Aprile:2017bgs,Aprile:2017qoy,Aprile:2019rep}, 
which we referred to as the ``double-particle bootstrap'', see section \ref{section_analitic_boot}.

It turns out that the tree-level holographic correlators in this class are given by 
sums of four-point contact diagrams, namely D-functions, and that the OPE predictions for 
exchanged double-particle operators fix the ansatz uniquely. A similar result holds 
for the tree-level single-particle correlators, $\langle {\cal O}_{p_1}{\cal O}_{p_2} {\cal O}_{p_3} {\cal O}_{p_4}\rangle$ 
\cite{Rastelli:2016nze,Rastelli:2017udc}, which were studied from the point of 
view of the OPE in \cite{Dolan:2006ec}. However, while it is always possible to bootstrap 
a given correlator of interest, the insertion of the double-particle operators brings 
new features when we study the dependence on the external charges, in comparison
with the known results for $\langle {\cal O}_{p_1}{\cal O}_{p_2} {\cal O}_{p_3} {\cal O}_{p_4}\rangle$.
In section \ref{sec_MellinAmps} we carried out the aforementioned comparison 
from the point of view of AdS Mellin amplitudes (since the relevant OPE analysis 
is obviously different in the two types of correlators)  and highlighted the novelties 
by considering two classes: first, we kept the double-particle operator fixed and equal to the 
simplest choice, ${\cal O}_2^2$,  and we studied  $\langle {\cal O}_2^2\, {\cal O}_{p_2} {\cal O}_{p_3} {\cal O}_{p_4}\rangle$ 
for arbitrary values of $p_2,p_3,p_4$, then, we considered varying the double-particle 
operator in the simplest infinite family of N$^2$E correlators, 
$\langle {\cal O}_{r}^2 \,{\cal O}_{2} {\cal O}_{p} {\cal O}_{p}\rangle$.

Let us recall that the tree-level Mellin amplitude~\eqref{singlep_treelevel_AdSxS} for 
single particle correlators, $\langle {\cal O}_{p_1} {\cal O}_{p_2} {\cal O}_{p_3} {\cal O}_{p_4}\rangle$, 
is simply related to the  flat space result. This relation is explained by the hidden conformal 
symmetry \cite{Caron-Huot:2018kta}, and it is manifest in the AdS$\times$S formalism \cite{Aprile:2020luw}. 
In the case of $\langle {\cal O}_2^2 \,{\cal O}_{p_2} {\cal O}_{p_3} {\cal O}_{p_4}\rangle$ 
we derived a formula for the corresponding Mellin amplitude ${\cal M}_{[2^2]p_2p_3p_4}$ 
in the AdS$\times$S formalism, and showed that the amplitude has more structure as 
it involves a sum of terms of the form $P(p_i)/(a+{\bf s})(b+{\bf t})(c+{\bf u})$ where $P(p_i)$ is 
a cubic polynomial in the KK modes $p_i$ and the shifts $a,b,c$ are either zero, one or minus one, 
see \eqref{ansatzM2p2p3p4}. Then we showed that, as a consequence 
of non-trivial cancellations, the leading term in the limit of large AdS Mellin variables 
is subleading compared to $1/{\bf s}{\bf t}{\bf u}$.

In the case of ${\cal M}_{[r^2]2pp}$ we obtained a formula with explicit dependence on $r$ and $p$. 
The dependence on the single-particle operators is similar to what we have seen already, but 
the dependence on the double-particle operator is quite non-trivial.
The structure of the amplitude is again a sum over shifted triple poles, $1/(a+{\bf s})(b+{\bf t})(c+{\bf u})$, 
weighted by $r$-dependent coefficients, but now the sum over poles is not restricted to $a,b,c=0,1,-1$ 
and instead the number of terms grows quadratically with $r$.

Finally, in section \ref{doublelimitsection}, through the so-called double-particle limit, we have discussed 
the direct implications of our results for the tree-level five-point correlators with external single-particle 
operators, recently conjectured in \cite{Fernandes:2025eqe}. The double-particle limit is the limit 
engineered by considering the OPE of two of the external single-particles and projecting onto the 
half-BPS sector. In this limit, a five-point correlator reduces to a four-point correlator with one 
double- and three single-particle operators. For all families of correlators that we bootstrapped, 
we have found a perfect match between our results and the double-particle limit taken 
on the five-point correlators. Thus, our results provide independent evidence for the formula 
and the solution technique put forward in \cite{Fernandes:2025eqe}.

The double-particle bootstrap with one external double- and three single-particle operators is quite flexible, 
and there are several future directions that would be worth exploring. First, it is quite straightforward 
to modify the algorithm so as to include $\alpha'$-corrections. The inputs are the free theory 
and now the Virasoro-Shapiro amplitude for the single-particle correlators. It would be very interesting 
to develop the $\alpha'$ expansion for $\langle [{\cal O}_{r_1}{\cal O}_{r_2}]{\cal O}_s {\cal O}_p {\cal O}_q\rangle$, 
as done in \cite{Drummond:2019odu,Drummond:2020dwr,Aprile:2020mus}, and at the same time generalize the worldsheet approach of 
\cite{Alday:2022uxp,Alday:2022xwz,Alday:2023jdk, Alday:2023mvu,Alday:2024ksp} and \cite{Wang:2025pjo}.
Similarly, it would be very interesting to explore our bootstrap approach 
for all maximally supersymmetric CFTs in diverse dimensions, 
and for AdS$_3\times$S$^3$.

{It should also be possible to apply the techniques discussed in this paper to the case where the double-particle operator in the 4-point correlator is $1/4$-BPS or semi-short. A first study of these correlators was presented in~\cite{Bissi:2021hjk} focusing on the case where the single-particle operators are ${\cal O}_2$ and the double-particle operator is $1/4$-BPS. It was shown that this correlator is protected and so it is determined by the free theory result, exactly as it happens for the corresponding $1/2$-BPS correlator. Thus, we expect that it is possible to carry out a similar analysis of four-point correlators with three single-particle operators and one $1/4$-BPS or semi-short double-particle, and that the results can be expressed in terms of contact diagrams, i.e. D-functions.}

On a different note is the question about the existence of a 10d 
generating function for $\langle [{\cal O}_{r_1}{\cal O}_{r_2}]{\cal O}_s {\cal O}_p {\cal O}_q\rangle$ 
at tree-level in supergravity. As a preliminary result in this direction, our AdS$\times$S Mellin 
space analysis shows that if such a generating function exists, its 10d seed is not as simple as a scalar amplitude. 
A posteriori, this could be argued as follows.  Consider $\phi$ to be
the 10d scalar field of twist 4 whose conformal four-point function $\langle \phi\phi\phi\phi\rangle$ generates 
the tree-level amplitudes $\langle {\cal O}_{p_1}{\cal O}_{p_2} {\cal O}_{p_3} {\cal O}_{p_4}\rangle$ \cite{Caron-Huot:2018kta}.
The OPE  $\phi\times \phi$ in 10d, which was successfully checked in the single-particle correlators, 
contains only the twist $8$ operators $\phi\partial^l\phi$ that lie at the 10d unitary bound. 
When a double-particle operator is inserted in the correlator, one could then conjecture a 10d ansatz given 
by a sum over correlators with a spinning field and three scalars,
of the form $\langle[\phi\partial^l\phi]\phi\phi\phi\rangle$.
The contribution with $l=0$, i.e.~$\phi^2$, corresponds to a scalar correlator in 10d.
However, $\langle\phi^2\phi\phi\phi\rangle$ at the lowest KK level is fixed so as to reproduce
the dynamical correlator for $\langle {\cal O}_2^2{\cal O}_2 {\cal O}_2 {\cal O}_2\rangle$, 
which vanishes because this correlator is protected. So the first non-trivial contribution
corresponds to a spin-two field in 10d. 
A preliminary analysis suggests that the simplest guess of restricting to a single spin-two field does not work, 
and thus the existence of a simple 10d conformal origin for correlators with multi-particle operators is far from obvious. It is interesting that also in perturbation theory \cite{Bargheer:2025uai} 
the question about how to implement the 10d symmetry is open.

Understanding the 10d origin of $\langle [{\cal O}_{r_1}{\cal O}_{r_2}]{\cal O}_s {\cal O}_p {\cal O}_q\rangle$ would be relevant for 
generalizing in our context the EFT approach of \cite{Abl:2020dbx} to the Virasoro-Shapiro 
amplitude and make a direct connection with flat space. As for the single-particle operators, 
the analytic structure of the amplitude at each order in the $\alpha'$ expansion is actually 
simpler than the supergravity term, so perhaps the proper generalization emerges 
already from bootstrapping the first $(\alpha')^3$ contribution.

Looking ahead, tree-level four point-correlators involving \emph{two} multi-particle operators cannot be obtained just 
by using the bootstrap approach discussed in this paper,
as the CFT data of double-particle operators does not provide enough information 
to fix the leading connected correlators. 
This is expected because the OPE predictions that would be needed are $O(\frac{1}{N^4})$ and 
generically these predictions involve multi-particle operators.
For the correlator $\langle {\cal O}_2^2 {\cal O}_2^2 {\cal O}_2{\cal O}_2\rangle$, 
 the analysis of \cite{Aprile:2025hlt} already demonstrated that this is the case. 
However, it turns out that the family of correlators $\langle {\cal O}_2^2 {\cal O}_2^2 {\cal O}_p{\cal O}_p\rangle$,
involving two ${\cal O}_2^2$ and two single-particle operators ${\cal O}_p$ with 
generic KK level, can be computed by the method of the coherent-state geometries mentioned 
in the Introduction. This is the subject of a companion paper~\cite{KKII} where we first re-derive 
the class of correlators discussed in section~\ref{sec:[22]pqr} and then extend the results  
of \cite{Aprile:2025hlt} to the family $\langle {\cal O}_2^2 {\cal O}_2^2 {\cal O}_p{\cal O}_p\rangle$ with $p>2$. 
It is currently more challenging to apply this geometric approach to correlators involving 
double-particle operators $[{\cal O}_{r_1}{\cal O}_{r_2}]$ with $r_1$ or $r_2$ larger than two 
since on the bulk side one would need to go beyond the supergravity consistent truncation 
of~\cite{Cvetic:2000nc} used in~\cite{Aprile:2025hlt}. Instead it is possible to work with the same 
truncation and construct geometries that capture the contribution of $1/4$-BPS states, 
see~\cite{Ganchev:2025dzn}. Then, by studying perturbations of these coherent-state geometries 
one derives both correlators with a single $1/4$-BPS and three single-particle operators, generalising 
those discussed in this paper, and correlators with two $1/4$-BPS and two single-particle operators. 
Work is in progress in this direction.

\subsection*{Acknowledgements}
It is a pleasure to thank Till Bargheer, Frank Coronado, James Drummond, Vasco Gon\c{c}alves, Paul Heslop, 
Hynek Paul, Michele Santagata, Bo Wang, Congkao Wen, Mitchell Woolley, and Xinan Zhou for discussions. 
FA would like to thank Frank Coronado for a private conversation about the results of \cite{Bargheer:2025uai} during BOOTSTRAP2025. 
FA would like to thank the organizers of the workshop \href{https://indico.global/event/14566/}{`New ways to higher points'},
for the possibility of presenting some results of this work prior to publication, and for stimulating conversations.

FA is supported by
RYC2021-031627-I funded by MCIN/AEI/10.13039/501100011033 and by the program NextGeneration EU/PRTR. 
FA also acknowledges support from the
``HeI" staff exchange program  
financed by the MSCA \href{https://cordis.europa.eu/project/id/101182937}{DOI:10.3030/101182937}.
JVB and RR are supported by the UK EPSRC grant ``CFT and Gravity: Heavy States and Black Holes" EP/W019663/1. RR is also supported by the Science and Technology Facilities Council (STFC) Consolidated Grant ST/X00063X/1 ``Amplitudes, Strings \& Duality'' and JVB by the FCT grant
2024.00230.CERN. No new data were generated or analysed during this study.

\appendix

\section{More on the double-particle bootstrap} \label{appendix_more}

In section \ref{section_analitic_boot} we described in great detail 
the double-particle bootstrap approach to ${\cal H}_{[2^2]2pp}$.
We demonstrated explicitly that there is a window of twists where only double-particle 
operators are exchanged and we gave formulae for the OPE predictions 
of exchanged double-particle operators, both long and protected, showing in particular how to 
deal with degenerate operators. 

For a generic correlator, ${\cal C}_{[r_1r_2]spq}$, 
 the schematic form of the predictions from the double-particle bootstrap, 
 written in the basis of superblocks \cite{Doobary:2015gia}, are those given in \eqref{intro_predictions}, namely
\begin{mdframed}
~\rule{0pt}{.3cm}\\
\begin{equation}\label{repeatedA1predictions}
\tfrac{1}{N^3}{\cal C}^{(\frac{3}{2})}_{[r_1r_2]spq}= \tfrac{1}{N}{\cal C}^{(\frac{1}{2})}_{ [r_1r_2] s; {\cal D}_{\tau,l}} \times 
 \left\{ \begin{array}{ll} \displaystyle  \tfrac{1}{N^2} \eta_{\cal D}\, {\cal C}^{(0)}_{pq; {\cal D} }\,\log(U)&\rule{.5cm}{0pt} \emph{if}\ \ \ \ p+q\leq \tau < r_1+r_2+s \\[.4cm] 
 				  \displaystyle \tfrac{1}{N^2} {\cal C}^{(1)}_{pq; {\cal D} } &\rule{.5cm}{0pt} \emph{if}\ \ \ \ \tau_{\emph unitary}\leq \tau< p+q \end{array}\right.
\end{equation}
~\\
\end{mdframed}
These predictions become predictions for the reduced correlator ${\cal H}_{[r_1r_2]spq}$.
In order to have a complete picture on how this works, there are
 two aspects of \eqref{repeatedA1predictions} that we want to illustrate, 
 in addition to the details given already in the main the text.  
These have to do with the ${\log^0(U)}{}$ sector and are quite crucial, e.g.~they 
have been used already in bootstrapping ${\cal H}_{[2^2],p,3,p+1}$ in section \ref{sec_22_NNNE}. 

{\bf The ${\log^0(U)}$ sector.}
At twist $\tau$, in the rep $[aba]$, the double-particle bootstrap predicts
\begin{equation}\label{generalA1}
{\cal H}^{(\frac{3}{2})}_{[r_1r_2]spq}  \Big|_{[a,b,a],\tau,l}
\equiv  -A^{(\frac{3}{2})}_{[r_1r_2]spq,[aba],\tau}\ \ + \sum_{ij }\frac{\left(  {\cal G}^{(\frac{1}{2})}_{[r_1r_2]sij}\Big|_{[aba],\tau} \right)\left( {\cal C}^{(1)}_{ij pq}\Big|_{[aba],\tau} \right)}{ \left( {\cal G}^{(0)}_{ ijij}\Big|_{[aba],\tau} \right)}
\end{equation}
The various elements in this formula are
\begin{itemize}
\item $A^{(\frac{3}{2})}_{[r_1r_2]spq}$ is the block coefficient of   
$\langle[{\cal O}_{r_1}{\cal O}_{r_2}]{\cal O}_s{\cal O}_p{\cal O}_q\rangle$ relative to
the $O(\frac{1}{N^3})$ connected tree-level free theory contribution to the correlator
at twist $\tau$ in the rep $[aba]$. 
\item ${\cal G}^{(\frac{1}{2})}_{[r_1r_2]sij}$ is the disconnected 
free theory contribution to the correlator $\langle[{\cal O}_{r_1}{\cal O}_{r_2}]{\cal O}_s{\cal O}_i{\cal O}_j\rangle$.
For this to be non-vanishing,  there must be (possibly a sum of) four-point diagrams that  individually, at order $O(\frac{1}{N})$, decompose into  
a (three-point function)$\times$(two-point function).
If we cut one such diagrams separating the pair $[{\cal O}_{r_1}{\cal O}_{r_2}]{\cal O}_s$ from the pair ${\cal O}_i{\cal O}_j$, 
the number of propagators across the cut, $\gamma$, has to be 
such that $\tau\ge \gamma$. This implies that the two-point function connects 
either ${\cal O}_{r_1}$ or ${\cal O}_{r_2}$ (in the double-particle operator) to one between 
${\cal O}_i$ or ${\cal O}_j$ (single-particle operators). 
\item
${\cal C}^{(1)}_{ij pq}$ is the tree-level contribution of a four-point single particle correlator $\langle{\cal O}_{i}{\cal O}_{j}{\cal O}_p{\cal O}_q\rangle$
and can be specified further, depending on whether $\tau$ 
equals the unitarity bound (therefore we are discussing the exchange of short operators) 
or it belongs to the window ${\rm min}(i+j,p+q)\leq \tau< {\rm max}(i+j,p+q)$
(therefore we are discussing the exchange of a long operator). 
\end{itemize}
As mentioned in the last point, to compute ${\cal C}^{(1)}_{ij pq}$ we distinguish
\begin{align}
\label{C1explained1}
\ \tau=2a+b+2,&\qquad {\cal C}^{(1)}_{ij pq}= {\cal G}^{(1)}_{ijpq} \\
\label{C1explained2}
\ 2a+b+4\leq\tau<{\rm max}(i+j,p+q),&\qquad {\cal C}^{(1)}_{ij pq}= {\cal G}^{(1)}_{ijpq}  + {\cal I}(U,V,\hat{\sigma},\tau) {\cal H}^{(1)}_{ijpq}\Big|_{\log^0(U)} 
\end{align}
On the RHS of \eqref{C1explained1} the value of ${\cal C}^{(1)}_{ij pq}$ is determined 
only by free theory, since we are at the 
unitarity bound and the exchanged double-particle operators are protected. 
On the RHS of \eqref{C1explained2} instead, we need to consider both free theory 
and the single-particle reduced correlator in its $\log^0(U)$ sector. 
In this case, the exchanged operators are long.

In the main text, we explained that predictions involving double-particle 
operators with quantum numbers at the unitarity bound,
are found by arranging a Gram matrix.
This Gram matrix has a block structure, determined by the following four blocks: 
\begin{equation}\label{gramA4}
\left(\begin{array}{cc} 
		\langle {\cal O}_{i} {\cal O}_{j} {\cal O}_{a}{\cal O}_{b} \rangle  & V^T_{pq} \\[.2cm] 
				V_{[r_1r_2]s} & \langle [{\cal O}_{r_1}{\cal O}_{r_2}]{\cal O}_s{\cal O}_p{\cal O}_q\rangle   \end{array}\right)
\end{equation}
where $i,j$ (and similarly $a,b$) label all pairs of single-particle operators such that the protected 
double-particle operators of interest are exchanged at leading order in the OPE 
$ {\cal O}_{i}\times{\cal O}_{j}$.
We would like to explain how \eqref{generalA1} is compatible with this Gram matrix. To this end, we point out that the block 
$\langle {\cal O}_{i} {\cal O}_{j} {\cal O}_{a}{\cal O}_{b} \rangle$ is diagonal 
at leading order -which is order  $O(1)$- 
therefore the constraint $\det\eqref{gramA4}=0$ gives a result for 
$\langle [{\cal O}_{r_1}{\cal O}_{r_2}]{\cal O}_s{\cal O}_p{\cal O}_q\rangle$ that at order $O(\frac{1}{N^3})$ 
yields the following simple formula for the corresponding block coefficient:
\begin{equation}
\sum_{ij} \frac{\left(  {\cal G}^{(\frac{1}{2})}_{[r_1r_2]sij}\Big|_{[aba],\tau} \right)\left( {\cal C}^{(1)}_{ij pq}\Big|_{[aba],\tau} \right)}{ \left( {\cal G}^{(0)}_{ ijij}\Big|_{[aba],\tau} \right)}\,.
\end{equation}
This turns into a prediction for the reduced correlator that is precisely what we claimed in \eqref{generalA1}.
Note that while the idea of assembling a Gram matrix is general, 
it is only at $O(\frac{1}{N^3})$ that we can restrict to double-particle operators exchanged.
Beyond this order we would need to enlarge  \eqref{gramA4} so as to 
take into account the exchange of multi-particle protected operators (when possible).

The two aspects of the general predictions in \eqref{generalA1} that we want to illustrate are: 
\begin{itemize}
\item The fact that two contributions appear on the RHS of \eqref{C1explained2}.
\item The interplay between multiplet recombination and the RHS of \eqref{C1explained1} 
for reps $[a,b,a]$ with $a\neq 0$.\footnote{This point was previously addressed in \cite{Aprile:2019rep} and more recently in \cite{Aprile:2025nta} }
\end{itemize}
 We will illustrate below these two points by giving additional details about the OPE predictions 
 of the N$^3$E correlators ${\cal H}_{[2^2],p,3,p+1}$ and the N$^4$E correlator ${\cal H}_{[2^2],4,4,4}$. ~\\[-.2cm]

\begin{center}
\underline{\bf The N$^3$E correlators $\langle {\cal O}_2^2 {\cal O}^{\phantom{2}}_p {\cal O}_3 {\cal O}_{p+1}\rangle$}
\end{center}

With reference to table \eqref{pred2dnontrivialN3E}, the reduced correlator for this family of correlators contributes to the following long reps
\begin{equation}
\begin{array}{|c|c|}
\hline
\rule{0pt}{.5cm}[a,b,a] & su(4)\ {\rm harmonic} \\[.15cm]
\hline
\rule{0pt}{.5cm}[0,p-2,0] & 1 \\[.15cm]
\hline
\rule{0pt}{.5cm}[0,p,0]& -\frac{2(p-2)}{p(1+p)}+\frac{2}{p}\hat{\sigma} +\frac{(p-2)}{p}\hat{\tau} \\[.15cm]
\hline
\rule{0pt}{.5cm}[1,p-2,1]&-\frac{p-4}{p+2}+ \hat{\sigma}-\hat{\tau}\\[.15cm]
\hline
\end{array}
\end{equation}

{\bf Free theory.} Upon extracting the prefactor
${\cal P}_{[2^2],p,3,p+1}=\sqrt{ 24 p(p+1) }N^{p+4} g_{12}^3 g_{14} g_{24}^{p-3} g_{34}^3$,  
the free theory has no $O(\frac{1}{N})$ disconnected term, and starts directly at $O(\frac{1}{N^3})$, where we find
\begin{equation}\label{free22p3pp1}
{\cal G}^{(\frac{3}{2})}_{[2^2],p,3,p+1} = \sqrt{ 24 p(p+1)} \Bigg[ 
 {\color{blue} (p-1)U\hat{\sigma}+ \frac{(p-2)(p-1)U\hat{\tau}}{2V}} + {\color{red}  2 U^2\hat{\sigma}^2 + \frac{2(p-1)U^2\hat{\sigma}\hat{\tau}}{V}} +\frac{U^3\hat{\sigma}^2\hat{\tau}}{V} \Bigg]
\end{equation}
The reason for the color coding will be explained in the next paragraphs.
In \eqref{free22p3pp1} we are considering the large-$N$ expansion without keeping track of the full $N$ dependence.
Some extra information would be needed in order to keep the full $N$ dependence. For example, the full result for the lowest value of $p$ is
\begin{equation}
{\cal G}_{[2^2],4,3,5}= \sqrt{ \frac{ 480(a-15)}{ (a-3)a^2(a+6)}} \Bigg[  \frac{U^3\hat{\sigma}^2\hat{\tau}}{V} 
 + \Bigg[ 3 U\hat{\sigma}+ \frac{3U\hat{\tau}}{V}\Bigg] +  \Bigg[ 2 U^2\hat{\sigma}^2 + \frac{6U^2\hat{\sigma}\hat{\tau}}{V}\Bigg] \Bigg]
\end{equation}
where we used the normalizations of $|{\cal O}_3|$ and $|{\cal O}_2^2|$ in \eqref{eq:O2nor} and also,
\begin{equation}
|{\cal O}_4|^2=\frac{4a(a-3)(a-8)}{(a+2)},\qquad|{\cal O}_5|^2=\frac{5a(a-3)(a-8)(a-15)}{(a+6)\sqrt{a+1}}\,.
\end{equation}
This is all known, but we will not need the full $N$ dependence for our purposes here, and therefore we can stick with \eqref{free22p3pp1}.

{\bf Prediction $[0,p-2,0]$ twist $p$ - {\color{blue} unitarity bound}.} We shall consider the terms in \eqref{generalA1}  separately. 
The prediction from the free theory correlator itself (in blue) reads
\begin{equation}
{ A^{(\frac{3}{2})}_{[2^2],p,3,p+1}}= \sqrt{ 24p(p+1)}\Bigg[ (p-1) M^{4,p,3,p+1}_{k=0,\gamma=p,[2+l]}+ \frac{(p-2)(p-1)}{2} M^{4,p,3,p+1}_{k=1,\gamma=p,[2+l]}\Bigg]\,.
\end{equation}
The other non-trivial term of  \eqref{generalA1} vanishes. To see this 
recall that a $n$-point single-particle correlator vanishes when it is near-extremal, 
thus $k\leq n-3$ \cite{Aprile:2020uxk}. The case ${\cal G}^{}_{[2^2]pij}$ when $i+j=p$, 
can be understood as a five-point near-extremal correlator of single-particle 
operators, for which $k=\frac{1}{2}(-p+(i+j)+4)\leq 2$. Thus it 
vanishes. Similarly, ${\cal G}_{ij,3,p+1}$ vanishes because the correlator 
can be understood as a four-point near-extremal correlator, for which 
$k=\frac{1}{2}(-p-1+(i+j)+3)\leq 1$. 

{\bf Prediction $[0,p-2,0]$ twist $p+2$ - {\color{red} long}.} In this case, the 
OPE prediction contains both contributions in \eqref{generalA1} and both contributions in \eqref{C1explained2}.
The contribution that comes from the 
free theory part of the correlator itself (in red) reads
\begin{equation}
A^{(\frac{3}{2})}_{[2^2],p,3,p+1,[0,p-2,0],\tau=p+2}= \sqrt{ 24p(p+1)}\Bigg[ 2 M^{4,p,3,p+1}_{0,p+2,[2+l,2]}+2(p-1) M^{4,p,3,p+1}_{1,p+2,[2+l,2]} \Bigg]\,.
\end{equation}
The non trivial contribution in  \eqref{generalA1} is obtained by gluing ${\cal C}_{[2^2],p,2,p}$ and ${\cal C}_{2,p,3,p+1}$.
For the first one, we are only interested in the disconnected contribution, i.e.~the first term in \eqref{free_22p2p},
and we find
\begin{equation}
 {\cal G}^{(\frac{1}{2})}_{[2^2]p2p}\Bigg|_{[0,p-2,0],p+2}=2 p M^{4,p,2,p}_{0,p+2,[2+l,2]}\,.
\end{equation}
The other term,  ${\cal C}^{(1)}_{2,p,3,p+1}$, requires more work. The novelty, compared to the main text, is the fact that 
this contribution is made by a term from the reduced correlator and a term from the free theory part of the correlator.
With the prefactor ${\cal P}=(\sqrt{ 6 p(p+1)} N^{p+3} +\ldots) g_{12}^2 g_{24}^{p-2} g_{34}^3$, 
the free theory and reduced correlators that we need are 
\begin{equation}\label{res2p3pp1}
{\cal G}_{2,p,3,p+1}= \frac{\sqrt{ 6 p(p+1)}}{a} \Bigg[ U\hat{\sigma} + (p-1) \frac{U\hat{\tau}}{V} + 2 \frac{U^2\hat{\sigma}\hat{\tau}}{V} \Bigg]\,,\qquad
{\cal H}^{(1)}_{2,p,3,p+1}= -\frac{ \sqrt{3(p+1)2 p}}{(p-2)!} U^3 \overline{D}_{3,p+3,2,p}.
\end{equation}
Combining the two terms, we find
\begin{equation}
\frac{ {\cal C}^{(1)}_{2,p,3,p+1}}{ \sqrt{6p(p+1)}} \Bigg|_{[0,p-2,0],p+2} = \frac{1}{(p-2)! }\left(2+ \frac{(-1)^l (l+3)! p! }{ (1+l+p)!}\right)\frac{(p+4)}{(p+2)}\frac{(l+p+1)!^2}{(2l+p+4)!} + 
2 M^{2,p,3,p+1}_{1,p+2,[2+l,2]}
\end{equation}
Finally, we need to consider ${\cal G}^{(0)}_{2p2p}\Big|_{[0,p-2,0],p+2}= M^{2p2p}_{0,p+2,[2+l,2]}$.

{\bf Prediction $[0,p,0]$ twist $p+2$ - {\color{red} unitarity bound}.}
In this case, the OPE prediction \eqref{generalA1} involves only free theory data. From the correlator itself, we read
\begin{equation}
A^{(\frac{3}{2})}_{[2^2],p,3,p+1,[0,p,0],\tau=p+2}= \sqrt{ 24p(p+1)}\Bigg[ 2 M^{4,p,3,p+1}_{0,p+2,[2+l]}+2(p-1) M^{4,p,3,p+1}_{1,p+2,[2+l]} \Bigg]
\end{equation}
and for the non-trivial contribution in  \eqref{generalA1} we input
\begin{align}
{\cal G}^{(\frac{1}{2})}_{[2^2]p2p}\Bigg|_{[0,p,0],p+2}& =2 p M^{4,p,2,p}_{0,p+2,[2+l]}\,, \qquad \qquad {\cal G}^{(0)}_{2,p,2,p}\Bigg|_{[0,p,0],p+2} = M^{2,p,2,p}_{0,p+2,[2+l]} \,, \\[.2cm]
{\cal G}^{(1)}_{2,p,3,p+1}\Bigg|_{[0,p,0],p+2}& =\sqrt{6 p(p+1)} (2) M^{2,p,3,p+1}_{1,p+2,[2+l]}\,.
\end{align}
The above computation is very similar to the ones explained in the main text. 

{\bf Prediction $[1,p-2,1]$ twist $p+2$ - {\color{red} unitarity} {\color{blue} bound}.} Again, 
the OPE prediction \eqref{generalA1} involves only free theory data. 
However, we have to be careful with what happened in $[0,p-2,0]$ and the fact that 
we used multiplet recombination. Schematically, multiplet recombination refers to the use of an identity in which
a semi-short rep $[0,b,0]$ with quantum numbers at the unitarity bound is written as a long rep $[0,b,0]$ 
and (for what we are concerned here) a semi-short rep $[1,b,1]$ with quantum numbers at the unitarity bound. 
This mechanism generates a cascade on the reps $[a,b,a]$ with non-zero 
$a$ that needs to be taken into account
at the moment of producing the predictions. 

For the case at hand, the value of $A^{(\frac{3}{2})}_{[2^2],p,3,p+1,[0,p,0],\tau=p+2}$ is obtained from the correlator itself, 
but comes from two contributions, 
\begin{align}
A^{(\frac{3}{2})}_{[2^2],p,3,p+1,[0,p,0],\tau=p+2}= \sqrt{ 24p(p+1)}\Bigg[ & {\color{blue} -(p-1) M^{4,p,3,p+1}_{0,p,[3+l]}- \frac{(p-2)(p-1)}{2} M^{4,p,3,p+1}_{1,p,[3+l]} } \\
&
\ \  {\color{red} + 2 M^{4,p,3,p+1}_{0,p+2,[2+l,1]}+2(p-1) M^{4,p,3,p+1}_{1,p+2,[2+l,1]} }\Bigg]\,.
\end{align}
The color coding shows the cascade. The Young diagram also helps understanding what is going on.
Next, for the non-trivial contribution in \eqref{generalA1} we have to consider the same cascading mechanism. However 
this only affects the value of  
\begin{align}
{\cal G}^{(1)}_{2,p,3,p+1}\Bigg|_{[1,p,1],p+2}& =\sqrt{6 p(p+1)}\Bigg[ - M^{2,p,3,p+1}_{0,p,[3+l]} -(p-1) M^{2,p,3,p+1}_{1,p,[3+l]} + 2 M^{2,p,3,p+1}_{1,p+2,[2+l,1]} \Bigg] \,.
\end{align}
The other inputs are
\begin{align}
{\cal G}^{(\frac{1}{2})}_{[2^2]p2p}\Bigg|_{[0,p,0],p+2}& =2 p M^{4,p,2,p}_{0,p+2,[2+l,1]}\,, \qquad \qquad {\cal G}^{(0)}_{2,p,2,p}\Bigg|_{[0,p,0],p+2} = M^{2,p,2,p}_{0,p+2,[2+l,1]} \,.
\end{align}
~\\

\begin{center}
\underline{\bf The N$^3$E correlators $\langle {\cal O}_2^2 {\cal O}^{\phantom{2}}_3 {\cal O}_p {\cal O}_{p+1}\rangle$}
\end{center}

With reference to table \eqref{pred2dnontrivialN3E}, the reduced correlator for this family of correlators contributes to the following long reps
\begin{equation}
\begin{array}{|c|c|}
\hline
\rule{0pt}{.5cm}[a,b,a] & su(4)\ {\rm harmonic} \\[.15cm]
\hline
\rule{0pt}{.5cm}[0,1,0] & 1 \\[.15cm]
\hline
\rule{0pt}{.5cm}[0,3,0]& \frac{1}{6}(-1+4\sigma+2\tau) \\[.15cm]
\hline
\rule{0pt}{.5cm}[1,1,1]&\frac{1}{5}(1+5\sigma-5\tau)\\[.15cm]
\hline
\end{array}
\end{equation}

{\bf Free theory.} Upon extracting the prefactor
${\cal P}_{[2^2],3,p,p+1}= \sqrt{ 24 p(p+1)} N^{4+p} g_{13}^3 g_{14} g_{34}^p$,  
the free theory contribution, at leading order, reads
\begin{equation}\label{free2d3ppp1}
{\cal G}^{(\frac{3}{2})}_{[2^2],3,p,p+1} = \sqrt{ 24 p(p+1)} \Bigg[  
2U\hat{\sigma} +\frac{U\hat{\tau}}{V} + {(p-1)U^2\hat{\sigma}^2}+\frac{2 (p-1)U^2\hat{\sigma}\hat{\tau}}{V} + \frac{(p-2)(p-1)U^3\sigma^2\tau}{2V}
 \Bigg]
\end{equation}

{\bf On the predictions at twist three}. The predictions in $[0,1,0]$ at 
twist three are straightforward to obtain, since there are no double-particle operators and 
one directly has to recombine the twist three contributions in \eqref{free2d3ppp1}.

{\bf On the predictions at twist five}. 
In this case, there is always the free theory part coming from \eqref{free2d3ppp1}, and 
in addition, there are non-trivial contributions from ${\cal G}_{[2^2]323}{\cal C}_{23p(p+1)}$. In the case of $[0,1,0]$ one has that 
${\cal C}_{23p(p+1)}$ is given as in \eqref{C1explained2}. For $[1,1,1]$ and $[0,3,0]$, since 
twist five is the unitarity bound, one only needs the free theory contribution.
For completeness, we quote the relevant result, which, however, can be deduced 
by applying a crossing transformation to \eqref{res2p3pp1}. 

Considering the prefactor ${\cal P}_{23p(p+1)}=(\sqrt{ 6p(p+1)} N^{3+p} +\ldots) g_{12}^2 g_{24} g_{34}^p$ we have
\begin{equation}
{\cal G}_{23p(p+1)}=\frac{1}{a}\Bigg[ U\hat{\sigma} + \frac{2U\hat{\tau}}{V} + \frac{(p-1)U^2\hat{\sigma}\hat{\tau}}{V} \Bigg]\qquad {\cal H}^{(1)}_{23p(p+1)}=\frac{\sqrt{6p(p+1)}}{(p-2)!}U^p \overline{D}_{p+2,p+1,4,1}
\end{equation}
The correlator ${\cal G}_{[2^2]323}$ was given in \eqref{free2332d}. 
~\\[-.2cm]

\begin{center}
\underline{\bf The N$^4$E correlator $\langle {\cal O}^2_2 {\cal O}_4 {\cal O}_4 {\cal O}_{4}\rangle$}
\end{center}

This correlator has degree of extremality ${\kappa}=2$, and therefore is a polynomial in $\hat{\sigma}$ and $\hat{\tau}$
up to degree two.
This correlator is also invariant under the exchange of points $3\leftrightarrow4$, $2\leftrightarrow4$, $2\leftrightarrow 3$, and for the reduced correlator this implies,  
\begin{align}
\label{crossing4444}
{\cal H}(U,V,\hat{\sigma},\hat{\tau})&=\frac{1}{\,V^2}{\cal H}\left(\frac{U}{V},\frac{1}{V},\hat{\tau},\hat{\sigma}\right)\\
{\cal H}(U,V,\hat{\sigma},\hat{\tau})&=\left(\frac{U}{V}\right)^4\!\!\hat{\tau}^2 {\cal H}\left(V,U,\frac{\hat{\sigma}}{\hat \tau},\frac{1}{\hat\tau}\right)\notag
&{\cal H}(U,V,\hat{\sigma},\hat{\tau})=\left(U\hat\sigma\right)^2 {\cal H}\left(\frac{1}{U},\frac{V}{U},\frac{1}{\hat{\sigma}} ,\frac{\hat\tau}{\hat\sigma}\right)\notag
\end{align}
Because of these crossing relations, there are only two independent scalar functions that parametrize ${\cal H}_{[2^2]444}$, which in fact can be written as
follows
\begin{equation}
{\cal H}_{[2^2]444}= 
{\cal F}(U,V) + \hat{\sigma}^2 U^2 {\cal F}(\tfrac{1}{U},\tfrac{V}{U}) + \frac{\hat{\tau}^2U^4}{V^4}{\cal F}(V,U) 
+ {\hat\sigma \hat\tau}{}\tilde{\cal F}(U,V) + \hat\tau U^2 \tilde{\cal F}(\tfrac{1}{U},\tfrac{V}{U})+ \frac{\hat\sigma U^4}{V^4} \tilde{\cal F}(V,U)
\end{equation}
where 
\begin{equation}
{\cal F}(U,V) = \frac{1}{V^2} {\cal F}\left( \frac{U}{V}, \frac{1}{V} \right),\qquad \tilde{\cal F}(U,V) = \frac{1}{V^2} \tilde{\cal F}\left( \frac{U}{V}, \frac{1}{V} \right)
\end{equation}

The reduced  correlator contributes 
to the following long reps
\begin{equation}
\begin{array}{|c|c|}
\hline
\rule{0pt}{.5cm}[a,b,a] & su(4)\ {\rm harmonic} \\[.15cm]
\hline
\rule{0pt}{.5cm}[0,0,0] & 1 \\[.15cm]
\hline
\rule{0pt}{.5cm}[0,2,0]& -\frac{1}{6}+\frac{1}{2} (\hat{\sigma}+\hat{\tau})\\[.15cm]
\hline
\rule{0pt}{.5cm}[1,0,1]& \hat{\sigma}-\hat{\tau}\\[.15cm]
\hline
\rule{0pt}{.5cm}[0,4,0]& \frac{1}{60}-\frac{2}{15}(\hat{\sigma}+\hat{\tau}) + \frac{1}{6} (\hat{\sigma}+\tau)^2  +\frac{1}{3} \hat{\sigma}\hat{\tau}\\[.15cm]
\hline
\rule{0pt}{.5cm}[1,2,1]& \frac{1}{4}(\hat{\sigma}-\hat{\tau})(-1+2\hat{\sigma}+2\hat{\tau})\\[.15cm]
\hline
\rule{0pt}{.5cm}[2,0,2]&\frac{1}{10} -\frac{1}{2}(\sigma+\tau) +(\sigma+\tau)^2-4 \sigma\tau \\[.15cm]
\hline
\end{array}
\end{equation}

The threshold for the exchange of triple-particle operators is twist eight, therefore the double-particle 
bootstrap only makes predictions strictly below twist eight. For this particular correlator, there will not 
be predictions in the  $\log^1(U)$ sector, but only in the $\log^0(U)$ sector

{\bf Free theory.}
 Upon extracting the prefactor
${\cal P}_{[2^2],4,4,4}=|{\cal O}_4|^3 |{\cal O}_2^2| g_{12}^4 g_{34}^4$,  
the free theory reads 
\begin{equation}\label{free224444}
{\cal G}_{[2^2],4,4,4} = \frac{16\sqrt{2}(1-\frac{5}{a}-\frac{15}{a+2}) \Bigg[  
U^2\hat{\sigma}^2 +\frac{U^2\hat{\tau}^2}{V^2} + 
4 \Bigg[\frac{U^2\hat{\sigma}\hat{\tau}}{V}+ \frac{U^3\hat{\sigma}^2\hat{\tau}}{V} +\frac{U^3\hat{\tau}^2\sigma}{V^2}\Bigg] + \frac{U^4\hat{\sigma}^2\hat{\tau}^2}{V^2}\Bigg]
}{\sqrt{(-8+a)(-3+a)(1+a)}}
\end{equation}

{\bf [0,0,0].} The free theory has no twist two contribution, therefore we move on to the long 
twist four and twist six predictions in  \eqref{generalA1}.
From the correlator itself, we find the free theory contributions,
\begin{align}
A^{(\frac{3}{2})}_{[2^2],4,4,4,[0,0,0],\tau=4} & = 16\sqrt{2}\Bigg[ M^{4,4,4,4}_{0,4,[2+l,2]} +M^{4,4,4,4}_{2,4,[2+l,2]} + 4M^{4,4,4,4}_{1,4,[2+l,2]}  \Bigg] \\
A^{(\frac{3}{2})}_{[2^2],4,4,4,[0,0,0],\tau=6} & = 16\sqrt{2}\Bigg[ M^{4,4,4,4}_{0,4,[3+l,3]} +M^{4,4,4,4}_{2,4,[3+l,3]} + 4M^{4,4,4,4}_{1,4,[3+l,3]}  + 4M^{4,4,4,4}_{1,6,[2+l,2,2]} +  4M^{4,4,4,4}_{2,6,[2+l,2,2]}\Bigg]\label{000twist622444}
\end{align}
The non trivial contributions in \eqref{generalA1} vanish: They involve correlators for which $i,j=2,2$ and $i,j=3,3$, however, 
the correlator ${\cal C}_{[2^2]422}$ vanishes and  ${\cal C}_{[2^2]433}$ does not have 
a disconnected contribution.

{\bf [1,0,1].} We have twist four operators at the unitarity bound and long twist six operators. 
The twist four OPE prediction \eqref{generalA1} are found only 
from the free correlator \eqref{free224444} because the correlator ${\cal C}_{[2^2]422}$ vanishes, as observed above. 
Then, since the twist two sector was absent in $[0,0,0]$ there is no cascade 
on the block coefficient in $[1,0,1]$, which simply reads
\begin{align}
A^{(\frac{3}{2})}_{[2^2],4,4,4,[1,0,1],\tau=4} & = 16\sqrt{2}\Bigg[ M^{4,4,4,4}_{0,4,[2+l,1]} +M^{4,4,4,4}_{2,4,[2+l,1]} + 4M^{4,4,4,4}_{1,4,[2+l,1]}  \Bigg] 
\end{align}
Regarding the OPE predictions  \eqref{generalA1}  at twist six, the non-trivial contribution vanishes because it involves the correlator ${\cal C}_{[2^2]433}$, which 
however does not have a disconnected contribution.
Therefore, the twist six prediction is determined only by
\begin{align}
A^{(\frac{3}{2})}_{[2^2],4,4,4,[1,0,1],\tau=6} & = 16\sqrt{2}\Bigg[ 4M^{4,4,4,4}_{1,6,[2+l,2,1]} + 4M^{4,4,4,4}_{1,6,[2+l,2,1]}  \Bigg] 
\end{align}
Note that only $\gamma=6$ propagator structures contribute, differently from \eqref{000twist622444}.

{\bf [0,2,0].} Similarly to the previous case, we have twist four operators at the unitarity bound and long twist six operators. 
For the same reasons as in $[1,0,1]$, the twist four prediction depends only on the free theory coefficient found from \eqref{free224444}, 
\begin{align}
A^{(\frac{3}{2})}_{[2^2],4,4,4,[0,2,0],\tau=4} & = 16\sqrt{2}\Bigg[ M^{4,4,4,4}_{0,4,[2+l]} +M^{4,4,4,4}_{2,4,[2+l]} + 4M^{4,4,4,4}_{1,4,[2+l]}  \Bigg] \,.
\end{align}
For the twist six contribution, we need to consider first
\begin{equation}
A^{(\frac{3}{2})}_{[2^2],4,4,4,[0,2,0],\tau=6} =16\sqrt{2}\Bigg[ 4 M^{4,4,4,4}_{1,6,[2+l,2]} + 4M^{4,4,4,4}_{2,6,[2+l,2]}  \Bigg] \,.
\end{equation}
and then, we need to add the non-trivial part which comes from summing over correlators with $i,j=2,4$ and $i,j=3,3$. 
The combination ${\cal C}_{[2^2]424}{\cal C}_{2444}$ is the only one that contributes. 
This is a case in which we will have the two terms, as in \eqref{C1explained2}.
With the prefactor ${\cal P}=( 4\sqrt{ 8} N^{7} +\ldots) g_{12}^2 g_{23}^{} g_{24}g_{34}^3$, 
the free theory and the reduced correlator that we need are 
\begin{equation}\label{free2444}
{\cal G}_{2,4,4,4}= \frac{16\sqrt{2}}{a} \Bigg[ U\hat{\sigma} + \frac{U\hat{\tau}}{V} + \frac{U^2\hat{\sigma}\hat{\tau}}{V} \Bigg]\,,\qquad
{\cal H}^{(1)}_{2,4,4,4}= -8 \sqrt{2} \,U^3 V \overline{D}_{5,5,5,1}.
\end{equation}
Combining the two terms, we find
\begin{equation}
{ {\cal C}^{(1)}_{2,4,4,4}}{ } \Bigg|_{[0,2,0],6} =16\sqrt{2}\Bigg[ 8(l+5)\frac{(l+4)!^2}{(2l+8)!}\frac{1+(-1)^l}{2} + 
M^{2,4,4,4}_{1,6,[2+l,2]}\Bigg]
\end{equation}
Finally, we need 
\begin{equation}\label{numerical8}
{\cal G}^{(0)}_{2424}\Bigg|_{[0,2,0],6}= M^{2424}_{0,6,[2+l,2]},\qquad
{\cal G}^{(\frac{1}{2})}_{[2^2],4,2,4}\Bigg|_{[0,2,0],6}= 8 M^{4424}_{0,2,[2+l,2]}
\end{equation} 
The  factor of $8$ in the disconnected contribution ${\cal G}^{(\frac{1}{2})}_{[2^2],4,2,4}$ follows 
from large $N$ counting, since it comes from $2\times \langle {\cal O}_2{\cal O}_2\rangle \langle {\cal O}_2{\cal O}_4 {\cal O}_4\rangle$.

{\bf [2,0,2].}  From this point on, we will only have twist six predictions at the unitarity bound.
 For this rep in particular, 
we need to consider the cascade from multiplet recombination in $[1,0,1]$ (and only $[1,0,1]$ because there was no twist two sector in $[0,0,0]$.)
Thus, the first contribution in \eqref{generalA1} is
\begin{equation}
A^{(\frac{3}{2})}_{[2^2],4,4,4,[2,0,2],\tau=6} =16\sqrt{2}\Bigg[ - M^{4,4,4,4}_{0,4,[3+l,1]} -M^{4,4,4,4}_{2,4,[3+l,1]} - 4M^{4,4,4,4}_{1,4,[3+l,1]} 
+ 4  M^{4,4,4,4}_{1,6,[2+l,1,1]} + 4M^{4,4,4,4}_{2,6,[2+l,1,1]} \Bigg] 
\end{equation}
The second contribution in \eqref{generalA1} vanishes: The relevant combination 
would be ${\cal C}_{[2^2]433}{\cal C}_{3344}$, but we pointed out already that ${\cal C}_{[2^2]433}$ does not have disconnected contributions.
Let us note that ${\cal C}_{[2^2]433}$ entered the discussion in previous reps, and it will also enter the discussion in the next two reps.  
Below, we will ignore its contribution for the same reason.

{\bf [1,2,1].}
The first contribution in \eqref{generalA1} reads
\begin{equation}
A^{(\frac{3}{2})}_{[2^2],4,4,4,[1,2,1],\tau=6} =16\sqrt{2}\Bigg[ - M^{4,4,4,4}_{0,4,[3+l]} -M^{4,4,4,4}_{2,4,[3+l]} - 4M^{4,4,4,4}_{1,4,[3+l]} 
+ 4  M^{4,4,4,4}_{1,6,[2+l,1]} + 4M^{4,4,4,4}_{2,6,[2+l,1]} \Bigg] 
\end{equation}
In this expression, we took into account multiplet recombination from $[0,2,0]$. Then, the second contribution in \eqref{generalA1} is non trivial and involves 
${\cal C}_{[2^2]424}{\cal C}_{2444}$. The free theory part
of ${\cal C}_{2444}$ was given already in \eqref{free2444} and we find
\begin{equation}
{\cal G}^{(1)}_{2444}\Bigg|_{[1,2,1],6} = 16\sqrt{2}\Bigg[ - M^{2,4,4,4}_{0,4,[3+l]} - M^{2,4,4,4}_{1,4,[3+l]} + M^{2,4,4,4}_{1,6,[2+l,1]} \Bigg] 
\end{equation}
where again we took into account multiplet recombination from $[0,2,0]$. Finally the disconnected 
contributions, for ${\cal C}_{[2^2]424}$ and ${\cal C}_{2424}$ are 
\begin{equation}
{\cal G}^{(0)}_{2424}\Bigg|_{[1,2,1],6}= M^{2424}_{0,6,[2+l,1]},\qquad
{\cal G}^{(\frac{1}{2})}_{[2^2],4,2,4}\Bigg|_{[1,2,1],6}= 8 M^{4424}_{0,2,[2+l,1]}
\end{equation} 
The factor of $8$ is explained as in \eqref{numerical8}.

{\bf [0,4,0].} Lastly, the first contribution in  \eqref{generalA1} is
\begin{equation}
A^{(\frac{3}{2})}_{[2^2],4,4,4,[0,4,0],\tau=6} =16\sqrt{2}\Bigg[ 4  M^{4,4,4,4}_{1,6,[2+l]} + 4M^{4,4,4,4}_{2,6,[2+l]} \Bigg] 
\end{equation}
The non trivial contribution involves again ${\cal C}_{[2^2]424}$ and ${\cal C}_{2424}$ and we find
\begin{equation}
{\cal G}^{(1)}_{2444}\Bigg|_{[0,4,0],6} = 16\sqrt{2} M^{2,4,4,4}_{1,6,[2+l]},\qquad 
{\cal G}^{(0)}_{2424}\Bigg|_{[0,4,0],6}= M^{2424}_{0,6,[2+l]},\qquad
{\cal G}^{(\frac{1}{2})}_{[2^2],4,2,4}\Bigg|_{[0,4,0],6}= 8 M^{4424}_{0,2,[2+l]}.
\end{equation}

{\bf Final result.} After collecting all predictions, and imposing 
the constraints from crossing symmetry we find the result 
\begin{align}
{\cal F} & = 8\sqrt{2}\oiint_{-i\infty}^{+\infty}\!\!\frac{dsdt}{(2\pi i)^2} U^{s+4} V^t \Gamma[-s]^2\Gamma[-t]^2\Gamma[-u]^2 \frac{ 1 }{(1+s)(2+s)(1+t)(u+1)}\\
\tilde{\cal F} & = 8\sqrt{2} \oiint_{-i\infty}^{+\infty}\!\!\frac{dsdt}{(2\pi i)^2} U^{s+4} V^t \Gamma[-s]^2\Gamma[-t]^2\Gamma[-u]^2 \frac{ (8+3s) }{(1+s)(1+t)(2+t)(u+1)(u+2)}
\end{align}
where $u=-s-t-6$.

\section{The double particle limit on the four-point correlators}

Through the double-particle limit we can 
reduce a $n$-point correlator of single-particle operators to a four-point correlator, 
and this is useful since, at the very least, we reduce the $n$-point kinematics to 
that of a four-point correlator for which the partial non-renormalization theorem holds.
The resulting four-point multi-particle correlator has less information
than the original $n$-point correlator but nevertheless it provides new CFT data that is
 simpler to analyze, e.g.~to decompose in superblocks.  While six and higher-point correlators 
 will be our next target, the double-particle limit also works if we start from a four-point correlator. 
However, this case is trivial in the sense that the double-particle limit yields a three-point 
coupling that is protected, so it will not provide dynamical information.

In this appendix, we will demonstrate that at tree level in supergravity,
\begin{equation}
\lim_{\vec{X}_0\rightarrow \vec{X}_1}\lim_{\vec{Y}_0\rightarrow \vec{Y}_1} \langle {\cal O}_{r_1}(z_0){\cal O}_{r_2}(z_1)  {\cal O}_p(z_2) {\cal O}_q(z_3) \rangle
= 
\langle [{\cal O}_{r_1}{\cal O}_{r_2}]\!(z_1)  {\cal O}_p(z_2) {\cal O}_q(z_3) \rangle
\end{equation}
The three-point function on the RHS is protected since all operators 
involved are half-BPS. In particular, 
\begin{equation}\label{free_bridges_app0}
\langle [{\cal O}_{r_1}{\cal O}_{r_2}]\!(\vec{X}_1,\vec{Y}_1)  {\cal O}_p(\vec{X}_2,\vec{Y}_2) {\cal O}_q(\vec{X}_3,\vec{Y}_3) \rangle= C_{[r_1r_2]pq} \prod_{i\neq j} g_{ij}^{b_{ij}}
\end{equation}
where the bridge lengths $b_{ij}\ge 0$ are
\begin{equation}\label{free_bridges_app}
 b_{12}=\frac{r_1+r_2}{2} +\frac{p-q}{2} \qquad;\qquad b_{13}=\frac{r_1+r_2}{2}+\frac{q-p}{2}\qquad;\qquad b_{23}= \frac{p+q}{2}-\frac{r_1+r_2}{2}\,.
\end{equation}
The bridge lengths only depend on the total charge $r_1+r_2$. 
The three-point coupling $C_{[r_1r_2]pq}$ is a non trivial function of 
$N$ and $r_1,r_2$. It can be computed from 
Wick contractions in the free theory. 
Here we shall focus on tree-level supergravity, thus $O(\frac{1}{N^2})$.

Considering the splitting that follows from the
partial non-renormalization theorem \cite{Eden:2000bk}, 
\begin{equation}
{\cal C}_{\vec{p}}(U,V,\hat{\sigma},\tau) = {\cal G}^{\emph free}_{\vec{p}}(U,V,\hat{\sigma},\hat{\tau})\, +\, {\cal I}(U,V,\hat{\sigma},\hat{\tau}) {\cal H}_{\vec{p}}(U,V,\hat{\sigma},\hat{\tau})
\end{equation}
we deduce that the limit is saturated by the free theory contribution, and therefore the factor 
${\cal I}(U,V,\hat{\sigma},\hat{\tau})\times$ ${\cal H}_{\vec{p}}(U,V,\hat{\sigma},\hat{\tau})$ must give a vanishing result 
in the limit. We will now show this by considering the reduced correlator written in AdS$\times$S Mellin space. 

First of all, it is useful to fix some conventions. Without loss of generality, 
we shall assume that $q\ge p$, $r_1\ge r_2$ and $(r_1-r_2)\leq (q-p)$. Let us introduce
\begin{equation}
c_s=\frac{r_1+r_2}{2}-\frac{p+q}{2}\qquad;\quad c_t=\frac{r_1-r_2+q-p}{2} \ge 0\qquad;\qquad c_u=\frac{-r_1+r_2+q-p}{2}\ge 0
\end{equation}
and the prefactor 
\begin{equation}
{\cal P}_{r_1,r_2,p,q}= g_{01}^{c_s} g_{03}^{c_t} g_{13}^{c_u} (g_{01}g_{23})^p
\end{equation}
In our conventions, we automatically get $c_t,c_u\ge 0$ but the sign of $c_s$ is not fixed. 
The dynamical function is
\begin{equation}\label{app_sugratree}
{\cal H}_{r_1,r_2,p,q}= \sum_{ {\tilde s},{\tilde t} }  {\tilde U}^{{ \tilde s} - p+2} \tilde{V}^{\tilde t}  \varoiint \frac{ds dt}{(2\pi i)^2} U^{s+p} V^{t}\,\, \frac{ \Gamma_{\textrm{AdS}\times \textrm{S}}  }{({\bold s}+1)({\bold t}+1)({\bold u}+1)}
\end{equation}
where $\tilde s+\tilde t+\tilde u=p-2$, $s+t+u=-p-2$, and
\begin{equation}\label{GammaAdSxS_ap}
\Gamma_{\textrm{AdS}\times \textrm{S}} = \frac{\Gamma[-s]\Gamma[-s+c_s] }{ \Gamma[\tilde{s}+1]\Gamma[\tilde{s}+1+c_s] }\times \big( \emph{t-channel}\big)\times \big( \emph{u-channel}\big)
\end{equation}
The cross ratios that we are using here are defined as follows,
\begin{equation}
U=\frac{\vec{X}^{2}_{01}\vec{X}^{2}_{23}}{  \vec{X}^{2}_{02} \vec{X}^{2}_{13} } \qquad;\qquad 
V=\frac{ \vec{X}^2_{03} \vec{X}_{12}^2 }{ \vec{X}_{02}^2 \vec{X}_{13}^2 }\qquad;\qquad 
\tilde{U}=\frac{ \vec{Y}_{01}^2 \vec{Y}_{23}^2 }{ \vec{Y}_{02}^2 \vec{Y}_{13}^2 }\qquad;\qquad 
\tilde{V}=\frac{ \vec{Y}_{03}^2 \vec{Y}_{12}^2 }{ \vec{Y}_{02}^2 \vec{Y}_{13}^2 }
\end{equation}
The relation with $\hat{\sigma},\hat{\tau}$ is $\hat{\sigma}=1/{\tilde U}$ and $\hat{\tau}={\tilde V}/{\tilde U}$. 

In the double particle limit we first consider $\vec{Y}_0\rightarrow \vec{Y}_1$, which implies $\tilde{U}\rightarrow 0$
and $\tilde{V}\rightarrow 1$. Let us deal with the $\sum_{\tilde s} \tilde U^{\tilde s}$ in \eqref{app_sugratree} first. This 
sum vanishes unless the arguments $\Gamma[\tilde{s}+1]\Gamma[\tilde{s}+1+c_s]$, 
are positive.
Now, from \eqref{free_bridges_app} and \eqref{free_bridges_app} 
we recognize that $-c_s=\frac{p+q}{2}-\frac{r_1+r_2}{2}=b_{23}\ge 0$, therefore the conditions $\tilde s\ge 0$ and
$\tilde{s}\ge -c_s$ are satisfied when $\tilde{s}\ge -c_s$. In this range, we need to solve for the value of $\tilde s$ such that
\begin{equation}
\lim_{y_0\rightarrow y_1}  {\cal P}_{r_1,r_2,p,q}\, {\cal I}(U,V,\hat{\sigma},\tau)  {\tilde U}^{{ \tilde s} - p+2} 
=   (\vec{Y}^2_{12})^{\frac{r_1+r_2}{2} +\frac{p-q}{2} } (\vec{Y}^2_{13})^{\frac{r_1+r_2}{2}+\frac{q-p}{2}} (\vec{Y}^2_{23})^{ \frac{p+q}{2}-\frac{r_1+r_2}{2}}\,.
\end{equation}
where the RHS is taken from \eqref{free_bridges_app0}-\eqref{free_bridges_app}.

Let us note that in the double particle limit ${\cal I}(U,V,\hat{\sigma},\hat{\tau}) \rightarrow {U^2}/{\tilde U^2}-2{U}/{\tilde U}+1$.
For each power of $\tilde{U}$ we obtain a value of $\tilde{s}$, but recalling that $\tilde{s}\ge -c_s$, it turns out that the only possible 
value is precisely $\tilde s= -c_s$ for which ${\cal I}(U,V,\hat{\sigma},\hat{\tau})\rightarrow {U^2}/{\tilde U^2}$. This result has a simple diagrammatic explanation:
Since the only diagrams that do not vanish in the double particle limit are those with no propagators going from ${\cal O}_{r_1}(X_0,Y_0)$ to ${\cal O}_{r_2}(X_1,Y_1)$,
the number of propagators on the vertical cut, let us call it $\gamma$,  of the propagator structure
\begin{equation}\label{diagrammatic_arg_app}
{\cal P}_{r_1,r_2,p,q}\, {\tilde U}^{{ \tilde s} - p}\propto (\vec{Y}^2)_{01}^{c_s} (\vec{Y}^2)_{03}^{c_t} (\vec{Y}^2)_{13}^{c_u} (\vec{Y}^2_{01}\vec{Y}^2_{23})^p 
\left( \frac{ \vec{Y}_{01}^2 \vec{Y}_{23}^2 }{ \vec{Y}_{02}^2 \vec{Y}_{13}^2 }\right)^{\!\!\tilde s-p}
\end{equation}
must be equal to $r_1+r_2$. In fact, we find that $\gamma=(c_t+c_u) +2(- \tilde s+p)=q+p-2\tilde s$. 
Thus the equation $\gamma=r_1+r_2$ is solved by $\tilde{s}=\frac{-r_1-r_2+p+q}{2}=-c_s$. 

To finalize, we need to consider the limit $\vec{X}_0\rightarrow\vec{X}_1$, which implies $U\rightarrow 0$ and $V\rightarrow 1$.
The crucial identity is again the reduction formula for D-correlators from four to three points, which is obvious in position space. 
It will not be needed here because it is clear that, by considering the power $U^{s+p}$, the least we can do 
is to consider the simple pole ${\bold s}+1=0$ evaluated at $\tilde s= -c_s$. 
Namely, the simple pole at $s=-1+c_s$.
By taking this residue, we find
\begin{equation}
\mathop{\mathrm{Res}}_{\ \rule{0pt}{.2cm}s = -1+c_s}
\Bigg[ \lim_{\vec{Y}_0\rightarrow \vec{Y}_1} {\cal P}_{r_1,r_2,p,q}\, {\cal I}(U,V,\hat{\sigma},\hat{\tau}) {\cal H}_{r_1,r_2,p,q}\Bigg] 
\propto U\,  (g_{12})^{\frac{r_1+r_2}{2} +\frac{p-q}{2} } (g_{13})^{\frac{r_1+r_2}{2}+\frac{q-p}{2}} (g_{23})^{ \frac{p+q}{2}-\frac{r_1+r_2}{2}}\,
\end{equation}
which therefore vanishes in the limit $U\rightarrow 0$.

In the derivation above we used the knowledge of ${\cal I}(U,V,\hat{\sigma},\hat{\tau})$. If one ignores this structure and 
writes the correlator as an expansion over free propagator structures, each one multiplied by a non-trivial function of $U,V$, 
as it happens for the Witten diagram expansion a la Rastelli-Zhou \cite{Rastelli:2017udc}, then the 
double-particle limit provides non-trivial information.  

Let us discuss the example of ${\cal C}_{22pp}$, 
by taking the result in \cite{Uruchurtu:2008kp} computed from the AdS$_5$ effective action. 
With 
\begin{equation}
\langle {\cal O}_2(X_0,Y_0) {\cal O}_2(X_1,Y_1) {\cal O}_p(X_2,Y_2) {\cal O}_p(X_3,Y_3) \rangle =\sqrt{2 N^2.2 N^2.p N^p.pN^p} \ g_{01}^2 g_{23}^p\  {\cal C}_{22pp}
\end{equation}
and quoting from  \cite[(61)]{Uruchurtu:2008kp}, we find
\begin{equation}
{\cal C}_{22pp}= a(U,V)
+ \Bigg[ \hat{\sigma} U b_1(U,V) + \frac{\hat{\tau} U}{V} b_2(U,V)\Bigg] 
+ \Bigg[ \hat{\sigma}^2 U^2 c_1(U,V) +\frac{\hat{\tau}^2 U^2}{V^2} c_2(U,V) + \frac{\hat{\sigma}\hat{\tau} U^2}{V} d(U,V)\Bigg]
\end{equation}
where $a,b_i,c_i,d$ are non trivial function written as combination of $\overline{D}$ correlators. The original Witten diagram representation  
is lengthy and given in \cite[(53)-(54)]{Uruchurtu:2008kp}. By the same diagrammatic argument in \eqref{diagrammatic_arg_app},  
the only terms that survive in the double-particle limit are
\begin{equation}
g_{01}^2 g_{23}^p\Bigg[ \hat{\sigma}^2 U^2 c_1(U,V) +\frac{\hat{\tau}^2 U^2}{V^2} c_2(U,V) + \frac{\hat{\sigma}\hat{\tau} U^2}{V} d(U,V)\Bigg]
\end{equation}
Since $\hat{\tau}=\hat{\sigma}$ and $V=1$ in the double-particle limit, we find
\begin{equation}
\lim_{z_0\rightarrow z_1} \frac{ \langle {\cal O}_2(z_0) {\cal O}_2(z_1) {\cal O}_p(z_2) {\cal O}_p(z_3) \rangle }{ \sqrt{2 N^2.2 N^2.p N^p.pN^p} } = 
g_{12}^2 g_{13}^p g_{23}^{p-2}\times \lim_{X_0\rightarrow X_1} \Bigg[ c_1(U,V) + c_2(U,V) + d(U,V)\Bigg]
\end{equation}
Thus, the double-particle limit  implies the constraint 
\begin{align}
g_{12}^2 g_{13}^p g_{23}^{p-2}\lim_{X_0\rightarrow X_1} \Bigg[ c_1(U,V) + c_2(U,V) + d(U,V)\Bigg]_{\frac{1}{N^2}} &= 
\Bigg[ \frac{ \langle {\cal O}_2^2 {\cal O}_p {\cal O}_p\rangle }{ \sqrt{2 N^2.2 N^2.p N^p.pN^p} } \Bigg]_{\frac{1}{N^2}}
\end{align}
which is non trivial w.r.t.~the Witten diagram representation of $c_1,c_2,d$.
In particular, it must be the case that 
\begin{align}\label{Linda_dp}
 g_{12}^2 g_{13}^p g_{23}^{p-2}\times \lim_{X_0\rightarrow X_1} \Bigg[ c_1(U,V) + c_2(U,V) + d(U,V)\Bigg]=g_{12}^2 g_{13}^p g_{23}^{p-2}\times \frac{2p(p-1)}{N^2}
\end{align}
For the RHS, call ${C}_{ijk}$ the (un-normalized) three-point coupling in $\langle {\cal O}_i{\cal O}_j{\cal O}_k\rangle$, 
then, one can check the first few values of $p$ explicitly, from Wick contractions:
\begin{align}
&
{C}_{{\cal O}_3,{\cal O}_3, {\cal O}_2^2}={ C}_{{T}_3,{T}_3, {T}_2^2}= \frac{72(a-3)(a)}{\sqrt{a+1}} \label{C33d2}\\
&
{C}_{{\cal O}_4, {\cal O}_4,{\cal O}_2^2}=
{C}_{{\cal O}_4,T_4, T_2^2}=  { C}_{T_4,T_4, T_2^2} + \alpha_4\,{C}_{T_2^2, T_4 , T_2^2} = \frac{192(a-8)(a-3)(a)}{a+2}  \label{C44d2} \\[.2cm]
&
{ C}_{{\cal O}_5, {\cal O}_5,{\cal O}_2^2}=
{C}_{{\cal O}_5,T_5, T_2^2} = {C}_{T_5,T_5, T_2^2}+\alpha_5\, {C}_{T_2T_3, T_5 , T_2^2} = \frac{ 400(a-15)(a-8)(a-3)a}{(a+6)\sqrt{a+1}} \label{C55d2}
\end{align}
In \eqref{C44d2}-\eqref{C55d2}, the first equality (from left to right) follows from 
 ${\cal O}_p=T_p\, +\sum \alpha_i\emph{(multi-traces)}_i$, by noting that upon 
 point-splitting the multi-traces, the admixture part can be thought of as a 
near-extremal correlators that vanish by multi-point orthogonality \cite[section 3.2]{Aprile:2020uxk}. 
The only contribution is therefore 
\begin{equation}
{C}_{{\cal O}_p,T_p,T_2^2}
\end{equation}
Note that when $p\ge 4$, we have ${C}_{{\cal O}_p,T_p,T_2^2}\neq { C}_{T_p,T_p,T_2^2}$ already at leading order. 
By using the explicit definition of ${\cal O}_p$ 
and some explicit three-point correlators, see \cite{Aprile:2020uxk},
it is straightforward to verify \eqref{C33d2}-\eqref{C55d2}.

In fact, one can prove the general result
\begin{equation}
{C}_{p,p,[2^2]}  = \big(2p\big)\big(2(p-1)\big)\langle {\cal O}_p {\cal O}_p\rangle\,.
\end{equation}
First, note that the bridge lengths are
\begin{equation}\label{free_bridges_appB}
 b_{12}=p-2 \qquad;\qquad b_{13}=2\qquad;\qquad b_{23}=2\,,
\end{equation}
but these do not fully specify the topology of the contractions and there are two possibilities. Since there cannot 
be contractions in between the double particle, $[T_2T_2]$, we can point split the trace structure. Then, 
the first possibility is the dumbbell shape ``$T_2(\vec{X}_3)={\cal O}_p(\vec{X}_1)=T_p(\vec{X}_2)=T_2(\vec{X}_3)$", 
which is vanishing thanks to the orthogonality properties of the single-particle operators \cite{Aprile:2020uxk}. 
The second possibility is
\begin{equation}\label{digramppd2}
\begin{array}{c}
 \begin{tikzpicture}[scale=1.5]
 
\def\latoxuno{0}
\def\latoyuno{0}
\def\latoxdue{1}
\def\latoydue{0}
\def\latoxtre{0.5}
\def\latoytre{1}

\draw[red!40] (\latoxuno,\latoyuno-.025) rectangle (\latoxdue, \latoydue+.025);
\draw (\latoxuno,\latoyuno) --  (\latoxtre, \latoytre);
\draw (\latoxdue,\latoydue) --  (\latoxtre, \latoytre);
\draw (\latoxuno,\latoyuno) --  (\latoxtre, -\latoytre);
\draw (\latoxdue,\latoydue) --  (\latoxtre, -\latoytre);

     \draw[fill=white!60,draw=white] (\latoxuno, \latoyuno) circle (2.5pt);
     \draw[fill=white!60,draw=white] (\latoxdue, \latoydue) circle (2.5pt);
     \draw[fill=white!60,draw=white] (\latoxtre, \latoytre) circle (2.5pt);
     \draw[fill=white!60,draw=white] (\latoxtre, -\latoytre) circle (2.5pt);

\draw(\latoxuno-.2, \latoyuno) node[scale=.8] {${\cal O}_p$};
\draw(\latoxdue+.2, \latoydue) node[scale=.8] {$T_p$};
\draw(\latoxtre+.05, \latoytre+.1) node[scale=.8] {$T_2$};
\draw(\latoxtre+.05, -\latoytre-.125) node[scale=.8] {$T_2$};      
     
\end{tikzpicture}   
\end{array}
\end{equation}
The diagram in \eqref{digramppd2} can be understood by performing Wick contractions between 
$T_p$ and $[T_2T_2]$ to produce again a $T_p$ (in double line notation), which gives $\langle {\cal O}_p T_p\rangle$ in the end. 
This is a version of the argument 
given in \cite[section 4.1.1]{Aprile:2020uxk}. When we glue $T_p$ and $[T_2T_2]$, in total, there are $2p$ contractions 
to glue $T_p$ with the first $T_2$ operator and $2(p-1)$ contractions to glue the other $T_2$. 
The rest of the combinatorics is the same as that of the one for the two-point function $\langle {\cal O}_p T_p\rangle$, 
which equals $\langle {\cal O}_p {\cal O}_p\rangle$.

Coming back to \eqref{Linda_dp} and the supergravity computation \cite{Uruchurtu:2008kp}, as we mentioned
 at the beginning of this appendix, once the Witten diagram computation is simplified 
in view of the partial non-renormalization theorem, the end result should become obvious. In fact, it does. 
The simplified results are given in \cite[(66)]{Uruchurtu:2008kp}. We quote them here below
\begin{equation}\label{simplified_Linda}
c_1\Bigg|_{\frac{1}{N^2}}\!\!\!=\ \frac{V}{U} {\cal H}_{},\qquad 
c_2\Bigg|_{\frac{1}{N^2}}\!\!\!\!=\ \ \frac{V^2}{U}{\cal H}_{},\qquad 
d\Bigg|_{\frac{1}{N^2}}\!\!\!\!=\ \ {2p(p-1)} + \frac{V}{U}(U-V-1){\cal H}_{}
\end{equation}
where ${\cal H}_{}=-\frac{2p}{(p-2)!} U^2 \overline{D}_{2,p+2,2,p}$. Collecting the terms proportional to ${\cal H}_{}$, and 
recalling that $V\rightarrow1$ in the double-particle limit, it is immediate to see from \eqref{Linda_dp} that ${\cal H}_{}$ 
does not contribute in the $U\rightarrow 0$ limit.  

\section{D-functions, Mellin space and the double-particle limit}
\label{appendix_toolkit}
A generic D-function of $n$ external points is given by the contact diagram
\begin{equation}
\label{eq:Dfuncdef}
    D_{\Delta_1,\dots,\Delta_n} = \int \frac{dZ_0\, d^d Z}{Z_0^{d+1}} \prod_{i=1}^{n} \left( \frac{Z_0}{Z_0^2 + (\vec{Z} - \vec{X}_i)^2} \right)^{\Delta_i}\,,
\end{equation}
where the integration is performed over the Poincar\'{e} coordinates of the contact point in Euclidean AdS$_{d+1}$.
Using the Feynman parameterization of \eqref{eq:Dfuncdef}, one can derive the relation
\begin{equation}
\label{eq:Dfuncrel}
    D_{\Delta_1,\dots,\Delta_i+1,\dots,\Delta_j+1,\dots,\Delta_n}
    = \frac{d/2 - \Sigma}{\Delta_i \Delta_j} \frac{\partial}{\partial X_{ij}^2} D_{\Delta_1,\dots,\Delta_n}\,.
\end{equation}
This relation is particularly useful since it allows us to find integer-weight D-functions from seed integrals. In fact, it follows from
~\eqref{eq:Dfuncrel} that all four-point D-functions can be obtained from the scalar one-loop box integral in four dimensions, $D_{1111}$, 
and all five-point D-functions can be obtained from the seed integral $D_{11112}$ \cite{Bern:1992em,Bern:1993kr,Goncalves:2019znr}.  

At four points, after the extraction of a kinematic factor, a D-function defines a $\overline{D}$-function that only depends on cross-ratios
\begin{equation}
\label{eq:DfunctoDbar}
\frac{\prod_{i=1}^4 \Gamma(\Delta_i)}{\Gamma\!\left(\Sigma - \tfrac{1}{2}d\right)} 
\frac{2}{\pi^{d/2}}
D_{\Delta_1 \Delta_2 \Delta_3 \Delta_4}(\vec{X}_1, \vec{X}_2, \vec{X}_3, \vec{X}_4)
= 
\frac{(X_{14}^2)^{\Sigma - \Delta_1 - \Delta_4} \, (X_{34}^2)^{\Sigma - \Delta_3 - \Delta_4}}
{(X_{13}^2)^{\Sigma - \Delta_4} \, (X_{24}^2)^{\Delta_2}}
\, 
\overline{D}_{\Delta_1 \Delta_2 \Delta_3 \Delta_4}(U, V)\,,
\end{equation}
where we use $\Sigma = \frac{1}{2} \sum_{i=1}^{n} \Delta_i$ and the cross ratios $U$ and $V$ defined in~\eqref{def_cross_ratiosFA}.
We quote the result for the seed function mentioned above, $\overline{D}_{1111}$,
\begin{equation}
\overline{D}_{1111}= \frac{1}{x_1 - x_2} \left[ 2 \operatorname{Li}_2(x_1) - 2 \operatorname{Li}_2(x_2) + \log(x_1 x_2) \log \left( \frac{1 - x_1}{1 - x_2} \right) \right]\,,
\end{equation}
where we use the change of variables $U=x_1 x_2$, $V=(1-x_1)(1-x_2)$.
To generate higher-weight D-functions
it is also helpful to use the differential relations
\begin{align}
    \partial_{x_1} \overline{D}_{1111}&= \frac{\overline{D}_{1111}}{x_2 - x_1} + \frac{\log[(1 - x_1)(1 - x_2)]}{x_1(x_2- x_1)} + \frac{\log(x_1 x_2)}{(x_1 - 1)(x_1 - x_2)}\,, \\
    \partial_{x_2} \overline{D}_{1111} &= \frac{\overline{D}_{1111}}{x_1 - x_2} + \frac{\log[(1 - x_1)(1 - x_2)]}{x_2(x_1 - x_2)} + \frac{\log(x_1 x_2)}{(x_2- 1)(x_2 - x_1)}\,.
\end{align}

In section~\ref{doublelimitsection}, we quoted the result 
\begin{equation}
\label{eq:OPEinDfuncs}
\lim_{\vec{X}_0\to \vec{X}_1} D_{\Delta_0,\Delta_1,\dots,\Delta_4}= D_{\Delta_0+\Delta_1,\dots,\Delta_4}\,.
\end{equation}
The proof of this follows immediately from the definition~\eqref{eq:Dfuncdef}.

In section~\ref{doublelimitsection}, we also considered Mellin space. 
In this regard, it is useful to use the identity between D-functions and Mellin amplitudes
\begin{equation}
\label{eq:DfuncstoMellin}
\prod_{i < j} (x_{ij}^2)^{-\alpha_{ij}} 
D_{\tilde{\Delta}_1 \ldots \tilde{\Delta}_n}
\;\longleftrightarrow\;
\widetilde{\cal{M}}(\tilde\delta)=\pi^{\tfrac{d}{2}}
\frac{
\Gamma\!\left( \tfrac{\sum_i \tilde{\Delta}_i - d}{2} \right)
}{
\prod_i \Gamma(\tilde{\Delta}_i)
}
\prod_{i<j}
\frac{\Gamma(\tilde \delta_{ij} - \alpha_{ij})}{\Gamma(\tilde \delta_{ij})}\,,
\end{equation}
where the $\tilde{\Delta}_i$ in general are not the values corresponding to scaling dimensions $\Delta_i$ of the external operators in the correlation function, but they are related 
to them by shifts as determined by
\begin{equation}
\tilde{\Delta}_i+\sum_j \alpha_{ij}=\Delta_i\,.
\end{equation}
Let us note that the RHS of \eqref{eq:DfuncstoMellin} is a rational function, so the Mellin transform of a D-function is a rational function.

It should be stressed that the map~\eqref{eq:DfuncstoMellin} has a kernel 
since the Mellin transform of free-like terms is zero. This means that 
all identities in which a combination of D-functions equals a free-like term 
are mapped in Mellin space to identities in which a combination of rational functions  
gives a vanishing result. This subtlety can be traced back to contour 
manipulations~\cite{Rastelli:2017udc}. It is an important subtlety 
in the case in which a correlator is bootstrapped as a sum over exchange (and contact) Witten diagrams. 
This was the case at four-points \cite{Rastelli:2016nze} and it is the case at five points, 
see the Witten diagrams of Figure~\ref{fig:Witten_diagrams_5pts}.
All of them are reduced to linear combinations of D-functions by exploiting 
the vertex identities studied in~\cite{DHoker:1999mqo}. 

However, the double-particle bootstrap at four points is free from the 
subtlety about the kernel of \eqref{eq:DfuncstoMellin} and free theory, because from the start 
the correlator is organized as the free theory contribution and the reduced correlator contribution, 
according to the partial non-renormalization theorem.
Then, the weight-0 stratum 
of the reduced correlator is uniquely fixed by the absence of the $x_1\rightarrow x_2$ singularities. 
For what concerns us, this implies that matching the double-particle limit from five to four-points can 
be entirely conducted in Mellin space,  handling more compact expressions.

Here, for concreteness, we provide more details about the discussion of 
section~\ref{doublelimitsection} and consider the Mellin amplitude $\widetilde{\cal{M}}_{222pp}$ 
as written in~\cite{Goncalves:2023oyx}. As discussed in~\eqref{expression5pt}, after taking 
the limit $\vec{Y}_0\to\vec{Y}_1$ on $\widetilde{\cal{M}}_{222pp}$ and removing a 
R-symmetry prefactor, we obtain the following $\widetilde{\cal{M}}_{\hat\sigma^m \hat\tau^n}$ components
\begin{align}\label{expression5pttau2_app}
\!\!\!\!\!\!\!\!\!\!\!\!\!\!\widetilde{\cal M}_{\hat\tau^2}&=\frac{2p}{(p-3)!}\Bigg[ 
\frac{\dlta_{34}(-1-2p+p(\dlta_{02}+\dlta_{04}-\dlta_{13}+\dlta_{23}) -\dlta_{34})}{(-1+\dlta_{04})(-1+\dlta_{23})}
-\frac{p\,\dlta_{24}\dlta_{34}}{(-1+\dlta_{03})(-1+\dlta_{14})} + (0\leftrightarrow 1)\Bigg] \,,
\end{align}
\begin{align}
\!\!\!\!\!\! \widetilde{\cal M}_{\hat\sigma\hat\tau}=\frac{2p}{(p-3)!}
\Bigg[ 
& \frac{ \dlta_{34}(-1-2p+p^2 +p(\dlta_{01}+\dlta_{04}+\dlta_{13})-\dlta_{34})}{(-1+\dlta_{04})(-1+\dlta_{13})} +\Bigg[ \frac{p(p-\dlta_{13})\dlta_{34}}{(-1+\dlta_{03})(-1+\dlta_{24})} + (3\leftrightarrow 4)\Bigg]+ (0\leftrightarrow 1) \Bigg]\notag\\
& \!\!\!\!+  \frac{2p}{(p-3)!}\Bigg[ \frac{2p \dlta_{34}}{(1-\dlta_{03})} + \frac{2p \dlta_{34}}{(1-\dlta_{04})} + (0\leftrightarrow 1) + (0\leftrightarrow 2) \Bigg] \,,
\end{align}
with $\widetilde{\cal M}_{\hat\sigma^2}= \widetilde{\cal M}_{\hat\tau^2}\Big|_{ 3 \leftrightarrow 4}$, and  then
\begin{align}
\label{Mtau}
\widetilde{\cal{M}}_{\hat\tau}=\frac{2p}{(p-2)!} \Bigg[ \widetilde{\cal{M}}^{\rm single}_{\hat\tau}  + \widetilde{\cal{M}}^{\rm double}_{\hat\tau}  + (0\leftrightarrow 1) +2p(p-2)\Bigg]\qquad;\qquad 
\widetilde{\cal{M}}_{\hat\sigma}=\widetilde{\cal M}_{\hat\tau}\Big|_{ 3 \leftrightarrow 4} \,,
\end{align}
where
\begin{align}
\!\!\!\!\!\! \widetilde{\cal{M}}^{\rm single}_{\hat\tau} =&\  
\frac{p(p-1)\dlta_{2,4}}{(1-\dlta_{03} ) } + \frac{2(p+1)\dlta_{3,4}}{(1-\dlta_{02} )} - \frac{p\dlta_{01}+p(p-2)(1+\dlta_{34})}{1-\dlta_{23}}  + \notag\\
 & \ \ - \frac{ 2 (\dlta_{04}+\dlta_{34}) +p ( 2(p-2) -(p-1) (\dlta_{13}+\dlta_{34}) -\dlta_{04}+\dlta_{12}) }{ (1-\dlta_{04}) } 
 \end{align}
 \begin{align}
\!\!\!\!\!\!  \widetilde{\cal{M}}^{\rm double}_{\hat\tau} =&\ 
 \frac{2p \dlta_{24}\dlta_{34} }{(-1+\dlta_{02})(-1+\dlta_{13})}
 + \frac{ p(\dlta_{02}+\dlta_{03})(\dlta_{24}+\dlta_{34}) +p(p-2)\dlta_{34}(\dlta_{14}-p) +p^2(\dlta_{34}+\dlta_{24})^2}{(-1+\dlta_{04})(-1+\dlta_{23})} + \\[.2cm]
 &\ \ 
  \frac{2p\dlta_{13}\dlta_{34}-2( \dlta_{01}+\dlta_{02}+p-2)(\dlta_{14}+\dlta_{24})}{(-1+\dlta_{04})(-1+\dlta_{12})} 
 +\frac{p(\dlta_{01}+\dlta_{13})(\dlta_{34}+p\dlta_{13}-(p-1)\dlta_{04})+p(p-2)\dlta_{34}\dlta_{24}}{(-1+\dlta_{03})(-1+\dlta_{14})} \notag
\end{align}
Finally,
\begin{align}
\label{Mone}
\widetilde{\cal{M}}_{1}=\frac{2p}{(p-2)!}&\left[\left[\frac{2 p\left(\tilde{\delta }_{01}+\tilde{\delta }_{04}\right) \left(\tilde{\delta }_{04}+\tilde{\delta
   }_{24}\right)}{\left(\tilde{\delta }_{02}-1\right) \left(\tilde{\delta }_{14}-1\right)}+(0\leftrightarrow 1,3\leftrightarrow 4)\right]+\frac{p \tilde{\delta
   }_{24}}{\tilde{\delta }_{03}-1}+\frac{p \tilde{\delta }_{23}}{\tilde{\delta }_{14}-1}-7p+6\right.\\
   &\left.+(p-2)
   \left[\frac{(3-p) \tilde{\delta }_{01}}{\tilde{\delta }_{34}-p+3}+\frac{2 \left(\tilde{\delta
   }_{23}+\tilde{\delta }_{24}-1\right) \left(\frac{\tilde{\delta }_{04}+\tilde{\delta }_{24}}{\tilde{\delta
   }_{02}-1}+\frac{\tilde{\delta }_{13}+\tilde{\delta }_{23}}{\tilde{\delta }_{12}-1}-1\right)}{\tilde{\delta
   }_{34}-p+2}\right]\right.\nonumber\\
   &\left.+\frac{2 \left(\tilde{\delta }_{01}+2 \left(\tilde{\delta }_{23}+\tilde{\delta
   }_{24}\right) \left(\frac{\tilde{\delta }_{04}+\tilde{\delta }_{24}}{\tilde{\delta
   }_{02}-1}+\frac{\tilde{\delta }_{13}+\tilde{\delta }_{23}}{\tilde{\delta
   }_{12}-1}\right)\right)}{\tilde{\delta }_{34}-p+1}+\frac{p^2 \left(\tilde{\delta }_{01}+\tilde{\delta
   }_{04}\right) \left(\tilde{\delta }_{01}+\tilde{\delta }_{13}\right)}{\left(\tilde{\delta }_{03}-1\right)
   \left(\tilde{\delta }_{14}-1\right)}\right.\nonumber\\
   &\left.+2 (p-1) \left[\frac{\tilde{\delta }_{13}}{\tilde{\delta
   }_{02}-1}+\frac{\tilde{\delta }_{04}}{\tilde{\delta }_{12}-1}\right]+(3\leftrightarrow 4)\right]+\frac{2p}{(p-2)! (1-\tilde{\delta }_{02})}\left[\frac{2 \left(\tilde{\delta }_{03}+\tilde{\delta }_{04}\right) \left(\tilde{\delta }_{23}+\tilde{\delta
   }_{24}\right)}{\tilde{\delta }_{34}-p+1}\right.\nonumber\\
   &\left.+(p-1) p \left[\frac{\left(\tilde{\delta }_{01}+\tilde{\delta }_{03}\right) \left(\tilde{\delta
   }_{12}+\tilde{\delta }_{23}\right)}{\tilde{\delta }_{13}-1}+(3\leftrightarrow 4)\right]+(p-2) \Bigg[2 \left(p \left(1-\tilde{\delta }_{02}\right)+1\right)\right.\Bigg.\nonumber\\
   &\left.\Bigg.+\frac{4 \left(\tilde{\delta
   }_{03}+\tilde{\delta }_{04}-1\right) \left(\tilde{\delta }_{23}+\tilde{\delta }_{24}-1\right)}{\tilde{\delta
   }_{34}-p+2}+\frac{(p-3) \left(\tilde{\delta }_{03}+\tilde{\delta }_{04}-2\right) \left(\tilde{\delta
   }_{23}+\tilde{\delta }_{24}-2\right)}{\tilde{\delta }_{34}-p+3}\Bigg]+(0\leftrightarrow 1)\right]\,\nonumber.
   \end{align}
It is straightforward to check that after the limit $\vec{Y}_0\to\vec{Y}_1$ in the R-symmetry polarization vectors, there are no poles in $\tilde{\delta}_{01}$ in $\widetilde{\cal{M}}_{\hat\sigma^m \hat\tau^n}$.

When considering the double-particle limit from the five-point Mellin amplitude,  it is necessary to integrate over three of the five independent Mellin variables, $\tilde{\delta}_{ij}$. To perform these integrations, we first massage the expressions~\eqref{expression5pttau2_app} to \eqref{Mone} in such a way that they can be written as linear combinations of ratios of gamma functions as the ones appearing on the right hand side of~\eqref{eq:DfuncstoMellin}. Note that each of those terms will then have a natural correspondence with a D-function in position space in which it is straightforward to take the OPE limit $\vec{X}_0\to \vec{X}_1$ with~\eqref{eq:OPEinDfuncs}. We find that also in Mellin space the integrations are simpler to do when the terms of the Mellin amplitude are split in this way.

To do so, as stated before, we use the identities~\eqref{eq:denomtogammas} and \eqref{eq:numtogammas} in the expanded version of the Mellin components. Note, furthermore, that this manipulation can be done in terms of the unconstrained Mellin variables with the advantage of dealing with simpler expressions. For concreteness, let us consider the component ${\widetilde{\cal M}}_{\tau^2}$. For the case at hand, where the reduced correlator is only a function of cross ratios $U$ and $V$, we note from~\eqref{notation4pt_intro} and \eqref{intriligator} that this component completely determines the ${\cal H}_{[2^2]2pp}$ up to an extra power of $U$.~\footnote{It is however interesting to repeat the strategy for all the Mellin components. We check the  consistency of this procedure with \eqref{notation4pt_intro}. } We find
 \begin{align}
 \label{appMtoDfunc_testcase}
&
\!\!\!\!\!\!\!\!\!\frac{(p-3)!}{2p}\prod_{i<j}\Gamma[ \dlta_{ij} ] {\cal M}_{\hat\tau^2}=\nonumber  \\[.2cm]
&\
 + p \Gamma[ \dlta_{01}] \Gamma[ \dlta_{12}]\Gamma[ \dlta_{03} ]( \Gamma[1+\dlta_{02}]\Gamma[-1+\dlta_{04}]+\Gamma[\dlta_{02}]\Gamma[ \dlta_{04} ])\Gamma[\dlta_{14}]\Gamma[\dlta_{13}]\Gamma[-1+\dlta_{23}]\Gamma[\dlta_{24}]\Gamma[1+\dlta_{34}] + \notag\\[-.0cm]
&\  +p\Gamma[ \dlta_{01}] \Gamma[ \dlta_{12}] \Gamma[ \dlta_{03} ] \Gamma[ \dlta_{02} ]\Gamma[ -1+\dlta_{04} ]\Gamma[\dlta_{14}]( \Gamma[\dlta_{13}]\Gamma[ \dlta_{23} ]-\Gamma[1+\dlta_{13}]\Gamma[-1+\dlta_{23}])\Gamma[\dlta_{24}]\Gamma[1+\dlta_{34}] + \notag\\
&\ -p \Gamma[ \dlta_{01}] \Gamma[ \dlta_{12}]\Gamma[ -1+\dlta_{03} ] \Gamma[ \dlta_{02} ]\Gamma[\dlta_{04}]\Gamma[-1+\dlta_{14}]\Gamma[\dlta_{13}]\Gamma[\dlta_{23}]\Gamma[1+\delta_{24}]\Gamma[1+\dlta_{34}] +\notag\\[-.15cm]
&\ -\,\Gamma[ \dlta_{01}] \Gamma[ \dlta_{12}] \Gamma[\dlta_{03}]\Gamma[\dlta_{02}]\Gamma[-1+\dlta_{04}]\Gamma[\dlta_{14}]\Gamma[\dlta_{13}]\Gamma[-1+\dlta_{23}]\Gamma[\dlta_{24}]\Gamma[2+\dlta_{34}]+(0\leftrightarrow 1) \,,
\end{align}
where we note that full expression takes the form of a sum of $6+6$ five-point D-functions. Using~\eqref{on_shell_constraints_5}, we restrict to 5 independent Mellin variables, namely ${\dlta}_{01},{\dlta}_{12},{\dlta}_{14}$ and ${\dlta}_{34},{\dlta}_{23}$,  and we proceed to integrate over the first three of them.
The initial integration is straightforwardly performed by capturing the residue of the simple pole of $\Gamma(\tilde{\delta}_{01})$ at $\delta_{01}=0$. As discussed, this residue represents the leading contribution in the OPE limit $\vec{X}_0\to \vec{X}_1$ after selecting $\vec{Y}_0=\vec{Y}_1$.  After the above manipulations, it is trivial to check that the remaining integrations in $\dlta_{12},\dlta_{14}$ can be done by using the first Barnes' lemma
\begin{equation}
\label{eq:Barneslemma}
\frac{1}{2\pi i}
\int_{-i\infty}^{i\infty}
\Gamma(a+s)\,\Gamma(b+s)\,\Gamma(c-s)\,\Gamma(d-s)\, ds
=
\frac{
\Gamma(a+c)\,\Gamma(a+d)\,\Gamma(b+c)\,\Gamma(b+d)
}{
\Gamma(a+b+c+d)
}\,.
\end{equation}
In the discussion above, it may seem that we neglected the contributions of $X_{ij}^{-2\dlta_{ij}}$ appearing in~\eqref{expression5pt} in the integrations over $\dlta_{01},{\dlta}_{12},{\dlta}_{14}$ but in fact it is easy to check that after multiplying by a prefactor one gets
\begin{equation}
\left(\prod_{0\le i<j\le 4} X_{ij}^{-2{\delta}_{ij}}\right)\times X_{12}^4 X_{13}^2 X_{14}^2 X_{34}^{2p-2} \to U^{p-1-\dlta_{34}} V^{-\tilde{\delta}_{23}}\,,
\end{equation}
where to get from one side to the other, we take $\tilde{\delta}_{01}\to 0, \vec{X}_0\to \vec{X}_1$ and impose the on-shell constraints on the Mellin variables. After redefining $\dlta_{23}=-t$ and $\dlta_{34}=-s-4+p$, one recovers the powers of the cross ratios observed in \eqref{sampeldbarnescalMtau2} and \eqref{eq:Hfrom5pts}, while the reduced Mellin amplitude follows from the integration procedure explained before.

\providecommand{\href}[2]{#2}\begingroup\raggedright\endgroup


\end{document}